\documentclass{aa}  

\usepackage{graphicx}
\usepackage{txfonts}
\usepackage{color}
\usepackage{needspace}
\usepackage{placeins}
\begin{document} 

 \newbox\grsign \setbox\grsign=\hbox{$>$} \newdimen\grdimen \grdimen=\ht\grsign
 \newbox\simlessbox \newbox\simgreatbox
 \setbox\simgreatbox=\hbox{\raise.5ex\hbox{$>$}\llap
     {\lower.5ex\hbox{$\sim$}}}\ht1=\grdimen\dp1=0pt
 \setbox\simlessbox=\hbox{\raise.5ex\hbox{$<$}\llap
      {\lower.5ex\hbox{$\sim$}}}\ht2=\grdimen\dp2=0pt
 \def\simgreater{\mathrel{\copy\simgreatbox}}
 \def\simless{\mathrel{\copy\simlessbox}}
 \newbox\simppropto
 \setbox\simppropto=\hbox{\raise.5ex\hbox{$\sim$}\llap
     {\lower.5ex\hbox{$\propto$}}}\ht2=\grdimen\dp2=0pt
 \def\simpropto{\mathrel{\copy\simppropto}}

 \def\lg{{${\rm \log g}$}}
\newcommand{\logg}{$\log$ $(g)$}
\newcommand{\drvm}{$\Delta \text{RV}_{\text{max}}$}
\newcommand{\kms}{km s$^{-1}$}

\def\red{\color{red}}
\def\cyan{\color{cyan}}
\def\green{\color{green}}

\title{Carbon enrichment in APOGEE disk stars as evidence of mass transfer in binaries}


\author{{Steve Foster}\inst{1}\fnmsep\thanks{steve.foster161@gmail.com} \and
{Ricardo P. Schiavon} \inst{1} \and
{Denise B. de Castro} \inst{1,2} \and
{Sara Lucatello} \inst{3} \and
{Christine Daher} \inst{4} \and
{Zephyr Penoyre} \inst{5,6} \and
{Adrian Price-Whelan} \inst{7} \and
{Carles Badenes} \inst{8} \and
{J. G. Fernández-Trincado} \inst{9} \and
{D. A. García-Hernández} \inst{10,11} \and
{Jon Holtzman} \inst{12} \and
{Henrik J\"onsson} \inst{13} \and
{Matthew Shetrone} \inst{14}
}

\institute{
{Astrophysics Research Institute, Liverpool John Moores
University, 146 Brownlow Hill, Liverpool L3 5RF, United Kingdom} \and
{Funda\c c\~ao Get\'ulio Vargas - Rio de Janeiro - 22250-900, RJ, Brazil} \and
{INAF – Osservatorio Astronomico di Padova, vicolo dell’Osservatorio 5, I-35122 Padova, Italy} \and
{Center for Cosmology and AstroParticle Physics (CCAPP), The Ohio State University, 191 W. Woodruff Ave., Columbus, OH 43206, USA} \and
{Institute of Astronomy, Madingley Rd, Cambridge CB3 0HA, United Kingdom} \and
{Leiden Observatory, Leiden University, PO Box 9513, 2300 RA, Leiden, the Netherlands} \and
{Center for Computational Astrophysics, Flatiron Institute, 162 5th Ave, New York, NY 10010, USA} \and
{Department of Physics and Astronomy, University of Pittsburgh,Allen Hall, 3941 O'Hara St, Pittsburgh PA 15260, United States} \and
{Instituto de Astronom\'ia, Universidad Cat\'olica del Norte, Av. Angamos 0610, Antofagasta, Chile} \and
{Instituto de Astrofísica de Canarias, 38205 La Laguna, Tenerife, Spain} \and
{Universidad de La Laguna (ULL), Departamento de Astrofísica, E-38206 La Laguna, Tenerife, Spain} \and
{Department of Astronomy, New Mexico State University, PO BOX 30001, Las Cruces, NM 88003United States} \and
{Materials Science and Applied Mathematics, Malm\"o University, SE-205 06 Malm\"o, Sweden} \and
{University of California Observatories
University of California Santa Cruz
Santa Cruz, CA 95064, USA}
}


   \date{Received March 18, 2024; accepted July 14, 2024}

 
  \abstract
   {Carbon abundances in first-ascent giant stars are usually lower
than those of their main-sequence counterparts.
At moderate metallicities, stellar evolution of single stars cannot account for the existence of red-giant branch stars with enhanced carbon abundances. The phenomenon is usually interpreted as resulting from past mass transfer from an evolved binary companion now in the white dwarf evolutionary stage.}
   {We aim to confirm the links between [C/O] enhancement, $s-$process element enhancement and binary fraction using large-scale catalogues of stellar abundances and probable binary stars.}
   {We use a large data set from the 17$^{th}$ data release of the SDSS-IV/APOGEE~2 survey 
to identify carbon-enhanced stars in the Galactic disk. We identify a continuum of carbon enrichment throughout three different sub-populations of disk stars and explore links between the degree of carbon enrichment and binary frequency, metallicity and chemical compositions. }
   {We verify a clear correlation between binary frequency and enhancement in the abundances of both carbon and cerium, lending support to the scenario whereby carbon-enhanced stars are the result of mass transfer by an evolved binary companion. In addition, we identify clustering in the carbon abundances of high-$\alpha$ disk stars, suggesting that those on the high metallicity end are likely younger, in agreement with theoretical predictions for the presence of a starburst population following the gas-rich merger of the Gaia-Enceladus/Sausage system.}
   {}


   \maketitle
%

\section{Introduction} \label{sec:intro}

The chemical properties of Galactic stellar populations provide
important constraints on scenarios for the formation and evolution
of the Galaxy.  During the main-sequence stage of stellar evolution,
the chemical composition of a star's atmosphere reflects that of
the interstellar medium it was formed from with some variation caused by atomic diffusion within the star \citep[e.g.,][]{Wachlin2011, Moedas2022}.  However, that situation changes
once the star reaches the red giant branch (RGB) when the products of core main sequence and/or shell H-burning are mixed with the photospheric material, changing its chemical composition.  

In the  case of low- and intermediate-mass ($\lesssim$ $8 M_{\odot}$) stars, the first dredge-up episode affects the surface composition of several light elements. In particular, the abundance of C has a modest decrease ($\sim$0.1 dex) and its isotopic ratio, $^{12}$C/$^{13}$C, lowers considerably, from 90 to
about 20-30 \citep[e.g.,][]{charbonnel98}. The effectiveness of this dredge-up decreases with increasing metallicity \citep[e.g.,][]{Salaris15}.
Following this episode, a poorly understood process known as extra mixing takes place, lowering the C abundance even more ([C/Fe]$\sim$ $-$0.4) as the star climbs up the RGB. As a result of these processes, first-ascent giants are normally expected to present sub-solar [C/Fe]-abundance ratios \citep[e.g.,][]{busso1995}.



There are two classes of enrichment process that can raise carbon abundance
ratios in stellar atmospheres to substantially super-solar values.
So-called intrinsic enrichment is a result of stellar evolution
whereby carbon enrichment occurs during the Asymptotic Giant
Branch (AGB) phase when the by-products of shell He-burning are
dredged-up to the surface of the stars. When this occurs, the
atmospheres of AGB stars are also enhanced in $s$-process elements
such as barium \citep[e.g.,][]{AllenandBarbuy2006}.

On the other hand, extrinsic enrichment is responsible for
producing stars in the main sequence or RGB phase that are enhanced in
both carbon and $s$-process elements.  These include various classes
of chemically peculiar objects such as Ba stars \citep[e.g.,][]{mcclure90},
with near Solar metallicities, CH stars \citep{keenan42} and CH-like
stars \citep[e.g.,][]{Yamashita, zamora2009, Sperauskas}, with metallicity
$-1.5 <$ [Fe/H] $< -0.5$  and, at lower metallicity (typically,
[Fe/H] $< -1.5$), CEMP-$s$\footnote{Where CEMP stands for Carbon-Enhanced Metal-Poor stars.} stars \citep[e.g.,][]{beers05,Hansen16}.
The physical mechanism invoked to alter the chemical compositions of the atmospheres of these types of stars is mass transfer in a binary system.
According to this scenario, the primary star of the binary system evolves into its AGB phase and, as a result of the third dredge-up (TDU), its atmosphere becomes
enriched intrinsically in carbon and $s$-process elements \citep[e.g.,][]{Karakas22}.  In
binary systems that meet the necessary physical conditions, transfer
of chemically altered material can take place from the extended atmosphere
of the primary AGB star onto the secondary member, polluting its
atmosphere \citep[e.g.,][]{jorissen16}.  Once the primary star evolves away
from the AGB to the post-AGB phase and then to the white dwarf
phase, the process leaves behind a binary system composed of a
chemically peculiar star with enhanced abundances of carbon and
$s$-process elements, and a much fainter  white dwarf companion \citep[e.g.,][]{mcclure90}.

Observational evidence in support of the above scenario comes primarily from 
detection of white dwarf/dwarf carbon pairs \citep[e.g.,][]{Liebert,Heber,Si15}, ultraviolet excess
in Ba stars \citep[e.g.,][]{bohmvitense85} and the presence of radial velocity (RV)
variability in different classes of chemically peculiar stars
\citep[e.g.,][]{mcclure80,mcclure84,udry98,lucatello05,deCastro16,jorissen16, Cseh2018,Cseh2022,Stancliffe21,Whitehouse} which
is indicative of the presence of a faint binary companion.

The different types of extrinsically enriched chemically peculiar
stars are each associated to specific stellar populations with some overlap.  For
instance, Ba stars are metal rich ([Fe/H] $\simgreater$ -0.5) and usually found in the disk \citep{deCastro16}, whereas CH stars
are typically found in the halo, with $-1.5\lesssim$ [Fe/H] $\lesssim-1$
\citep{keenan42}. CEMP-$s$ stars are also members of the halo population, exhibiting much lower metallicities \citep[e.g.,][]{beers05}.
Within the metal-poor halo, the fraction of carbon-rich stars has been found to be higher at very low metallicity \citep[e.g.,][]{norris97,lee13}.
It has been previously proposed that the increased frequency of carbon-rich
stars at lower-metallicity is caused by an increase in the
efficiency of carbon production in low metallicity AGB stars \citep{Marigo2007}.
However, recent results from the AEGIS survey point in a different direction. \citet{Yoon18} distinguished CEMP-$s$ stars from the other dominant class of carbon-enhanced objects in the halo \citep[CEMP-no stars;][]{beers05} on the basis of differences in their absolute carbon abundances, $A$(C), demonstrating that there appears to be little increase in the fractions of CEMP-$s$ stars with decreasing metallicity. They contend that the phenomenon is instead  related to the dramatic increase in the fraction of CEMP-no stars at low-metallicity.
\citet{Andersen16} and also \citet{Hansen16a} argue that the mechanism that produces CEMP-no stars may also be quite different from that which produces CEMP-$s$ stars and may not even involve mass transfer.

It has been argued that mass-transfer efficiency also may be a function
of metallicity \citep[e.g.][]{Hansen16, Karakas02}, being more efficient in lower metallicity stars. 
\cite{Moe19} report a strong anti-correlation between metallicity and binarity for solar-type stars.
However, the precise mass-transfer mechanism
responsible for the production of extrinsic carbon-enhanced stars is
still a matter of debate. 
Some authors claim that wind accretion
is more likely to occur in systems with higher orbital periods,
while in short-period binaries a thermally pulsing AGB primary may
fill the system's Roche lobe, causing more efficient mass transfer to happen
\citep[e.g.,][]{han95}.  
Other mechanisms such as pollution by carbon-enriched interstellar medium cannot be ruled out completely \citep[e.g.][]{Crossfield19}. 
Improved mass-transfer simulations
have been performed to try to reproduce the period and eccentricity distributions of carbon-rich stars \citep[e.g.,][]{starkenburg14,staff16, Abate18, Liu2017}.

Predominantly, and with the exception of Ba stars, studies of carbon-rich stars have focused on halo objects.
In contrast, in this paper we report the identification of a large population of stars in the Galactic disk by the APOGEE survey.  
By analysing the RV variability and residual astrometric motions in relation to the degree of carbon enrichment of stars in our sample, we conclude that there is a clear correlation between carbon enrichment and binarity.
We also confirm a corresponding trend in cerium abundance, consistent with mass transfer from an evolved star rich in $s$-process elements.

This paper is organised as follows. Sect.~\ref{sec:data} describes the
data and sample employed.  Sect.~\ref{sec:crich}  presents
the criteria adopted to define carbon-enhanced and carbon-normal stars.
Sect.~\ref{sec:analysis} discusses the 
binary incidence of carbon-enhanced stars.
Conclusions are summarised
in Sect. \ref{sec:conclusions}.

\section {Data and sample} \label{sec:data} 

\subsection{APOGEE parameters and initial selection}
The data employed in this analysis were obtained as part of the
Apache Point Observatory Galactic Evolution Experiment (APOGEE),
which was part of Sloan Digital Sky Surveys III and IV \citep[][henceforth, simply SDSS]{Eisenstein2011,Blanton2017}.
For a description of the APOGEE survey, see \cite{majewski17}; and
for further details on the data and the data-reduction
procedures, see \cite{Holtzman2018} and \cite{Jonsson2020}.  A description of the APOGEE
Stellar Parameters and Chemical Abundances Pipeline\footnote{{\tt ASPCAP} compares stellar spectra to a large library of synthetic spectra to determine chemical abundances.} ({\tt ASPCAP})
can be found in \cite{garcia16}.  We used a value-added version, AstroNN, of APOGEE's 17$^{th}$ data release (DR17). This contains additional data on stellar kinematics, calculated using the {\tt galpy} package \citep{Bovy2015} and assuming a \cite{McMillan2017} potential, as well as stellar ages \citep{Leung2019}. 
The specific data upon which our results are based consist of the measured stellar abundances of Fe, C, N, O, Mg, Al, Ce, and radial velocities (RVs). 
However, the DR17 catalogue also contains duplicate entries for around 10\% of stars which have been observed for different aspects of the APOGEE programme using different field plates and different fibre-optic connections and so de-duplication of the catalogue is a necessary first step.
For this exercise, the duplicate entry with the best S/N ratio was used and other entries discarded.
As described below, additional catalogues were used to add binarity information and to help eliminate globular cluster members, variable stars, and AGB stars from the sample.

This paper focuses on the stellar populations of the Galactic
disk, including thin- and thick-disk stars, and for that reason our
sample is restricted to stars within 2 kpc{\footnote{Distances used in this paper are the weighted distances given in the AstroNN version of the APOGEE DR17 catalogue}} of the Galactic mid-plane. To remove stars in the Galactic bulge, stars within 4 kpc of the Galactic centre were eliminated.
Also, stars outside of the Milky Way itself, i.e. stars more than 25 kpc from the Galactic centre and known GC members were excluded.
As these criteria are based on distances, any stars for which the parallax error in Gaia's early-third data release (EDR3) was more than 20\% of the parallax were also excluded. 
Finally, there are some stars belonging to Milky Way satellites which, due to inaccurate distances, are included in our disk sample. 
These have characteristic abundances in the [Mg/H]-[C/Fe] plane and were removed by excluding stars with [C/Fe] $<$ -0.25 and [Mg/H] $>$ +0.3.

To maximise the precision of the stellar abundances, we further
limited the sample to stars whose spectra have S/N $\geq 70$ per
pixel.  A further cut based on $T_{\rm eff}$ was imposed, $3250 < T_{eff} < 4700$~K.
On the warm end, the sample is limited because carbon abundances in DR17 are based on molecular lines which become
vanishingly small in the spectra of warm stars, especially at lower
metallicity, leading to large uncertainties.  
In addition, those stars whose [C/Fe] abundance carried uncertainties of 0.05 dex or more were eliminated.
Stars at the cool end of this range may produce significant amounts of dust, which is not included in the ASPCAP analysis.
We expect most of these stars to be AGB stars and have taken steps to eliminate them from the sample as described in Sec. \ref{AGB}.

\begin{figure}
\centering
\includegraphics [width= \columnwidth]
{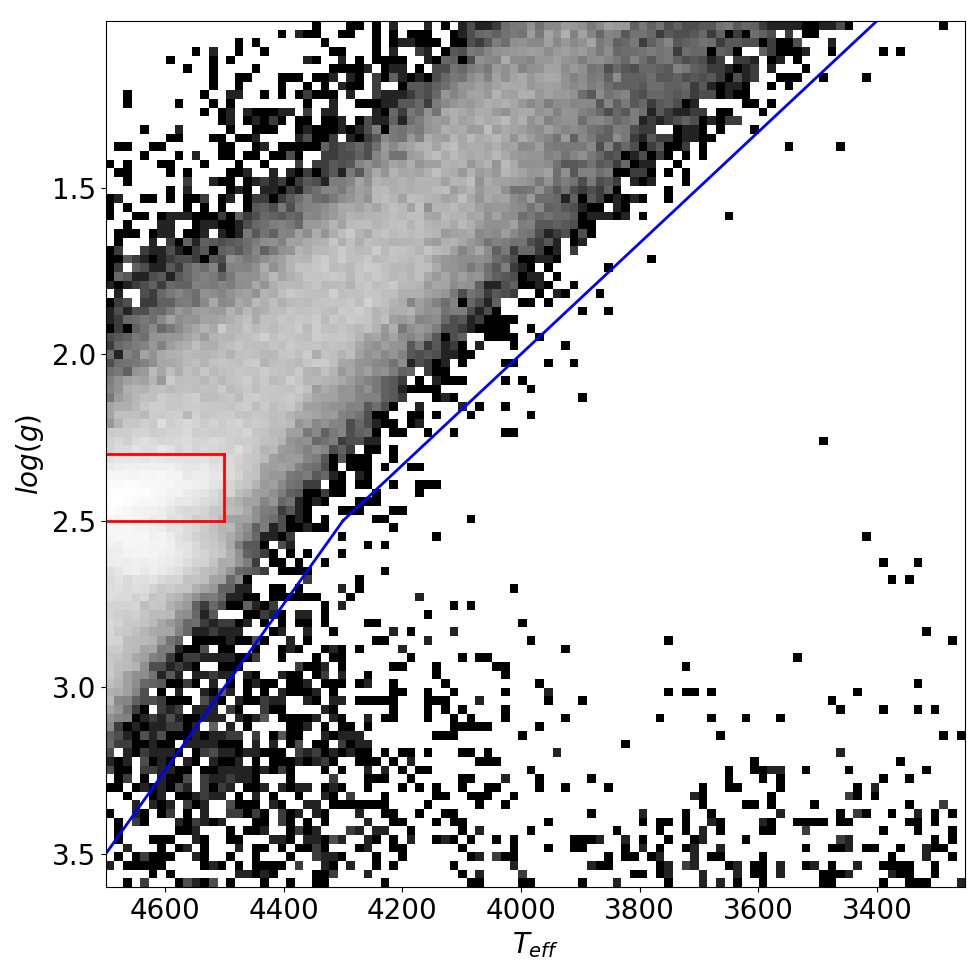}    
\caption{Distribution of {\lg} vs $T_{\rm eff}$ showing anomalous stars with suspected systematic ASPCAP errors. Stars below and to the right of the blue lines were subsequently excluded. Also excluded where Red Clump stars, seen as the white area near the centre of the y-axis between {\lg} = 2.3 to 2.5 and surrounded by the red lines. The white area below this is the ''red bump'' in the RGB.}
\label{fig:loggteff}
\end{figure}

To limit our sample to giants, we further restricted it to stars with $1.0 <$ {\lg}  $< 3.6$.
Consideration of the distribution (see Fig. \ref{fig:loggteff}) of {\lg} vs $T_{\rm eff}$ showed a number of very cool stars with gravities that do not correspond to the regions of interest in the H-R diagram.
These stars are suspected to be the result of systematic errors in the ASPCAP processing and were excluded from the final sample set.

Furthermore, it was clear from this plot that our sample contained a number of Red Clump (RC) stars, visible as the white area enclosed by the red lines near the centre of the y-axis in Fig. \ref{fig:loggteff}.
These stars are in a core helium burning phase and have deep convective envelopes and convective cores. (See e.g. \cite{Girardi}).
We had two concerns about RC stars. 
Firstly, convective overshoot may dredge up triple-$\alpha$ process  fusion products that affect the photospheric carbon abundance. 
Although \citet{Spoo22} find no evidence that this process leads to enhanced levels of carbon in the stellar envelope, our approach to defining carbon-enhanced stars is sensitive to even small levels of enhancement which may contaminate our sample with intrinsically enriched stars.
Secondly, the ages of RC stars in the value-added APOGEE catalogue show strong clustering which produces a marked secondary peak in the age distribution of our sample stars. The stellar ages in APOGEE are derived from C and N abundances and the onset of the He flash in RC stars changes those abundances, making the age determination for these stars unreliable. 
Consequently, as a precautionary measure to avoid these problems, we have chosen to remove all RC stars.
We did this by excluding those stars with {\lg} between 2.3 and 2.5 and a $T_{eff}$ $>$ 4,500K.

In addition to the above criteria (Table \ref{tab:criteria}), we removed from the sample stars
which were flagged for {\tt STAR\_BAD} in the {\tt ASPCAPFLAG} bitmask, or flagged for {\tt BAD\_PIXEL}, {\tt COMMISSIONING}, {\tt VERY\_BRIGHT\_NEIGHBOUR} or {\tt BAD\_RV\_COMBINATION} in the {\tt STARFLAG} bit mask in the APOGEE database, or which have any potential errors flagged against the elemental abundances being used.

\begin{table}
    \caption{Summary of initial selection criteria}
    \centering
    \begin{tabular}{l c c}
        Criterion & Min & Max \\
        \hline
        \lg & 1.0 & 3.6 \\
        $T_{eff}$ & 3250K & 4700K \\
        Gaia parallax error & & 20\% \\
        {S/N} & 70 & \\
        \tt{ASPCAPFLAG} &  \tt{STAR\_BAD}\\
        \tt{STARFLAG} & (see text) \\
        Galactocentric distance & 4 kpc & 25 kpc \\
        Solar distance & & 4 kpc \\
        Height above/below Galactic plane & -2.0 kpc & 2.0 kpc\\
        & & \\
    \end{tabular}
    \tablefoot{These criteria are used to filter stars in APOGEE DR17 prior to elimination of stars with unusable abundances, GC members, RC stars, AGB stars, and stars with suspected systematic errors.}
    \label{tab:criteria}
\end{table}

\subsection{Summary of APOGEE data and stellar populations} \label{DataSummary}

The net result of this and subsequent filtering of the 733,900 stars in the APOGEE catalogue produced a final sample of 58,842 stars as summarized in Table \ref{table:sum1}. Please refer to Sects. \ref{AGB} and \ref{Stellarpops} for further explanation of exclusions and stellar populations.

\begin{table}
\caption{Summary of rejections from the APOGEE DR17 data set.}
\begin{tabular}{l r}
{Reasons for rejection} &  {} \\
\hline
{Duplicate APOGEE entry:} & {76,461} \\
\hline
{{\lg} not in range:}	& {285,312} \\
{$T_{eff}$ not in range:} &	{166,013} \\
{More than 4kpc from Sun:} &	{36,682} \\
{\tt STAR\_BAD flag(s) set:}	& {35.151} \\
{Poor S/N:}	& {18,939} \\
{Probable Red Clump star:} & {17,173} \\
{Within 4.0 kpc of Galactic centre:} & 	{16,381} \\
{More than 2.0 kpc above/below Galactic plane:}	& {9,169}\\
{Known GC member:} & 	{6,012} \\
{Possible AGB/YSO/PNe (IR excess method):} &	{4,243} \\
{Inaccurate distance:} & 	{1,457} \\
{{\lg} vs. $T_{eff}$ forbidden zone:} &	{962} \\
{Too far from Galactic centre ($>$25.0 kpc):} & 	{684} \\
{Possible AGB in ASAS-SN variables:} & {202} \\
{Member of LMC/SMC/Fnx/Sgr dwarfs:} &	{109} \\
{[C/Fe] abundances invalid or unusable:} & 	{70} \\
{[Mg/Fe] abundances invalid or unusable:} &	{19} \\
{[C/Fe] abundance uncertainty $\geq$ 0.05 dex:} & {10} \\
{In AGB Star Catalogue:} & 	{9} \\
\hline
{Total rejections:} &  675,058\\
Stars in DR17: & 733,900 \\
Stars in sample: & 58,842\\ 
\hline
Low $\alpha$ stars: & 39,175 \\
Early high $\alpha$ stars: & 11,860 \\
Late high $\alpha$ stars: & 5,632 \\
Buffer zone stars: & 2,175 \\

\hline

& \\

\end{tabular}
\tablefoot{The final sample of 58,842 stars was used in the remainder of this paper. Also shown is a breakdown by the three populations used herein.}
\label{table:sum1}
\end{table}

\subsection {Binarity} \label {binaritybasis}
Historically, detection of binary systems depended on ground-based observations, the output of which were catalogues of a few hundred to a few thousand stars.
These existing catalogues are inadequate for this study because they constitute a small, inhomogeneous data set with little overlap with the APOGEE catalogue. 
For this exercise, we exploited the availability of large-scale databases to determine binarity of stars. 
Two complementary catalogues are used. 
The first catalogues uses an astrometric method based on residual position error (RUWE) from the Gaia EDR3 catalogue  \citep{Penoyre22}. 
This catalogue provides binarity probabilities determined by a modified RUWE parameter.
For this paper, we have adopted a minimum RUWE of 1.25 as a criterion for likely binarity.
We acknowledge three potential issues with this procedure.
Firstly, the accuracy of the RUWE method falls off with the accuracy of the Gaia parallax data.
For this reason, the data for this paper were further restricted to only those stars within 4 kpc of the Sun.
Secondly, the method is really only sensitive to shorter-period binaries, i.e.  with periods under 30 years, and so underestimates the true number of binary systems, biasing the sample against longer-period binaries. 
Thirdly, the movement of a foreground star near an unrelated and more distant background star can create a false appearance of binarity \citep{Holl2022}. 
However, such contaminants will not preferentially affect certain types of star. So, in the context of our analysis, the effect is to create a low, constant background "signal" which affects the precision but not accuracy of our results.

The second catalogue, which is also based on the APOGEE DR17 release, uses a radial velocity (RV) method \citep{PriceWhelan2020}.
The APOGEE survey visited each star in our sample between one and 35 times, with a mean of 2.9 visits. 
Since the RV errors reported by the APOGEE pipeline \citep{Nidever2015} are below 0.1 km/s, any star with peak-to-peak RV variability above a few km/s \citep[after the quality cuts described by][]{Mazzola2020} and far from the pulsational instability strip probably has, with a very high degree of confidence, a binary companion.

Though both of these catalogues are probabilistic, the complementary approaches provide an excellent check on the quality of the binarity determination.
The RUWE method preferentially detects binaries with high orbital inclinations as these have the highest residual astrometric motions.
The RV method preferentially detects those binaries with low orbital inclinations.
The number of stars considered binary systems by the combined approaches represented about 16.6\% of the sample and, encouragingly, about 12.5\% of these stars (i.e 2.1\% of the overall sample) were considered binary by both methods (See Table \ref{tab:binaries}).

\begin{table}
    \caption{Summary of binaries in sample}
    \centering
    \begin{tabular}{l r r c}
        & Count & Percentage& \\
        \hline
        Binaries found by RUWE method & 6.887  & 11.7\% & (1)\\
        Binaries found by RV method & 4,110 & 7.0\% &(2)\\
        Binaries found by both methods & 1,219 & 2.1\% & (3)\\ 
        Total binaries found (1)+(2)-(3) & 9,778 & 16.6\% &\\
        & & \\
    \end{tabular}
    \tablefoot{Number of stars in sample found to be binary by either the RUWE method, the RV method, or both. 
    Numbers are show both as counts and as a percentage of the total sample.
    }
    \label{tab:binaries}
\end{table}

We also considered the question of how complete this binarity information was. 
The fraction of stars which are binary, $f_b$,  varies with metallicity. \cite{2017MNRAS.469L..68G} note that $f_b$ varies from 25\%-50\% for dwarf stars in the temperature range covered by our sample.
There is a wide range of estimates of the binary fraction of disk stars, depending on both spectral type and metallicity.
\cite{Yuan2015} give an average of 39\% for FGK disk stars and \cite{Fischer1992} estimate a binary fraction of 42\% for M dwarfs. 
\citet{Badenes2018} find $f_b$ = 0.35 using APOGEE data. 
\citet{Moe19} give a range of 10\%-53\% depending on metallicity, mass and spectral class. 
With such wide ranges in the literature, we can only choose a broad average for $f_b$ and have selected 40\% as an indicative value.
Assuming that this value is typical also of our sample, we estimated that our combined methods were detecting approximately 41\% of the binaries that it contains.

The binary fraction of stars observed in each sub-population (see Sect. \ref{Stellarpops}) of the sample is summarized in Table \ref{tab:fb}.

\begin{table}
    \caption{Summary of observed binary fraction, $f_b$, by population.}
    \centering
    \begin{tabular}{l|r|r|l|l}
Population &	Base Count	& Binaries	& $f_b$	& ${\sigma}_{f_b}$\\
\hline
Low $\alpha$ &	39,174 &	6,610&	16.9\%	& 0.2\%\\
Early High $\alpha$ &	11,860 &	1,959	& 16.5\%  & 0.3\%\\
Late High $\alpha$ &	5,632 &	870 & 15.5\%  & 0.5\%\\
    \end{tabular}
    \tablefoot{See Sect. \ref{Stellarpops} for description of populations. The final column shows binomial standard deviation in binary fraction assuming a uniform distribution and expressed as a percentage of population for ease of comparison with $f_b$.}
    \label{tab:fb}
\end{table}

\subsection {AGB stars, planetary nebulae and young stellar objects} \label{AGB}
Our focus is on extrinsic carbon-enhanced stars, i.e. stars whose elevated carbon abundances cannot be explained by stellar evolution effects.
Thus, we have to eliminate as many objects as possible from our sample which may exhibit intrinsically elevated carbon abundances for well understood reasons not related to mass transfer.
These are, broadly, AGB stars and young stellar objects.

A great deal of the work on carbon-rich stars has concentrated on AGB stars \citep[e.g.,][]{Abia22}.
AGB stars can, depending on mass and evolutionary stage, have a numerical abundance of C/O $>1$ which meets the conventional definition of a carbon star.
These stars are still indirectly important to our study. We hypothesize that the more massive members of our binary systems must have, at some earlier epoch, passed through a carbon star AGB phase. 
It must have undergone considerable mass loss during the thermally-pulsing AGB (TP-AGB) stage and said mass must have had an elevated abundance of carbon.
Finally, the star must also by now have completed its AGB and planetary nebula phases and become a white dwarf.
If the secondary member of our binary pair has now reached the AGB stage then it, too, may well be carbon-enhanced and we will be unable to tell to what extent its carbon abundance is a result of extrinsic mass transfer as opposed to intrinsic nucleosynthesis. 
Purely on the ground of carbon abundance, we needed to ensure that our sample was contaminated by as few AGB stars as possible.

Moreover, AGB stars can also be long-period variables (LPVs) whose pulsations cause RV fluctuations that can mimic orbital motion. One of the binary catalogues that we use is based on RV fluctuations.
Consequently, we also wished to eliminate AGB contamination to minimize false positives when assessing binarity.

The APOGEE survey does target a number of YSOs (young stellar objects) within  the Perseus, Orion, and NGC 2264 star forming regions \citep[e.g.][]{Cody}.
Fortunately, the number of targeted objects is small and they can be identified by infrared excess, so we eliminated these from our sample as well.

Overall, we adopted a four-pronged approach to reduce contamination of the final data set by these objects as much as possible, the first three of which are as follows:

\begin{itemize}

\item The selection of stars with \lg~$\geq$~1 naturally eliminated many later-stage AGB stars.

\item Stars from a catalogue of known or suspected AGB stars from \citet{2021ApJS..256...43S} were removed.
However, the overlap of this catalogue with the APOGEE catalogue is very small, resulting in only 9 stars being removed from our sample.

\item Many suspected AGB stars were eliminated using the IR excess approach adopted by \cite{Schiavon2017}.  
This uses two colour excess methods, one of which is based on dereddened H, J and K magnitudes from APOGEE. 
The J, H and K magnitudes given in APOGEE are not dereddened, but the catalogue  lists the K-band extinction, $A_K$, for each star. The J- and H-band extinctions were calculated using the extinction law by \cite{Indebetouw}: 

$A_J$ = 2.50 $\times A_K$ and $A_H$ = 1.55 $\times A_K$.

\end{itemize}

\begin{figure}
\centering
\includegraphics[width= \columnwidth]{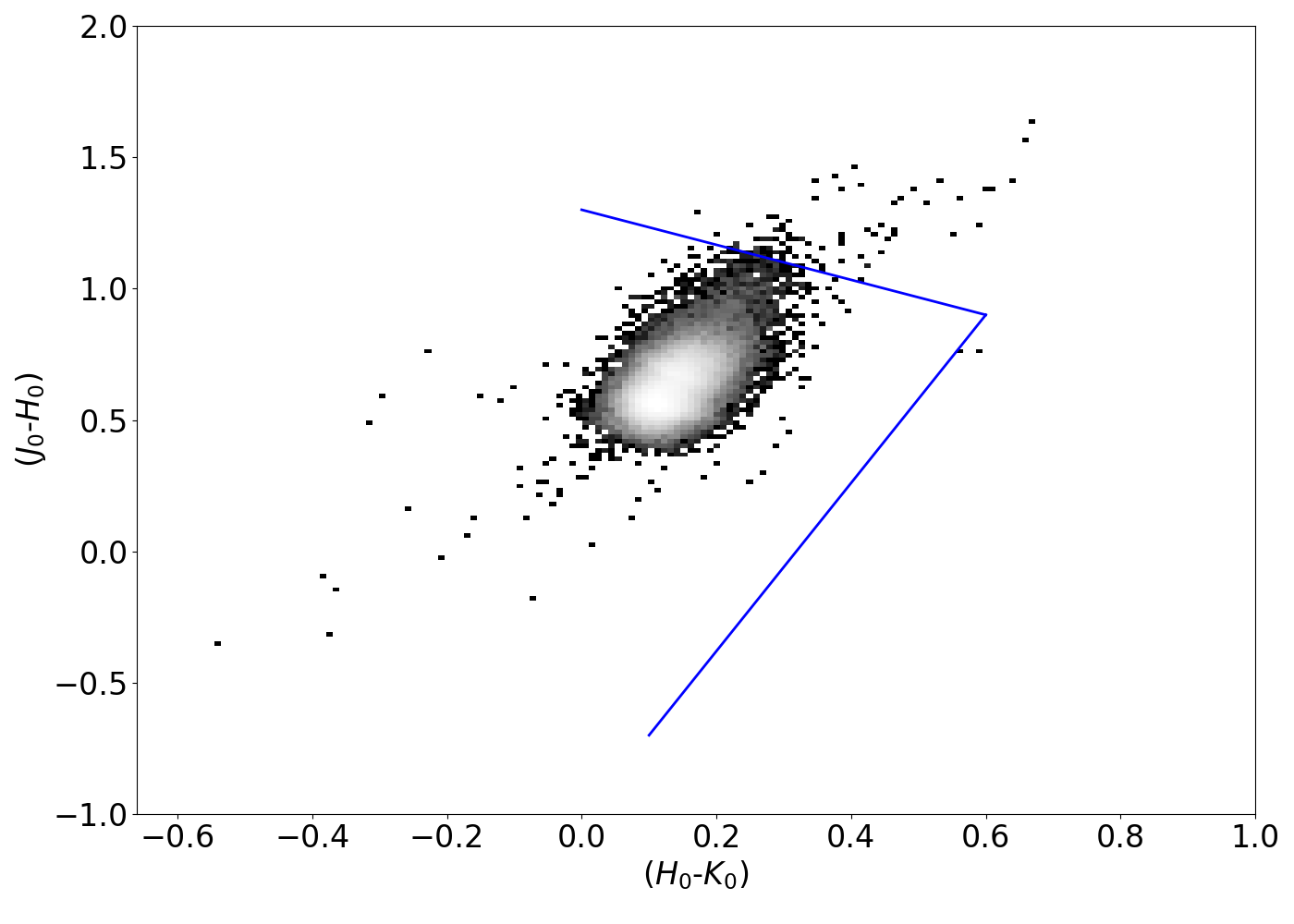}
\caption{Dereddened IR colour density plot of stars in sample based on method in \citet{Schiavon2017}. Stars above the upper blue line or to the right of the lower blue line are assumed to be AGB stars, YSOs or planetary nebulae and excluded from further analysis }
\label{fig:IRExcess}
\end{figure}

We used the dereddened magnitudes to plot a near infrared colour-colour diagram, shown in Fig. \ref{fig:IRExcess}.
The upper of the two blue lines shown in this plot separate AGB stars (above the line) from RGB+MS stars, whereas the lower blue line separates YSOs and PNe (to the right of the line) from RGB+MS stars. 
This method removed 4,478 stars from our sample, which should thus be unaffected by contamination by YSOs or PNe (planetary nebulae).

As a fourth check, we cross-matched the initial sample to the ASAS-SN catalogue of variable stars \citep{ASASSN} and noted stars which were marked as semi-regular or irregular variables. While stars on the RGB can be variable, the majority of semi-regular and irregular variables in this sample are likely to be AGB stars that have not been caught by the checks above. Consequently, as a conservative measure, a further 202 potential AGB stars were removed.

We discounted the use of IR colour excess in Spitzer IRAC (Infra-Red Array Camera) data as means of eliminating AGB stars owing to sparsity of the data in APOGEE. See Sec. \ref{App:IRAC} in the appendix for details.

\subsection{Stellar populations and chemical substructure in the high $\alpha$ disk} \label{Stellarpops}
\begin{figure}
    \includegraphics[width= \columnwidth]{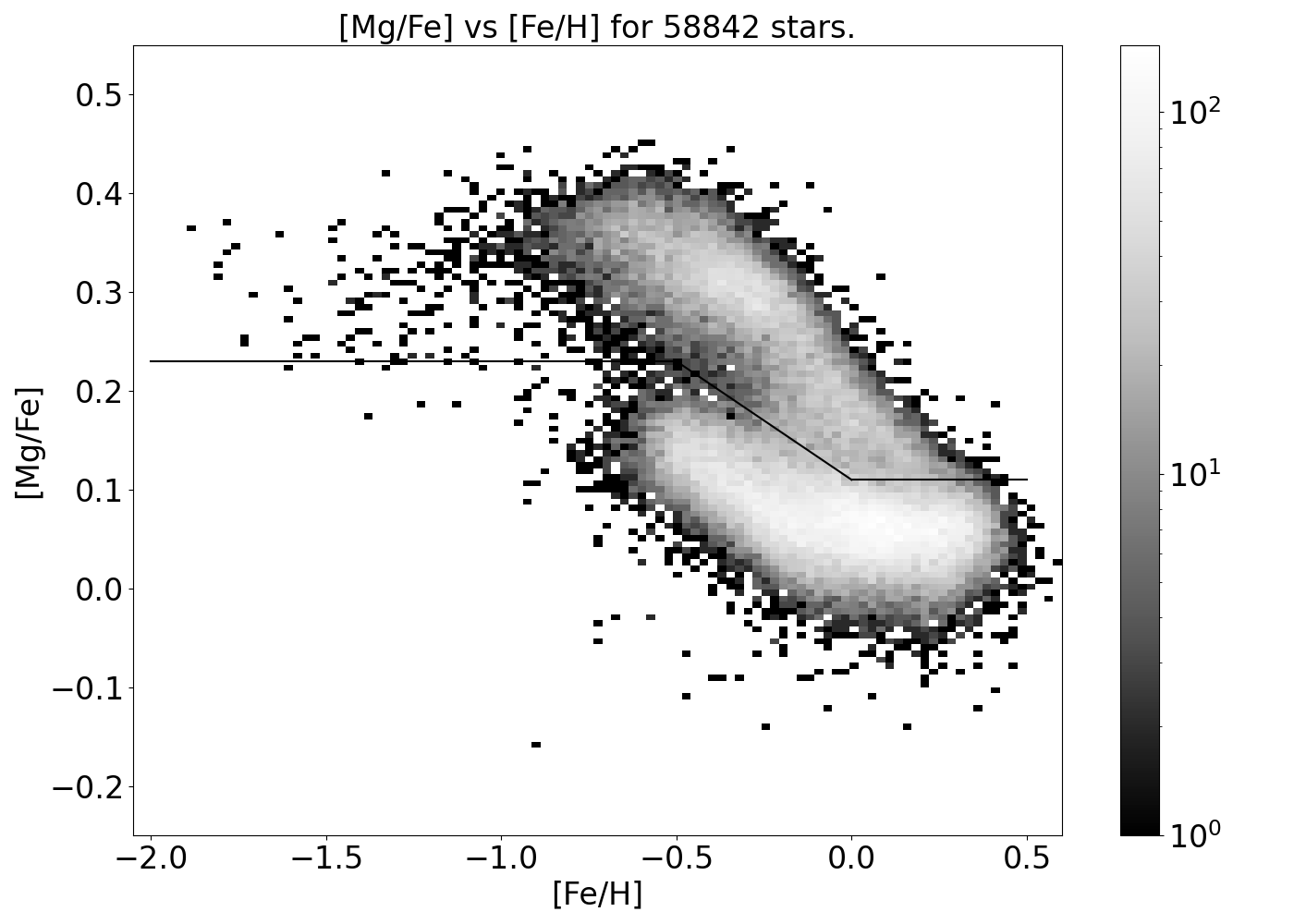}
    \caption{Stars in the sample plotted in the [Mg/Fe]-[Fe/H] plane. The lines show the division into high- and low-$\alpha$ components. A buffer zone 0.01dex either side of the lines (i.e. 0.02 dex total) was discarded from the separated samples to avoid population cross-contamination. There is faint evidence of clustering in the high-$\alpha$ disk which is more prominent in the [C/Fe]-[Fe/H] plane as show in Fig.\ref{fig:V14Bimodal}}
    \label{fig:MgvsFe}
\end{figure}

The sample was separated into high  and low~$\alpha$ disk populations adopting the boundaries in the [Mg/Fe]-[Fe/H] plane as shown in Fig. \ref{fig:MgvsFe}. 
The lines shown on the diagram are the adopted demarcations between the populations.
A buffer-zone of 0.01~dex in [Mg/Fe] was imposed either side of these lines, to minimize inter-sample contamination\footnote{This gives a total buffer zone width of 0.02 dex, cf. mean uncertainty of 0.012 dex in [Mg/Fe] for the sample.}.  
While stars in the buffer zones were excluded from population-specific analysis, they were included in the analysis of the whole sample.

\begin{figure}
\includegraphics[width= \columnwidth]{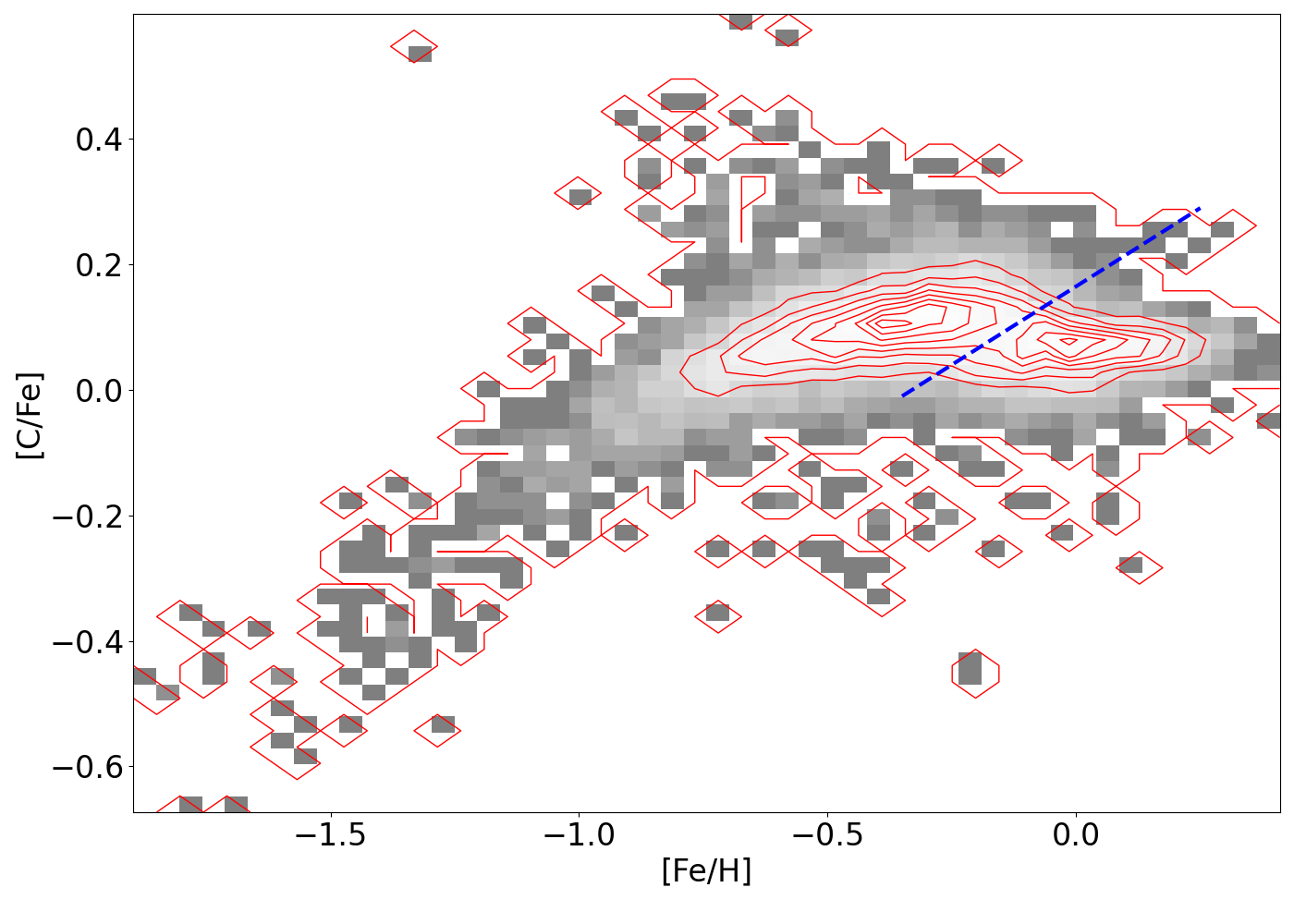}
\caption{Density plot of stars of the high-$\alpha$ population plotted in the [C/Fe] vs. [Fe/H] plane with overlaid contour map. There is a sub-population of late high-$\alpha$ stars in the high-$\alpha$ region below the blue dotted line. This clustering is also evident in the [O/Fe]-[Fe/H] plane.}
\label{fig:V14Bimodal}
\end{figure}

\begin{figure}
\includegraphics[width= \columnwidth]{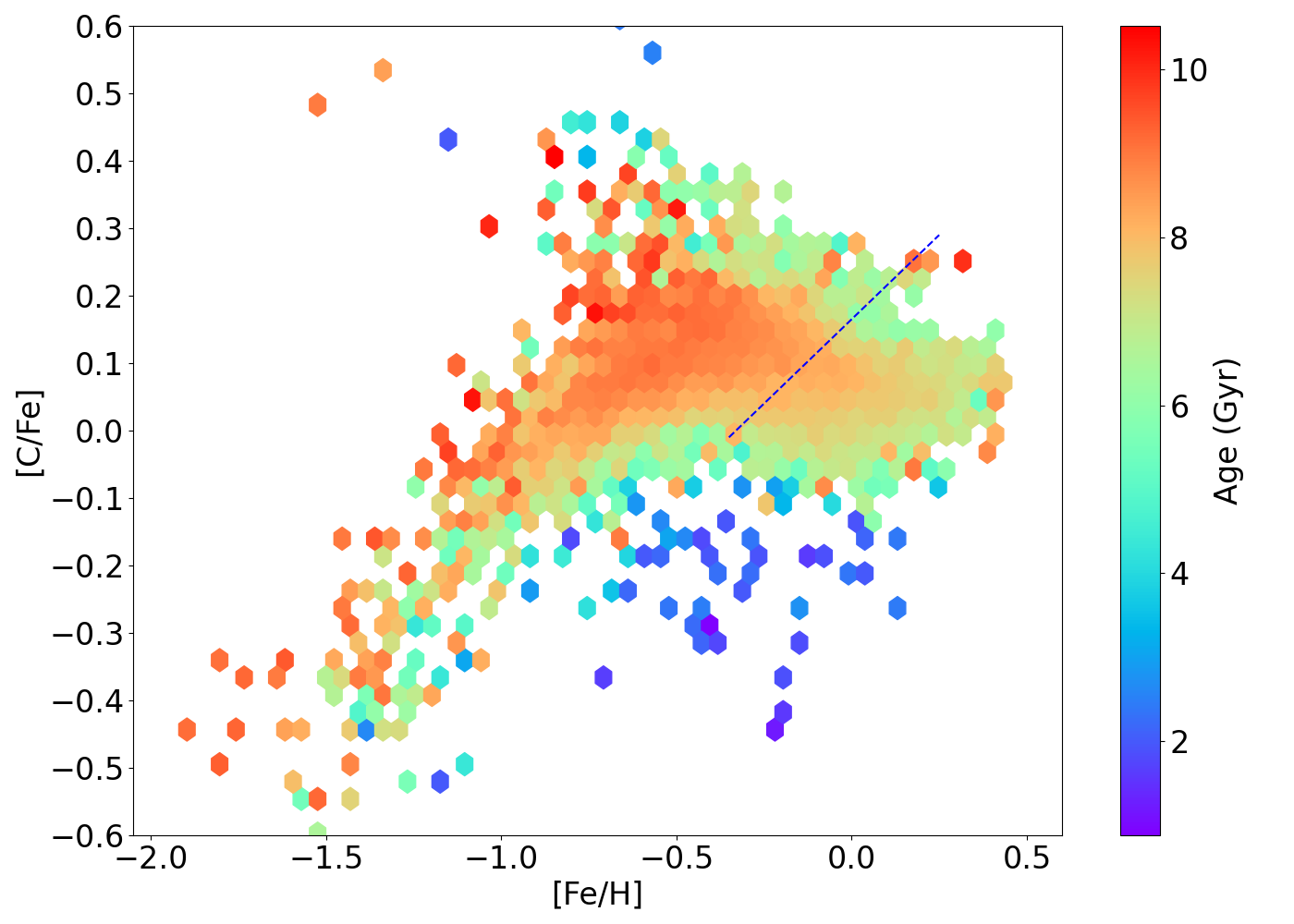}
\caption{Stars of Fig. \ref{fig:V14Bimodal} colour-coded by age. The blue line indicates the same separation as above. The clustering in the [C/Fe]-[Fe/H] plane corresponds to a difference in mean stellar age and shows that the High-$\alpha$ disk is divided into early and late populations.}
\label{fig:V14AgeBi}
\end{figure}

We noticed some evidence of clumpiness in the distribution of high $\alpha$ disk stars in the Mg-Fe plane. Indeed the distribution of the same population displays two clearly evident peaks in C-Fe plane (see Fig. \ref{fig:V14Bimodal}).  Interestingly, the peaks in the C-Fe plane map are characterised by different stellar age, as can be seen in Fig. \ref{fig:V14AgeBi}).  
Stars located in the peak with lower [C/Fe]/higher [Fe/H] are approximately 2-3~Gyr younger than their higher[C/Fe]/lower [Fe/H] counterparts. 

\begin{figure}
    \includegraphics[width= \columnwidth]{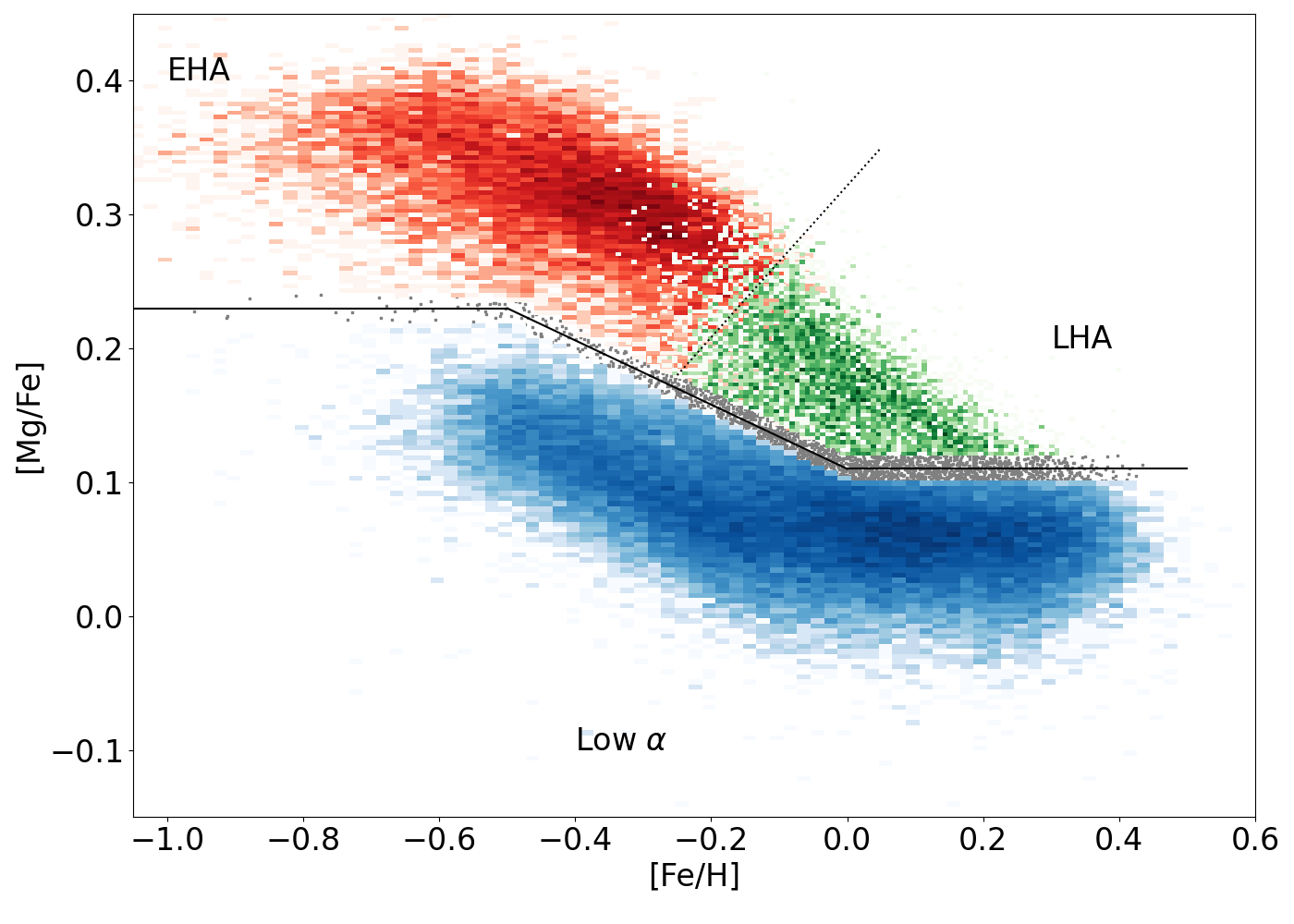}
    \caption{Stars from Fig. \ref{fig:MgvsFe} in the [Mg/Fe]-[Fe/H] plane. This time the three populations are identified by colour. Blue is the low-$\alpha$ population, red is early high-$\alpha$ and green is late high-$\alpha$. The grey stars are the buffer zone used to minimize contamination between the low and high-$\alpha$ disks. (The dotted line is a purely visual aid to show the approximate boundary of the two high-$\alpha$ populations.) }
    \label{fig:ColourMg}
\end{figure}

We show in Sect.~\ref{sec:age} that the two populations in the high-$\alpha$ disk have different stellar age distribution.  Therefore, we term the lower metallicity, higher carbon population the early~high-$\alpha$ (EHA) population and the higher metallicity, lower carbon population the late~high-$\alpha$ (LHA). 
Furthermore, because early-high, late-high and low~$\alpha$ stars have distinct age and metallicity distributions \citep[e.g.,][]{Mackereth2017}, it is sensible that they are studied separately, as those properties may influence the incidence as well as intrinsic properties of carbon-enhanced stars.
We show in Fig. \ref{fig:ColourMg} the three populations in the [Mg/Fe]-[Fe/H] plane.
There is some overlap between the early and late high $\alpha$ populations but the loci are clearly distinct in this plane also.

It was important that we established whether this clustering represents real chemical structure within the disk, or whether it was an artefact of our sample selection or an underlying selection function effect in the APOGEE parent sample. 
Consequently, we undertook a number of explorations to test the robustness of this detection of high~$\alpha$ disk substructure.

Firstly, we experimented with the distance limits of our stellar sample. The composition of the disk does vary with distance from the Galactic centre \citep[e.g.,][]{Nidever2014,Hayden2015,Mackereth2017}, so we reduced our sample initially to stars no farther than 2 kpc, and then 1 kpc. 
In each case, the two clumps seen in Fig. \ref{fig:V14Bimodal} could still be discerned. 
We also applied Galactocentric distance ${\rm R_{GC}}$) limits, examining the data within two annuli: 5.0$<{\rm R_{GC}}<$6.0~kpc and 10.0$<{\rm R_{GC}}<$11.0 kpc.
Again, the substructure is still evident in both annuli. We conclude that this structure was not an artefact of the sample's distance distribution or abundance gradients within the Milky Way disk.

In addition, we checked whether the substructure in the [C/Fe]-[Fe/H] plane was caused by  sensitivity to directional effects.  To do so, we looked at  $30 \degr \times30 \degr $ fields in the direction of both the Galactic centre and the Galactic anti-centre.
Once again, clustering was independently present in samples taken from both directions. We concluded that the effect was not caused by radial abundance gradients within the Milky Way disk, nor by line-of-sight effects directly towards or away from the Galactic centre.

Next, we looked at varying the $T_{\rm eff}$ range of our sample. The determination of carbon abundances by ASPCAP is known to become less precise towards warmer $T_{\rm eff}$. 
The upper limit that we adopted initially, $T_{\rm eff}$=4700K, was chosen as the highest temperature at which the possible clustering could be discerned given our other selection criteria. 
We reduced the limit to 4000K as a test and could still discern the bimodal structure. 
We concluded that the structure was not an artefact of the $T_{\rm eff}$ distribution of our sample. 

We also know that determination of [C/Fe] within ASPCAP is sensitive to {\lg} \citep{Jonsson2020}. 
We executed a number of tests with {\lg} restricted to narrow ranges: [1.0, 1.5], [1.5, 2.0], [2.0, 2.5], [2.5, 3.0] and above 3.0. 
For the latter range, the number of giant stars was too small for any determination  to be made.
For the other {\lg} bins, which all had sufficient numbers of stars, the structure was still evident.

\begin{figure}
\includegraphics[width=\columnwidth]{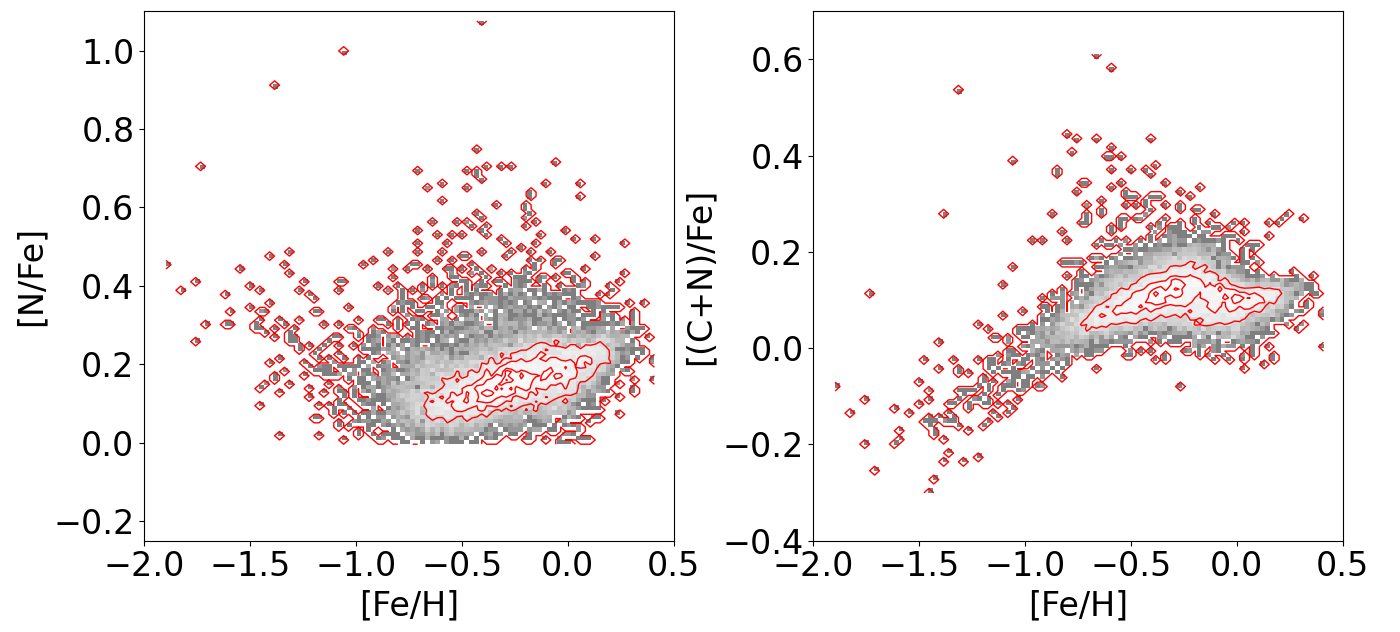}
\caption{Stars of Fig. \ref{fig:V14Bimodal} plotted on [N/Fe] vs. [Fe/H] and [(C+N)/Fe] vs. [Fe/H]. 
The clustering visible in [C/Fe] in Fig. \ref{fig:V14Bimodal} is not apparent here in [N/Fe] and persists in [(C+N)/Fe], which suggests that the [C/Fe] clustering is not the result of stellar evolution. (The seeming cut-off in the left hand plot at [N/Fe] = 0.0 is a visual illusion. There are 8 stars in the sample plotted here with [N/Fe] $\lesssim$ 0.0 but the lowest value is only -0.017, leading to an apparent cut-off.)
}
\label{fig:V14CN}
\end{figure}

We further checked for the possibility that the density clustering is caused by stellar evolution effects associated with mixing along the giant branch.
The combined abundance of carbon plus nitrogen, [(C+N)/Fe],  is expected to stay constant for low and intermediate mass stars during their pre-AGB lifespans \citep[e.g.][]{Masseron2015}, whereas the ratio C/N changes as the star evolves up the giant branch. 
For example, a population of carbon-deficient giants has been recently identified \citep{Maben2023} and explained on such evolutionary grounds\footnote{These predominantly were identified as red clump stars that have been eliminated from our sample.}.
Consequently, we looked at the abundance of nitrogen ([N/Fe]) and of carbon plus nitrogen ([(C+N)/Fe]) across the high $\alpha$ population (see Fig. \ref{fig:V14CN}).  
If the clustering in carbon were a result of chemical evolution then there should be corresponding, anti-correlated clustering in [N/Fe], where an excess in [C/Fe] corresponded to depletion in [N/Fe] and vice versa, and no clustering in [(C+N)/Fe].
Instead, here we saw no clustering in [N/Fe] while the clustering persisted in [(C+N)/Fe], which implied that the [C/Fe] clustering was independent of stellar evolution effects.

Our final test was to check for density clustering in other elements. We noted similar twin high~$\alpha$ populations in [C/N] and [O/Fe] but not [C/O]. 
From this we suspected that the underlying cause of the clustering must be a combination of increased iron abundance and a reduced abundance of carbon.

Having determined that the structure we saw was not caused by selection criteria, was not a distance or direction effect, and was not caused by known sensitivities of the ASPCAP processing, the next step was to look at the position and design purpose of each APOGEE field and whether these somehow cause the observed clustering.
One of the basic operational units of the survey is the field, corresponding to a telescopic field of view. 
Each observed APOGEE field and the targets within it are selected to fulfil a specific scientific or operational purpose, e.g. to observe disk stars, or halo stars, or a specific star cluster, or targets associated with ancillary science programs, or calibration targets.
Even in fields designed for a specific scientific purpose, there may be other stars that still fall into the overall target population for APOGEE so these will be observed alongside the principal targets provided sufficient fibres are available for that field. 
The result is that the objects observed in a specific field are a mixture of specific and general survey targets.
Moreover, the stellar populations included in a field are intentionally quite different according to whether that field is aimed at the bulge, disk, halo or particular object.
We therefore need to rule out two potential selection effects: that the clustering is caused by specific areas of the sky, or that it is somehow related to the specific scientific purposes of certain fields. 
In both cases, a selection effect would produce a similar result in that certain types of field would exhibit large variations in bimodality compared to others.

We examined these two possibilities slightly differently as not every field has a stated design purpose.
There are 2,068 APOGEE fields in our sample, 1,955 (95\%) of which contained one or more high $\alpha$ stars from either population and 1,478 (71\%) contained stars from both. 
These fields cover a range of sky areas including disk, halo and bulge as well as both Galactic and globular clusters \footnote{Our sample excludes known GC stars but not foreground stars within fields where a GC is the scientific target.}.
The bottom 29\% of the fields contained 8 or fewer stars from our sample so it is to be expected that such fields might not contain both populations. 
The percentage of high-$\alpha$ disk stars ranged from 8\% to 65\% but, as previously mentioned, this was expected due to the different targeting of different types of field.
Of more interest is the fraction of the two proposed high-$\alpha$ populations in each field.  
Because we have no model for the distribution of early and late high-$\alpha$ stars, our investigation was limited to checking that specific targeting strategies did not favour one population to the almost total exclusion of the other.
The vast majority of stars were in fields where the percentage of EHA stars as a proportion of the high alpha population was strongly clustered between 60\% and 71\%.
We concluded that there was no evidence that APOGEE spatial selection could be the source of the existence of these separate populations.

For the second possibility, we checked whether that clustering was only marked in fields with certain designated scientific or operational purposes.
Of the 58,842 stars in our sample, we removed those in fields whose operational purposes or science programs contained less than 100 stars in our sample to avoid unrepresentative random statistical variation. 
This left 38,733 stars in 30 design groupings, plus 20,071 for which no scientific or operational classification is given in APOGEE DR17.
We found the percentage of early high-$\alpha$ stars in the fields with a designated purpose was still very strongly clustered around 60\%-75\% (see Table \ref{tab:fields}.) 
The lowest percentages occurred in fields targeting YSOs and sub-stellar objects, and accounted for under 120 high-$\alpha$ disk stars, or 0.7\% of the total. 
On the basis of these numbers, we concluded that there was no evident strong bias by field targeting type that would explain the observed distribution.

\begin{table}
    \caption{Percentage of early high $\alpha$ stars as a fraction of all high $\alpha$ stars in the sample.}
    \centering
    \resizebox{\columnwidth}{!}{\begin{tabular}{l r r r r r r}
{Field} & {Exclusions} & {Low $\alpha$} & {LHA} & {EHA} & {Total} &  {\% EHA} \\
{Type} & {(Buffer)} & & & & & \\
\hline
{(None)} & 658 & 13394 & 1685 & 4334 & 20071 & 72\%\\
{disk2} & 195 & 5506 & 509 & 611 & 6821 & 55\%\\
{TeskeVanSaders\_18a} & 219 & 1840 & 578 & 1260 & 3897 & 69\%\\
{disk} & 135 & 2701 & 348 & 504 & 3688 & 59\%\\
{bulge} & 139 & 2303 & 436 & 665 & 3543 & 60\%\\
{kep\_apokasc} & 149 & 2208 & 325 & 535 & 3217 & 62\%\\
{odisk} & 108 & 2316 & 250 & 475 & 3149 & 66\%\\
{cvz\_btx} & 96 & 1089 & 212 & 517 & 1914 & 71\%\\
{disk1} & 69 & 1284 & 155 & 194 & 1702 & 56\%\\
{k2\_btx} & 65 & 908 & 173 & 505 & 1651 & 74\%\\
{manga} & 14 & 946 & 41 & 84 & 1085 & 67\%\\
{k2} & 51 & 442 & 145 & 407 & 1045 & 74\%\\
{halo} & 32 & 333 & 105 & 347 & 817 & 77\%\\
{kep\_koi} & 33 & 465 & 78 & 125 & 701 & 62\%\\
{cluster\_oc} & 22 & 560 & 31 & 49 & 662 & 61\%\\
{yso} & 18 & 443 & 28 & 48 & 537 & 63\%\\
{cluster\_gc} & 18 & 211 & 52 & 136 & 417 & 72\%\\
{weinberg\_17a} & 23 & 98 & 56 & 172 & 349 & 75\%\\
{anc} & 11 & 255 & 21 & 38 & 325 & 64\%\\
{substellar} & 4 & 249 & 12 & 9 & 274 & 43\%\\
{cal\_btx} & 3 & 206 & 15 & 31 & 255 & 67\%\\
{kollmeier\_19b} & 13 & 128 & 33 & 80 & 254 & 71\%\\
{halo\_stream} & 13 & 134 & 28 & 54 & 229 & 66\%\\
{yso\_btx} & 4 & 193 & 13 & 9 & 219 & 41\%\\
{cluster\_gc1} & 6 & 99 & 35 & 60 & 200 & 63\%\\
{halo\_btx} & 9 & 63 & 37 & 88 & 197 & 70\%\\
{TeskeVanSaders\_17b} & 9 & 92 & 28 & 57 & 186 & 67\%\\
{teske\_17a} & 13 & 65 & 28 & 68 & 174 & 71\%\\
{cluster\_gc2} & 4 & 53 & 30 & 73 & 160 & 71\%\\
{kollmeier\_17a} & 3 & 107 & 12 & 14 & 136 & 54\%\\
{sgr} & 5 & 42 & 17 & 42 & 106 & 71\%\\
{removed} & - & - & - & - & 861 & - \\
\hline
{Totals} & 2141 & 38733 & 5516 & 11591 & 58842 & 68\%\\
    \end{tabular}}
    \tablefoot{Column 1 is the APOGEE field design type. Column 2 is the number of stars in the fields that are in the excluded buffer zone between the low and high $\alpha$ disks. Column 3 is the number of low $\alpha$ stars in the fields. Columns 4 and 5 give the number of early and late high $\alpha$ stars. Column 6 is the total number of stars and column 7 gives the percentage of high $\alpha$ stars that are designated early high $\alpha$. As stated in the text, the percentages are clustered around the 60\%-75\% range. The ``sgr'' entry is related to fields positioned on the main body of the Sgr dSph galaxy, where only foreground stars are included in the statistics, whereas ``removed'' lists the statistics for fields containing very low numbers of stars.}
    \label{tab:fields}
\end{table}

As a result of all these tests, we concluded that our observation of clustering in the high $\alpha$ disk was robust and was not caused by sample selection, systematic processing or systematic target selection effects.

We conjecture that the late-high $\alpha$ population is associated with a burst of star formation. We elaborate further on this hypothesis in Sect.~\ref{sec:age}.

\section {Carbon-enhanced stars}  \label{sec:crich}

\subsection{Definition of $C_{enh}$ stars}

We start this section defining the main objects of our analysis.  Carbon-enhanced stars ($C_{enh}$) are those that have carbon abundances significantly above the average value of the population at the same metallicity.  
Because our sample encompassed three distinct stellar populations (namely, early-high~$\alpha$, late-high~$\alpha$, and low~$\alpha$), each displaying a  different correlation between abundance ratios and metallicity, we had to treat each population separately.
Specifically, we defined $C_{enh}$ stars as those having [C/O] ratios 3$\sigma$ or more above the mean value of the population at the same metallicity. 
Such a definition of enrichment also contemplates the possibility of carbon-poor stars, but consideration of these is outside the scope of this paper. 

To identify $C_{enh}$ stars, each population was divided into bins 0.025~dex-wide in [Fe/H] and a mean [C/O] for each bin was calculated by fitting a curve to the whole population using the LOWESS algorithm.
A python package, {\tt{statsmodels}}, was used for this purpose{\footnote{The {\tt{statsmodels}} software is available at \url{https://github.com/statsmodels/statsmodels}}}.  
We then calculated a single standard deviation against this mean for each entire population and each star was tagged with the number of standard deviations it lay from the mean for its [Fe/H] abundance bin.


\subsection{$C_{enh}$ stars vs carbon stars}

It is important that we draw a clear distinction between our $C_{enh}$ stars and classical carbon stars.  
Carbon stars were originally defined as a separate spectral class, whose optical spectra are dominated by carbon-bearing molecular bands, as opposed to oxygen-rich stars, whose spectra are dominated by TiO bands \citep{KM1941}.  
The separation between the two classes is dictated by the ratio of the number of carbon atoms, $n_C$, to the number of oxygen atoms, $n_O$ in the stellar atmosphere.  
Carbon stars are those for which $n_C/n_O~>~1$.

Given the wide and continuous range of carbon enhancement brought about by mass transfer, it was possible that some stars in our sample of $C_{enh}$ stars qualified for the status of carbon star. 
Based on values given in \cite{Lodders2019}, a [C/O] abundance ratio of +0.24 dex is equivalent to $n_C/n_0~=~1$.  
As we show below, the vast majority of our sample stars lie below that threshold, implying that carbon stars are a vanishingly small minority of our $C_{enh}$ star sample. 
We stress, therefore, that despite having enhanced carbon abundances, $C_{enh}$ stars fall squarely within the category of oxygen-rich stars.  

\begin{figure*}
\includegraphics[width= 0.95\textwidth]{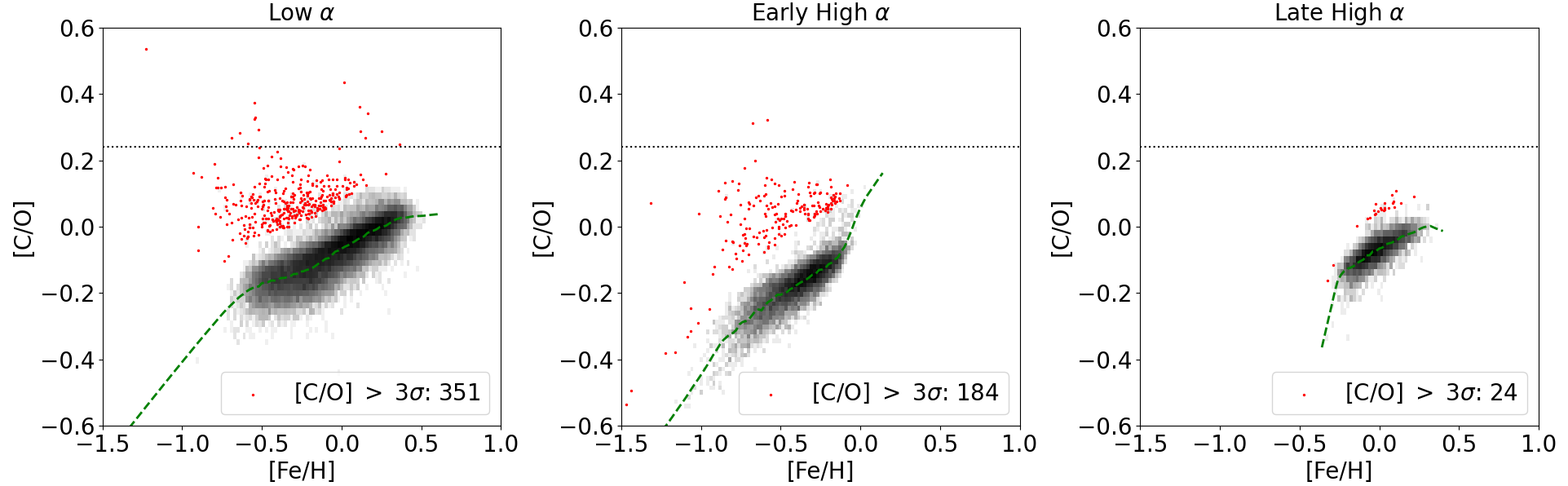}
\caption{Three populations shown in [C/O] vs. [Fe/H] plane. 
The green line is the calculated mean within each [Fe/H] bin using the LOWESS method. Stars which are 3$\sigma$ or more from this mean are shown in red. The horizontal dotted lines indicate the C/O parity criterion for traditional carbon stars.}
\label{fig:V203upLOWESS}
\end{figure*}

Our definition of $C_{enh}$ stars is shown in Fig. \ref{fig:V203upLOWESS}, where stars from our three sub-samples are displayed on the [C/O] vs [Fe/H] plane. 
The dotted horizontal line in these plots indicates the threshold above which the numeric atomic ratio, $n_C/n_O = 1$, the criterion for classical carbon stars.

\subsection{Cerium enrichment} \label{Cerium}
\begin{figure*}
    \centering
    \includegraphics[width=0.95\textwidth]{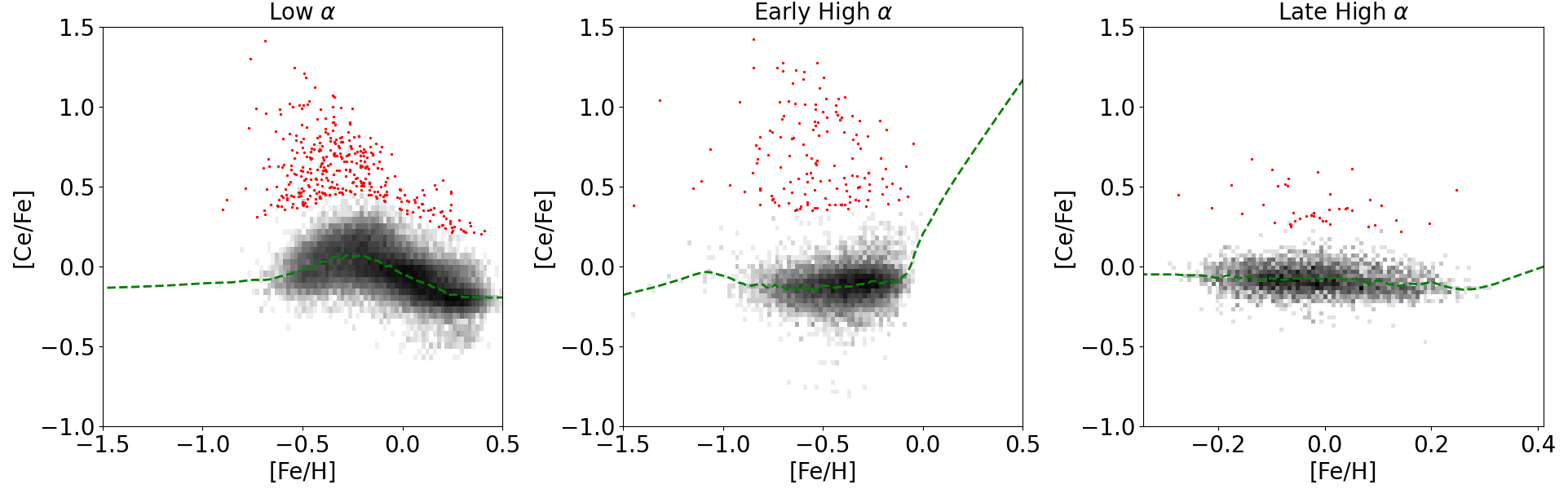}
    \caption{[Ce/Fe] abundances for the three populations in the sample against [Fe/H]. The green lines are the LOWESS mean curves and the red dots show Ce-enhanced stars.}
    \label{fig:CeLowess}
\end{figure*}

As AGB stars are the sources of $s$-process elements \citep[e.g.,][]{Karakas2014}, it is useful to examine the abundances of these elements in our sample. 
Cerium is the best represented of these elements in the APOGEE catalogue. 
Consequently, we analyzed the abundance of [Ce/Fe] in our sample in a similar way to the [C/O] abundance and define Ce-enhanced ($Ce_{enh}$) and Ce-normal stars on the same basis of being 3$\sigma$ or more from the LOWESS mean for stars with similar [Fe/H]. (See Fig. \ref{fig:CeLowess})

\subsection{Estimate of contamination by [O/Fe]-poor stars} \label{sec:contamination}
When considering $C_{enh}$ stars mentioned above, we needed to exclude contamination by stars which are relatively poor in oxygen rather than rich in carbon as this, too, could result in an anomalously high [C/O] abundance. 
We determined the standard deviations in abundances for [C/Fe], [O/Fe], [C/O] and [Ce/Fe] from their respective LOWESS means for each population. 
If a star is O-poor rather than C-enhanced it should show a negative deviation in [O/Fe].
Moreover, we should expect that the absolute value of its deviation from the mean trend in [O/Fe] is higher than the positive deviation in the [C/Fe] plane (if any).
On this basis we counted the number of O-poor but [C/O] rich stars in each population as shown in Table \ref{tab:contamination}. 
The contamination rate was 3.8\% and 6.0\% in the EHA and Low-$\alpha$ populations respectively but is 25.0\% in the LHA population. 
The higher value for this population may be a selection effect since the population is, by nature, low in carbon and thus the [C/O] abundance is more sensitive to variation in [O/Fe]. 
Overall, we noted that contamination of $C_{enh}$ by O-poor stars was 6.1\% and that this was unlikely to affect our analysis in Sect. \ref{sec:analysis}.

When considering the effects of O-poor stars on the wider populations, i.e. not just $C_{enh}$ stars, it was neither possible 
nor appropriate to eliminate contamination. 
Instead, we considered whether trends seen against [C/O] are also seen against [C/Fe] as discussed in Sect. \ref{sec:binarity}.

\begin{table*}
    \caption{Standard deviations of abundances by population}
    \centering
    \begin{tabular}{l|r|r|r|r|r|r|r|r|r}
    Population & Base Count & [C/Fe] & [O/Fe] & [C/O] & [Ce/Fe] & $\frac{\sigma([C/Fe])}{\sigma([O/Fe])}$ & No. O-poor & $C_{enh}$ Count & Contamination \% \\
    \hline
    Early High $\alpha$	& 11,860 &  0.051 & 0.035	& 0.055 & 0.158		& 1.45	&	7	&184	& 3.8\% \\
Late High $\alpha$	& 5,632 & 0.035 & 	0.029& 	0.036& 	0.11 &	1.21	& 6	& 24	& 25.0\% \\
Low $\alpha$	& 39,175 & 0.060 & 0.028	& 0.051 & 0.13 &		2.12 &		21	& 351	& 6.0\% \\    
\hline
Overall & ~ & ~ & ~ & ~ & ~ & 34 & 559 & 6.1\% \\
\end{tabular}
    \tablefoot{Columns 3-6 show the number of dex corresponding to 1$\sigma$ divergence from the LOWESS mean for each abundance for each population. Column 6 show the ratio of $\frac{\sigma([C/Fe])}{\sigma([O/Fe])}$. Column 8 shows the number of O-poor stars identified in the population. Column 9 shows the size of the population and column 10 shows the percentage contamination of $C_{enh}$ stars by O-poor stars.
    }
    \label{tab:contamination}
\end{table*}


\section{Analysis} \label{sec:analysis}
\subsection{Carbon and cerium enhancement by population} \label{sec:pops}
Analysing the incidence of C- and Ce-enhancement in each population, we found that both C- and Ce-enhanced stars were more common in the early high $\alpha$ population than in either the low $\alpha$ or late high $\alpha$ populations. 
These differences were statistically robust (See Table \ref{tab:richness}).

\begin{table*}
    \caption{Proportion of $C_{enh}$ and Ce-enhanced stars  by population }
    \centering
    \begin{tabular}{l|r|r|l|l|r|l|l}
Population &	Base Count	& $C_{enh}$	& Percentage	&$\sigma_{C_{enh}}$ &	Ce-enhanced &	Percentage	& $\sigma_{Ce_{enh}}$ \\
\hline
Low $\alpha$ &	39,175 &	351&	0.90\%	& 0.05\% &	376	& 0.96\% &	0.05\% \\
Early High $\alpha$ &	11,860 &	184	& 1.55\% &	0.11\%	& 155	& 1.31\%	& 0.10\%\\
Late High $\alpha$ &	5,632 &	24& 0.43\% &	0.09\% &	43	& 0.76\%& 	0.12\% \\
    \end{tabular}
    \tablefoot{Proportions are expressed as a percentage of each population. 
    The $\sigma_{C_{enh}}$ and $\sigma_{Ce_{enh}}$ columns show the expected natural statistical deviations based on a simple uniform binomial probability within each population and expressed as a percentage of the total.
    }
    \label{tab:richness}
\end{table*}

We hypothesize that this can be attributed to stellar ages; whatever processes create these $C_{enh}$ stars are essentially complete within the average age of early high $\alpha$ stars but are ongoing in the low $\alpha$ population where star formation was more prolonged.
It is unclear why the percentage of enriched stars is so much lower in the late rather than early high $\alpha$ population. 
Though we caution that the number of $C_{enh}$ stars in the late high-$\alpha$ population is small, the difference does appear to be statistically significant. 
The late high $\alpha$ population seems more closely to resemble the low $\alpha$ population in degree of enhancement, which suggests that the evolution and age of these stars may be significant factors.

\subsection{Binarity} \label{sec:binarity}
\subsubsection{Mass transfer processes}

A good starting point for a discussion is a consideration of the properties of systems where we expect mass transfer and carbon enrichment to occur.  
In this way we can inform our subsequent examination of any trends between abundance ratios and binarity.
These systems are, of necessity, ``asymmetric'' binaries where the originally more massive member has evolved through the AGB stage to become a white dwarf, and the secondary is usually at the subgiant or RGB stage.

For an AGB star to become carbon-enhanced, its main-sequence mass must fall within a limited range \citep{Straniero}.  
At the low mass end are stars that undergo a Third Dredge Up (TDU), since it is this event that brings the carbon resulting from He burning to the upper photosphere. 
The minimum stellar mass for this is still open to debate and may be dependent on metallicity, but we adopt Straniero's minimum value of 1.2 $M_\odot$.  
At the upper end, the limiting mass of 4 M$_\odot$ is governed by the  occurrence of hot bottom burning (HBB), where the temperature at the bottom of the convective layer exceeds \(80\times 10^6\)K with the result that any carbon that would be dredged up is first converted into nitrogen \citep{Sackmann1992,GH07, 2013A&A...555L...3G}. 
Thus, we expect our initially primary star to lie in the mass range \(1.2 M_\odot \leq M_\star \leq 4 M_\odot\).  
According to stellar evolution models \citep{FRUITY}, the primary experiences considerable mass loss during its TP-AGB phase, at rates up to \(10^{-4}\) $M_\odot$ per year, leaving a remnant WD with mass within the range \(0.6 \leq M_\star \leq 1.4 M_\odot\).
We use these mass ranges in Sect. \ref{sec:age} when considering stellar ages. 

The secondary star has a lower initial mass and, due to our sample definition, is most likely in the first ascent red giant branch.  
As a lower limit, we note that stars substantially less massive than the Sun will not have had sufficient time to evolve to this stage in a Hubble time.
The secondary must also be significantly less massive than the original primary or it too would have evolved to AGB or WD states in approximately the same time-frame. 
Since the requirement to avoid HBB mentioned above places an upper limit of 4 $M_\odot$ on the primary, we propose that the less massive secondary cannot be more massive than 3 $M_\odot$.
We can hypothesize therefore that the secondary star - which is the one now recorded in APOGEE - is in the mass range \(1 M_\odot \leq M_\star \leq 3 M_\odot\). 
If mass transfer has taken place, its current stellar mass may be higher than its original mass. 
The evolution of the secondary star will also constrain the timescale for the formation of $C_{enh}$ binaries. 
The secondary star must have moved off the main sequence so, using a Salpeter initial mass function with the above mass limits, and noting from simple homology that main sequence lifetime $\propto$ ($\frac{M}{ M_\odot}$)$^{-2.5}$, we calculated $\tau_C$, the average formation time to be approximately 5.9 Gyr. We assume here a MS lifespan for the Sun of 10Gyr \citep{Guinan2009} and a mean mass for the secondary of 1.25$M_\odot$. 
More massive secondaries correspond to shorter formation times but the large negative index in the Salpeter function means that such binaries  will become rapidly less common at ages down to $\sim$ 1 Gyr. 
 
Now we give consideration to the initial and final orbital separations of the pair.
The initial separation cannot have been too close. 
When the primary entered its AGB stage it would have had a diameter of at least 1~AU \citep{Chiavassa2020}. 
For a system with masses of 3 $M_\odot$ and 2 $M_\odot$, a separation of 1~AU corresponds to an orbital period of about 160 days. 
If the initial period were less than this, the system would have undergone a common envelope (CE) event.
Depending on the initial separation, the outcome of a CE can be a merger of the two stars or orbital shrinkage and a very rapid dissolution of the AGB star's envelope in about a decade \citep{Chamandy}, equivalent to a mass-loss rate of 0.1-0.2 $M_\odot$ \(yr^{-1}\). 
Where orbital shrinkage does occur \citep{Chen20}, resulting in shortening of the binary period, the currently observed period may be considerably different to that at the time of mass transfer.

The mass transfer between the binary members can happen by one of four mechanisms.
Firstly, for the smallest orbital separations, a Common Envelope event as described above can result in transfer of significant amounts of mass from the AGB primary onto the secondary in timescales of $\sim$10~yr. 
For larger separations, the AGB primary may fill the system's Roche lobe and Roche Lobe Overflow (RLOF) can transfer mass directly from its photosphere onto the secondary. 
At larger separations still, the AGB star does not fill the Roche lobe, but the star does eject copious amounts of mass at rates of \(10^{-5}\) to \(10^{-4}\) $M_\odot$ \(yr^{-1}\), in the form of a stellar wind.
It is this wind which overflows the primary's Roche lobe and leads to the phenomenon of Wind Roche Lobe Overflow (WRLOF). 
Finally, at the largest separations, the secondary is merely moving through relatively quiescent circumbinary material ejected by the AGB primary, leading to Bondi-Hoyle-Lyttleton (BHL) accretion \citep{Bondi}.

Assuming that mass transfer is indeed responsible for the excess C and Ce abundances reported in Sect.~\ref{sec:crich}, we still do not know which of the processes summarised above, contributed to mass transfer. 
We would in any case have reason to expect that the intrinsic spread in the parameters governing the physics of the various mass transfer processes would lead up to the presence of a continuum of carbon enhancement within our sample, as indeed observed.
Furthermore, if mass transfer is indeed the \emph{principal} route for increasing carbon enhancement in non-AGB stars, 
by whatever mechanism, then there should also be observed an increase in binary fraction within sub-populations of increasing degree of enhancement. 

\subsubsection{Binarity trends using RUWE and RV catalogues}\label{sec:bintrends}

\begin{figure*}
    \centering
    \includegraphics[width= 0.95\textwidth]{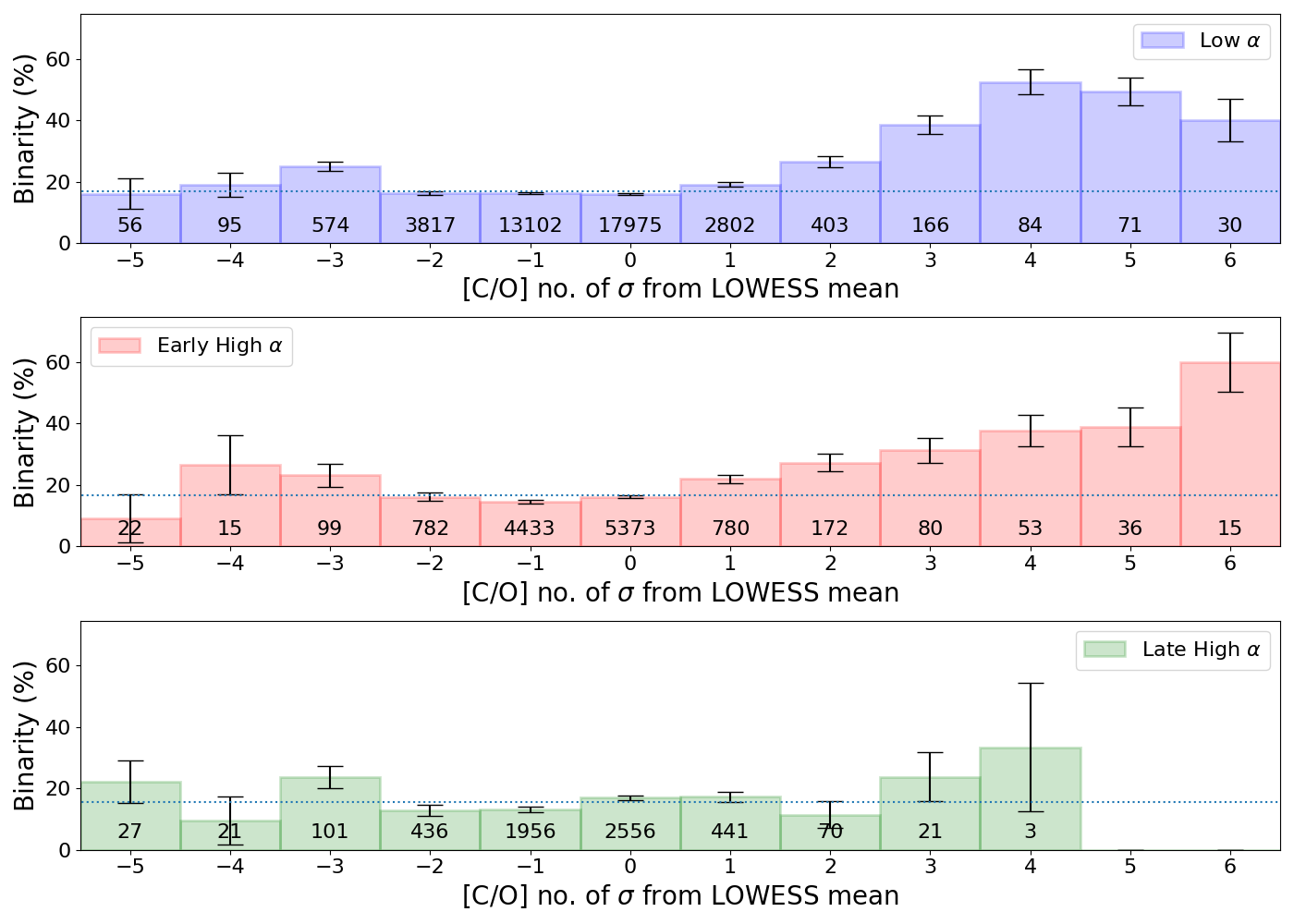}
    \caption{Percentage of stars which are binary against number of standard deviations  of [C/O] abundance from the mean for stars of the same population and similar [Fe/H]. The panels show the early high $\alpha$, low $\alpha$ populations and the late high $\alpha$ population. The dotted lines show mean binary percentage across each entire sub-population. Error bars are based on binomial distribution of sub-population with uniform binarity independent of [C/O] variation.}
    \label{fig:BinarityHistos}
\end{figure*}

In Fig. \ref{fig:BinarityHistos} the degree of carbon enhancement, as defined in Sect.~\ref{sec:crich}, is plotted against the percentage of stars that are binaries. 
The number of stars included in the statistics for each [C/O] bin is displayed at the bottom of the bin.  
Error bars on top of each bin are calculated following simple binomial statistics assuming uniform binarity across the whole population.  

Indeed, in agreement with our expectations, the early high-$\alpha$ and low-$\alpha$ populations showed a clear association between a higher degree of [C/O] enhancement and a higher binary frequency.  However, this trend was not visible in the late high-$\alpha$ population; a K-S test against uniform binarity gave a p-value of 0.0995 so there was no statistical evidence of a trend with carbon enhancement. 
It was unclear, however, whether the absence of trend was real, or simply due to limited data.  The late high-$\alpha$ population has higher metallicity and, 
as previously mentioned, \cite{Moe19} note that binary frequency tends to be lower towards higher metallicity.  Therefore, the lack of a visible trend may be attributable to the lower binary fraction amongst these more metal-rich stars.
In Table \ref{tab:fb} we can see that the binary fraction for this population is lower than those of the low-$\alpha$ and early high-$\alpha$ populations  by 1.4\% and 1.0\%, respectively.
Assuming a uniform binarity fraction and using a simple binomial calculation, these percentages correspond to deviations of about 2.9$\sigma$ and 2.0$\sigma$ in that population respectively.
Additionally, our sample for this population is considerably sparser than those for either the low-$\alpha$ or early high-$\alpha$ populations, which are about 2 and 7 times larger, respectively. 
The absence of a trend may be partly due to the small population sizes in the most carbon-enhanced bins.
By nature of its definition, the late high $\alpha$ population consists of stars from the overall high $\alpha$ disk population with lower [C/Fe] abundances and, for reasons that are not yet understood, has a tighter distribution of [C/O] than the early high $\alpha$ population. 
It seems to be intrinsically biased against C-enhancement so these small sample sizes are not a surprise and should not be construed as counter-evidence against the mass transfer hypothesis.

\begin{figure*}
    \centering
    \includegraphics[width= 0.95\textwidth] {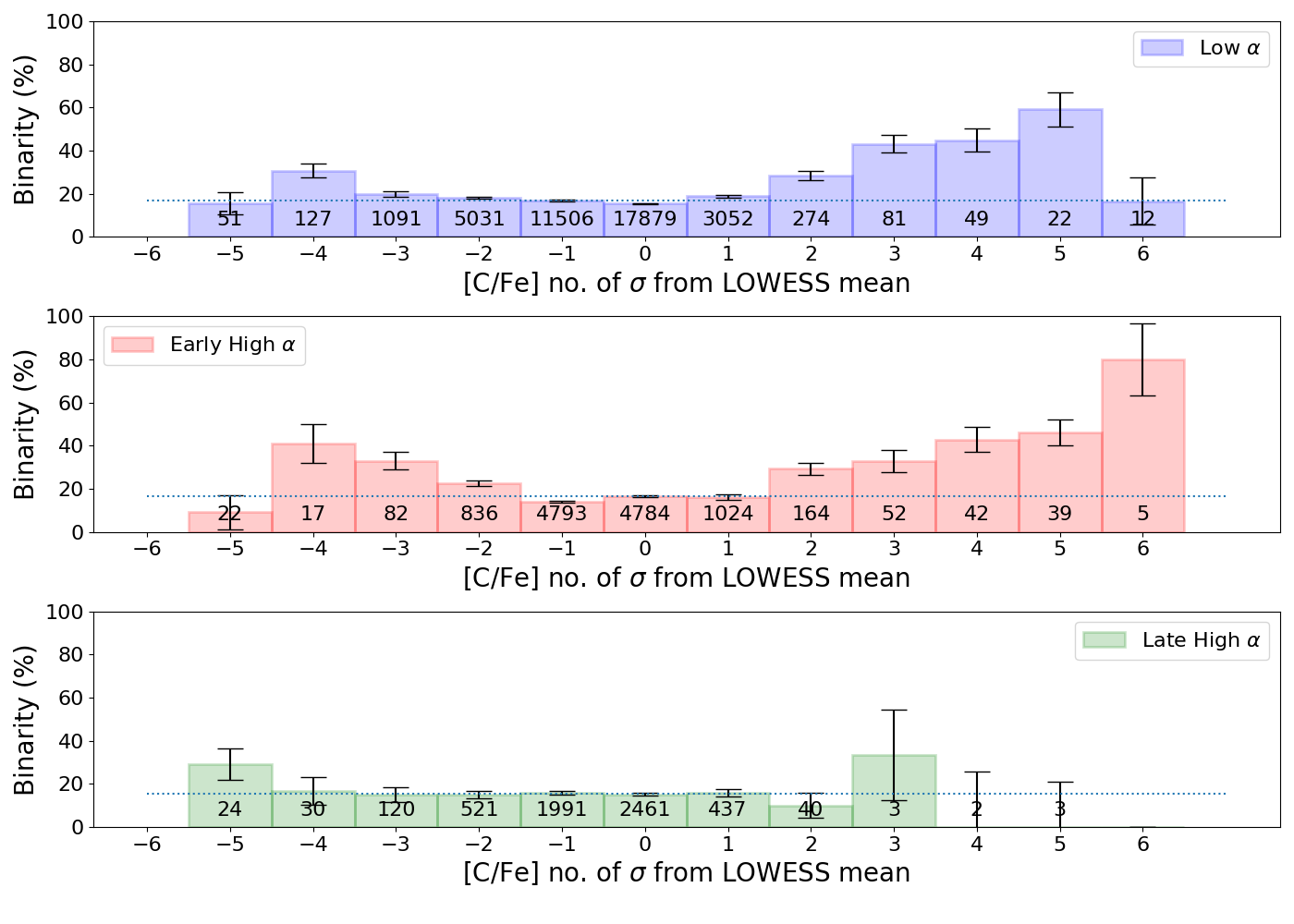}
    \caption{Same as Fig. \ref{fig:BinarityHistos}, this time displaying the percentage of stars which are binary against number of standard deviations  of [C/Fe] abundance ratio from the mean for stars of the same population and similar [Fe/H]. The panels show the early high $\alpha$, low $\alpha$ populations and the late high $\alpha$ population. The dotted lines show mean binary percentage across each entire sub-population. Error bars are based on binomial distribution of sub-population with uniform binarity independent of [C/Fe] variation. A similar trend in binarity is observed, ruling out the possibility that the trend against [C/O] is caused by O-poor stars.}
    \label{fig:CFeBinarityHistos}
\end{figure*}

To again address the concern of contamination by O-poor stars, we repeated the above analysis using [C/Fe] abundances.
If the observed [C/O] trend was due to a trend in oxygen depletion then we would not have expected to see this trend mirrored in [C/Fe].
As we see in Fig. \ref{fig:CFeBinarityHistos}, a similar trend to that against [C/O] was also present against [C/Fe].
Consequently, we ruled out the possibility that the trend against [C/O] was caused by [O/Fe] depletion rather than carbon enhancement. (See also Sec. \ref{app:OFe}.)

\begin{figure*}
    \centering
    \includegraphics[width=0.95\textwidth]{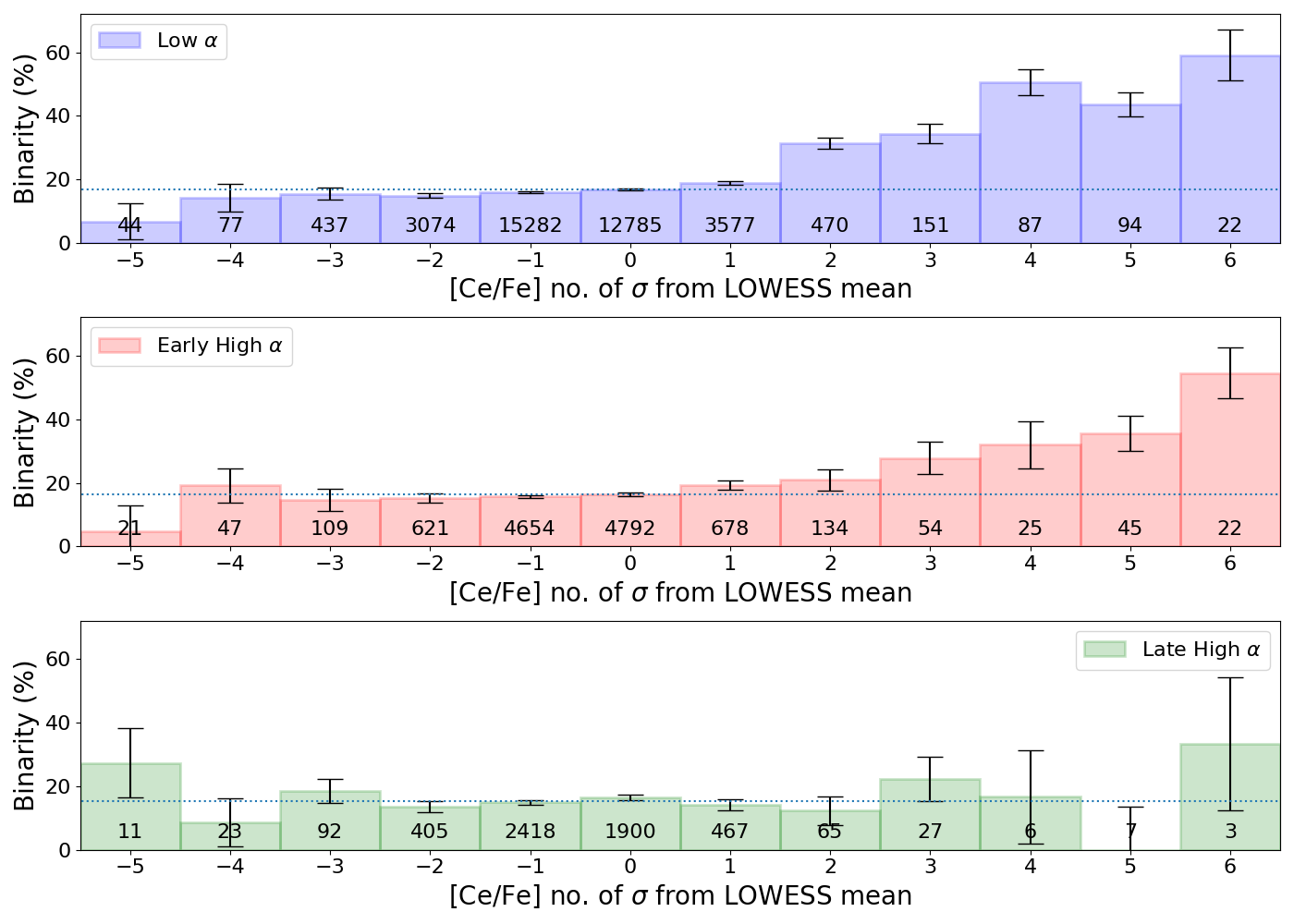}
    \caption{Percentage of stars are binary against number of standard deviations of [Ce/Fe] abundance for stars from the same population and similar [Fe/H]. The panels show the low early high $\alpha$ and late high $\alpha$ populations. Error bars are based on binomial distribution of overall population binarity. The dotted line represents the binarity fraction for the whole of each population.}
    \label{fig:CeBinarityHis}
\end{figure*}

\begin{figure}
    \centering
    \includegraphics[width=\columnwidth]{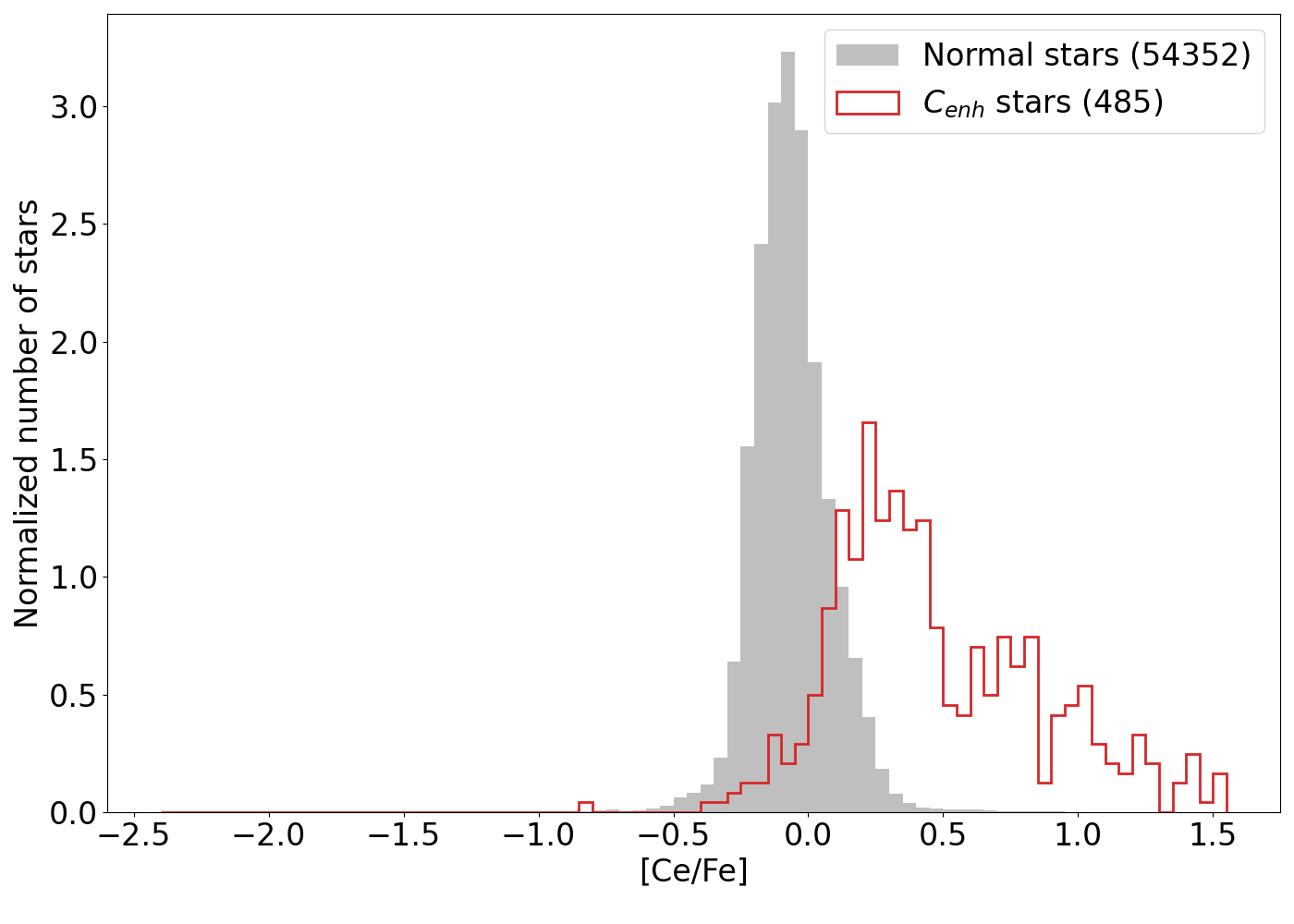}
    \caption{Comparison of $C_{enh}$ and C-normal stars against [Ce/Fe]. The $C_{enh}$ population is conspicuously richer in Ce as well, showing a clear link between carbon and cerium enhancement. This is evidence that the same mechanism underlies enhancement of both elements. (Stars with missing or invalid [Ce/Fe] have been eliminated from this plot.)}
    \label{fig:CRichCerium}
\end{figure}

If mass transfer occurs from AGB stars to a companion, then it is expected to also enrich the companion envelope in $s$-process elements such as cerium.  This hypothesis was verified by repeating the exercise from Figs. \ref{fig:BinarityHistos} and \ref{fig:CFeBinarityHistos}, this time looking at the [Ce/Fe] abundance ratio. This is shown in
Fig. \ref{fig:CeBinarityHis} where it can be seen that, similarly to the case of [C/O] and [C/Fe], a clear correlation between the binary frequency and the residuals relative to the mean [Ce/Fe]-[Fe/H] relation is present.
While Fig. \ref{fig:CeBinarityHis} shows evidence for a clear relation between Ce enhancement and binary frequency, it is important to establish whether stars that are enriched in C are also enriched in Ce and vice-versa.  To check for a correlation between C and Ce enhancement, we compared the [Ce/Fe] distribution of $C_{enh}$ and C-normal populations in Fig. \ref{fig:CRichCerium}. 
The two populations had  greatly different distributions of [Ce/Fe], with the $C_{enh}$ stars extending towards fairly large values of cerium enhancement, which shows that indeed there is a correlation between carbon and cerium enhancement, in further support of the mass-transfer hypothesis. 

\subsubsection{Binarity of $C_{enh}$ stars using \drvm}\label{DRVM}
APOGEE DR17 provides high-quality spectroscopic parameters for $\sim4\times10^{5}$ stars and sparsely-sampled radial velocity (RV) curves for the vast majority of them. 
Our $C_{enh}$ stars are no exception: 13.8\% have two RV observations; 42.9\% have three; and
22.7\% have four or more. 
Most of these sparsely sampled RV curves cannot be used to constrain full orbital solutions
\citep[e.g.][]{PriceWhelan2020}, but they are very effective for revealing the presence of a close (P~$\simless 10^3$ days) companion
via the maximum shift in the observed RVs, $\Delta \text{RV}_{\text{max}} = |\text{RV}_{\text{max}}-\text{RV}_{\text{min}}|$
\citep[for a discussion, see][]{Badenes2012}. To ensure the individual RVs are of sufficient quality, we required each visit spectrum to have S/N$\geq40$. Adopting a lower S/N threshold results in an unacceptably large number of outliers likely caused by catastrophic failures of the RV pipeline.  Following that procedure we managed to calculate \drvm\ for 430 of our $C_{enh}$ stars. Of these, 275 (64\%) are classified as low-$\alpha$, 143 (33\%) as early high-$\alpha$, and only 12 (3\%) as late high-$\alpha$. 
The number of visits for these stars ranged from 2 to 34, with an average of 3.2 and a median of 3 visits. The time between visits, $\Delta{T_{visit}}$, ranged from 0.93 to 3256.0 days, with a mean of 289.6 days and a median of 51.8 days.
(Obviously, \drvm\ cannot be assessed with fewer than 2 visits.)

The shape of a \drvm\ distribution strongly depends on the \lg\ distribution of the underlying sample, as lower {\lg} values
correlate with increased RV uncertainties and a smaller maximum value of \drvm, but in general any star with \drvm\ above a few \kms\ can be confidently identified as a short-period binary \citep[see][for discussions]{Badenes2018,Mazzola2020}. 
We thus selected a control sample of APOGEE DR17 giants with $1.0\leq \log(g)\leq3.6$ (N = 159801) that excluded the red clump stars identified by \citet{Bovy2014} and that passed the quality cuts outlined in \citet{Mazzola2020}.
(These cuts are S/N ratio $\geq$40 for individual measurements, elimination of stars without valid measurements for both $T_{\rm eff}$ and \logg. Stars with the APOGEE {\tt{STAR\_BAD}} flags were eliminated as well as those designated as commissioning or telluric calibration targets.)
 
\begin{figure*}
	\includegraphics[width=\textwidth]{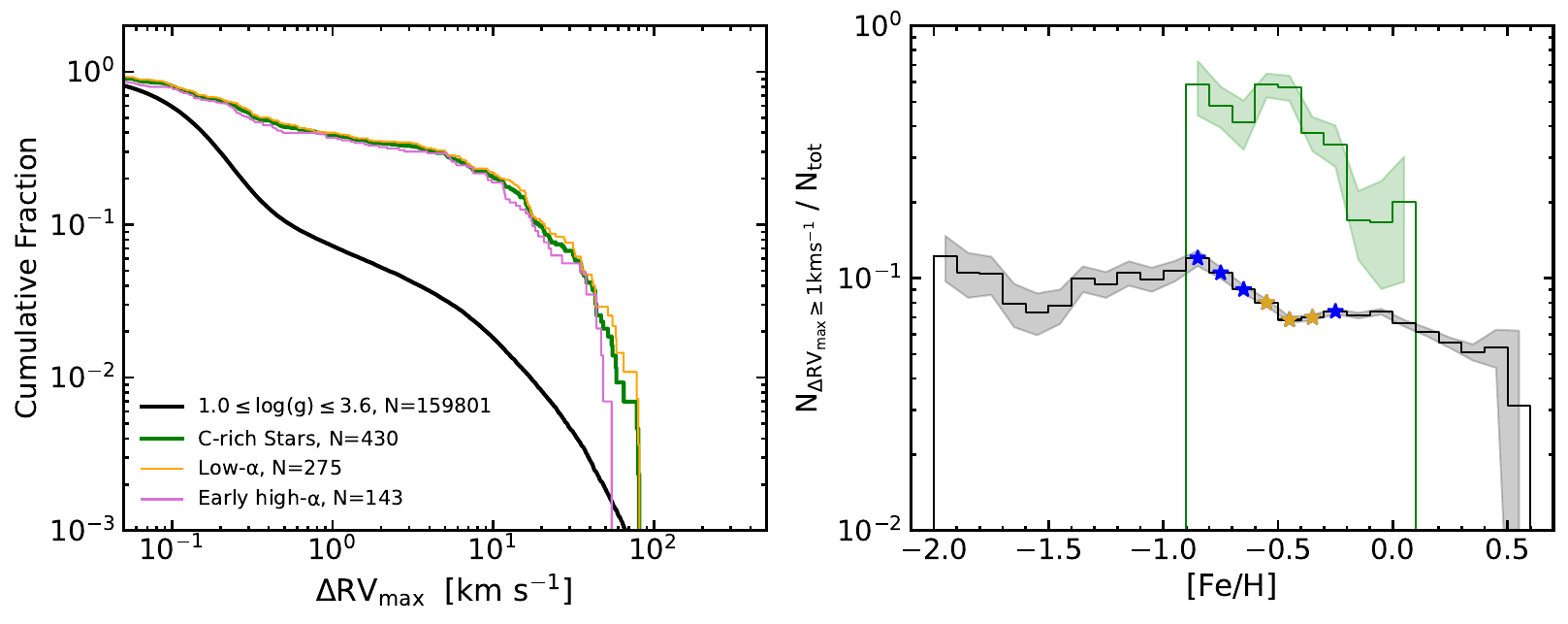}
    \caption{Left panel: Cumulative fraction histogram of \drvm\ for the $C_{enh}$ stars (green), \lg-controlled comparison sample from APOGEE DR17 (black), and the low-$\alpha$ (orange) and early high-$\alpha$ (pink) sub-samples. The $C_{enh}$ stars are skewed towards larger \drvm\ values compared to the control sample. Right panel: RV variability fraction as a function of metallicity for the $C_{enh}$ (green) and control sample (black). Uncertainties on the RV variability fractions are shown as the shading, and bins where there is a difference of between $2\sigma$ and $5\sigma$ in RV variability are marked with blue and golden stars, respectively.  Both samples demonstrate an anti-correlation with metallicity, but the $C_{enh}$ stars have consistently larger fractions than the control sample.}
    \label{fig:drvms}
\end{figure*}

In the left-hand panel of Fig. \ref{fig:drvms}, we compare the cumulative histograms of \drvm\ for the \lg-controlled sample (black), $C_{enh}$ stars (green), and the low-$\alpha$ (orange) and early high-$\alpha$ (pink) sub-samples. The $C_{enh}$ stars are strongly skewed towards high-\drvm\ relative to the control sample, indicating these systems are dominated by short-period binaries. The low-$\alpha$ sub-sample also appears to be skewed towards larger \drvm\ values than the early high-$\alpha$ stars, which may be a manifestation of the anti-correlation between the close binary fraction and $\alpha$ abundances observed by \citet{Mazzola2020}.

We further compared the fraction of stars that show RV variability significantly larger than the expected uncertainties, here
chosen to be $\Delta \text{RV}_{\text{max}}\geq1$ \kms\ based on our samples' properties. The right hand panel of
Fig. \ref{fig:drvms} shows the RV variability fractions for the $C_{enh}$ stars (green) and control sample (black) as a function of
metallicity. We required at least five stars per bin and the shading indicates the binomial process uncertainty,
\( \sigma_{f}\) = \(\sqrt{\frac{f \left( 1-f \right)}{N}}\) where $f$ is the RV variability fraction and $N$ is the total number of stars in that bin. Both
samples show the well-known anti-correlation between the close binary fraction and metallicity \citep[e.g.,][]{Moe19}, but the
$C_{enh}$ stars have significantly larger RV variability fractions than the control sample, with the bins that show 2$\sigma$ and $5\sigma$ discrepancies marked with blue and golden stars, respectively. Given that the RV uncertainties are not dramatically different between the two samples, this implied that the $C_{enh}$ stars had an enhanced frequency of close binaries with systematically shorter periods. 
This is an interesting result, as it suggests that mass-transfer binaries are characterised by tighter orbits than other binaries, which is what one would expect to happen as tighter orbits would favour the occurrence of mass transfer between companions.

\subsection{Age and metallicity distributions} \label{sec:age}
\subsubsection{Metallicity distribution function (MDF) and chemical composition of $C_{enh}$ stars}
A dependency between metallicity and carbon enhancement is to be expected because of the dependency of the efficiency of the TDU on metallicity \citep[e.g.,][]{Karakas02}. 
Empirically, \cite{Hansen16} and \cite{Andersen16} have previously noted a link between low metallicity and carbon enhancement in CEMP stars, but these are old halo objects, unlike the disk stars being studied in this work and the enhancement mechanisms may differ. 

\begin{figure}
    \centering
    \includegraphics[width=\columnwidth]{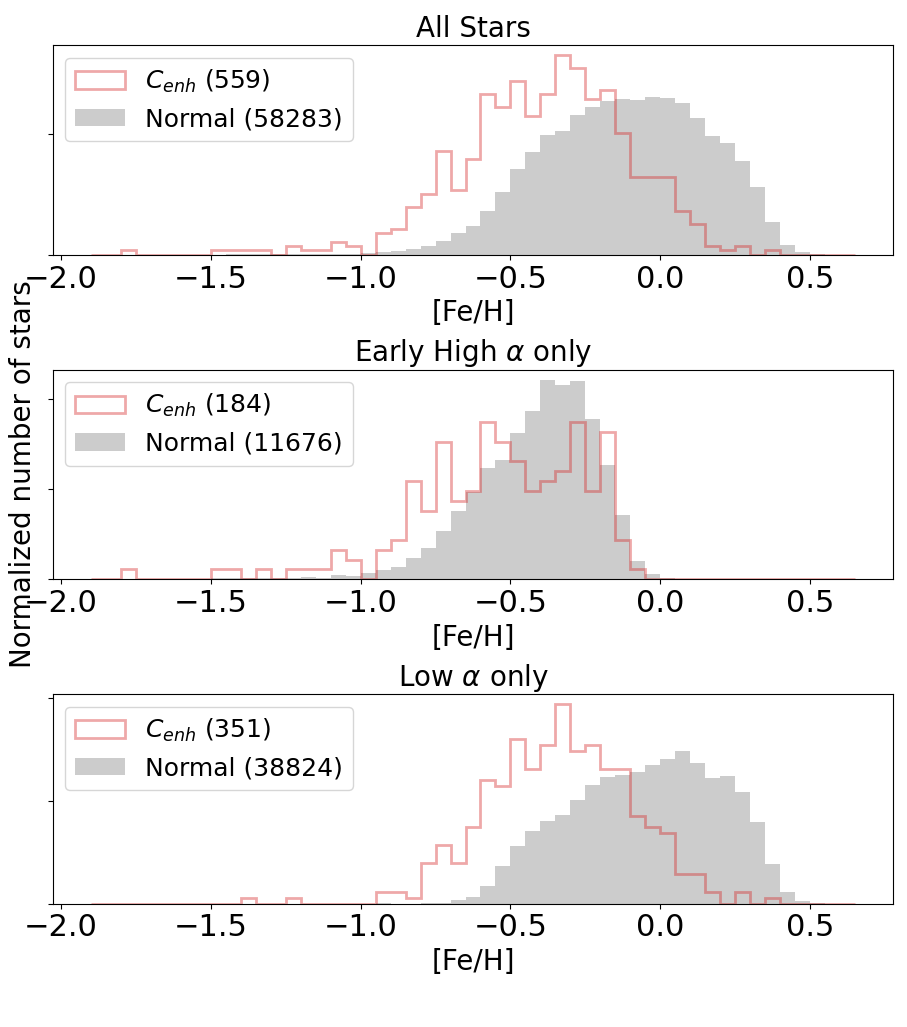}
    \caption{Histogram showing normalized metallicity distributions of $C_{enh}$ and C-normal stars. The top panel shows all stars combined, the middle panel shows high $\alpha$ stars only, and the bottom panel only low $\alpha$ stars. The differences between means is 0.26 dex in the top panel, 0.08 dex for high $\alpha$ stars and 0.30 dex for low $\alpha$ stars. No histogram for late high $\alpha$ stars is shown due to the small number of $C_{enh}$ stars in this sub-population.}
    \label{fig:NewFig4}
\end{figure}

Fig. \ref{fig:NewFig4} compares the MDFs of $C_{enh}$ and C-normal stars within our samples. The top panel displays the MDFs for the whole sample, while the middle and bottom panels limit the comparisons to the early high-$\alpha$ and low-$\alpha$ populations alone (the number of   $C_{enh}$ stars in the late high-$\alpha$ population is too low for a meaningful comparison).
As previously noted, we saw a clear difference between the MDFs of $C_{enh}$ and C-normal stars.  The mean of the distribution of $C_{enh}$ stars is lower in [Fe/H] by 0.26 dex than that of their C-normal counterparts.  
There is a clear difference between the low-$\alpha$ and high-$\alpha$ populations in this regard. In the low-$\alpha$ population, the mean [Fe/H] of the $C_{enh}$ stars is 0.30~dex lower in than that of C-normal stars, suggesting that the $C_{enh}$ stars are predominantly at the older end of the low $\alpha$ population. However, [Fe/H] does not translate into age in a  straightforward way, so we make this suggestion only tentatively. For the early high-$\alpha$ stars, there is only a small difference (0.08~dex) in mean metallicity between $C_{enh}$ and C-normal stars.

\begin{figure}
    \centering
    \includegraphics[width= \columnwidth]{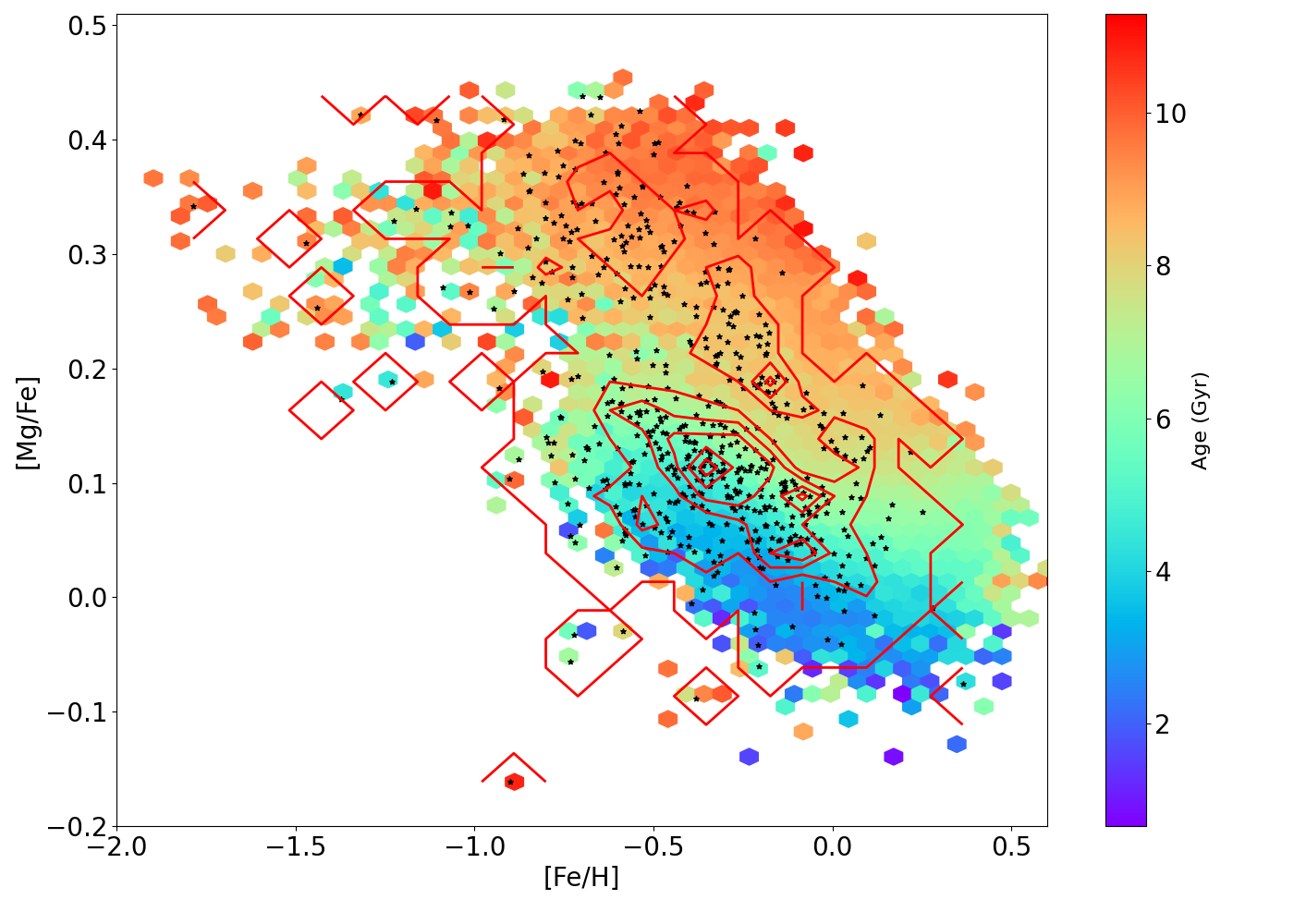}
    \caption{Stars from the selection in the [Mg/Fe]-[Fe/H] plane with colour-coding to represent age. Black dots represent $C_{enh}$ stars and the red contours show the density distribution of these. We note that there are fewer $C_{enh}$ stars in the low $\alpha$ disk with ages younger than $\sim$~4~Gyr. We hypothesize that this is explained by lack of time for the more massive member of an asymmetric binary to evolve to the AGB and WD stages. }
    \label{fig:agebinarity}
\end{figure}

To further illustrate this point, we show in Fig. \ref{fig:agebinarity} the distribution of $C_{enh}$ stars on the [Mg/Fe] vs. [Fe/H] plane, having colour-coded the overall sample by age using the bias-corrected ages from the AstroNN catalogue \citep{Mackereth2019}.
We note the existence of two areas with conspicuously lower numbers of $C_{enh}$ stars. 
Firstly, in both disk sub-populations shown, there are very few carbon-enhanced stars with [Fe/H]$\simgreater$ +0.20, and that numbers also peter out for ages $\simless$ 5 Gyr. 
We hypothesize that this ``zone of exclusion'' is associated with the timescale, $\tau_C$  $\sim 5.9$ Gyr, for the formation of extrinsic $C_{enh}$ stars, as discussed in Sect.~\ref{sec:binarity}. 
That result
implies that the low-$\alpha$ sequence at [Fe/H]$\simgreater$--0.1 is
dominated by stars younger than $\sim$~4-6 Gyr, in good agreement with
previous work on the ages of thin-disk stars \citep{Martig2016}.

There is a second\footnote{A third zone between the high and low $\alpha$ disks is an artifact of the buffer zone described in Sect. \ref{Stellarpops}}. ``zone of exclusion'' visible in Fig. \ref{fig:agebinarity} in the high $\alpha$ disk around --0.2 $\simless$ [Fe/H] $\simless$ +0.5 and corresponding to ages $>$ 8 Gyr.
The cause of this zone is currently unclear. 


\subsubsection{Ages of high-$\alpha$ stars}

\begin{figure} \centering \includegraphics[width=\columnwidth]{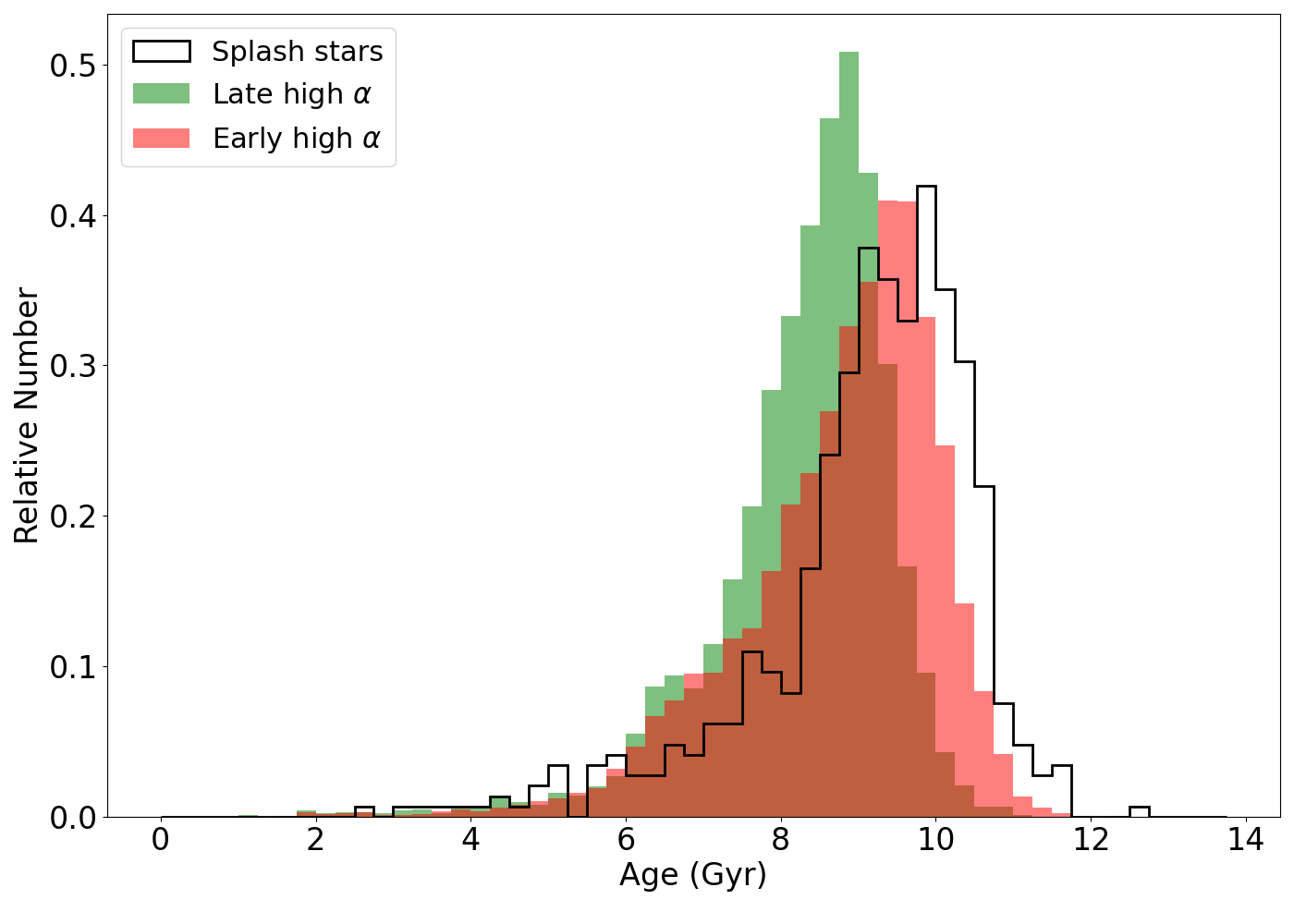}
\caption{Age distribution of stars in the late and early high $\alpha$ populations compared to the age distribution of potential Splash stars. The Splash stars are slightly older than the early high $\alpha$ disk stars and there is a gap of $\sim$ 0.5 Gyr between the early and late high $\alpha$ populations.} 
\label{fig:splash} 
\end{figure}

Theoretical predictions of sub-structure in the high-$\alpha$ disk have been made by \citet{2020MNRAS.497.1603G} and \citet{2022arXiv220402989C}, and other works have indicated the existence of a young $\alpha$-rich (YAR) population within the thick disk \citep{2015A&A...576L..12C,Martig2015,2023A&A...671A..21J}.
As discussed in Sect.~\ref{Stellarpops}, we identified chemical substructure in the high-$\alpha$ disk, in the form of two stellar populations with different characteristic values of [C/Fe].  In this subsection we investigate whether these chemical signatures map into age differences.  
With that goal in mind, we first extracted from the high-$\alpha$ sample stars with orbital eccentricities $>$0.6, which represent the Splash population \citep{2020MNRAS.494.3880B}. 
Fig. \ref{fig:splash} shows that time ordering of the populations. The early/high-$\alpha$ populations are defined according to the dividing line displayed in Fig.~\ref{fig:ColourMg}.  We note that our results are not changed by adoption of a slightly different dividing line.

Based on the peak of the age distributions, the ages of the three populations were determined to be 8.75~Gyr, 9.25~Gyr and 9.75~Gyr, which gave an age difference of~0.50 Gyr between the two high-$\alpha$ populations.
These ages are broadly consistent with a Splash model and a Gaia Sausage/Enceladus (GSE) event 8-10 Gyr ago \citep{2020MNRAS.494.3880B}. 

\begin{figure} \centering \includegraphics[width=\columnwidth]{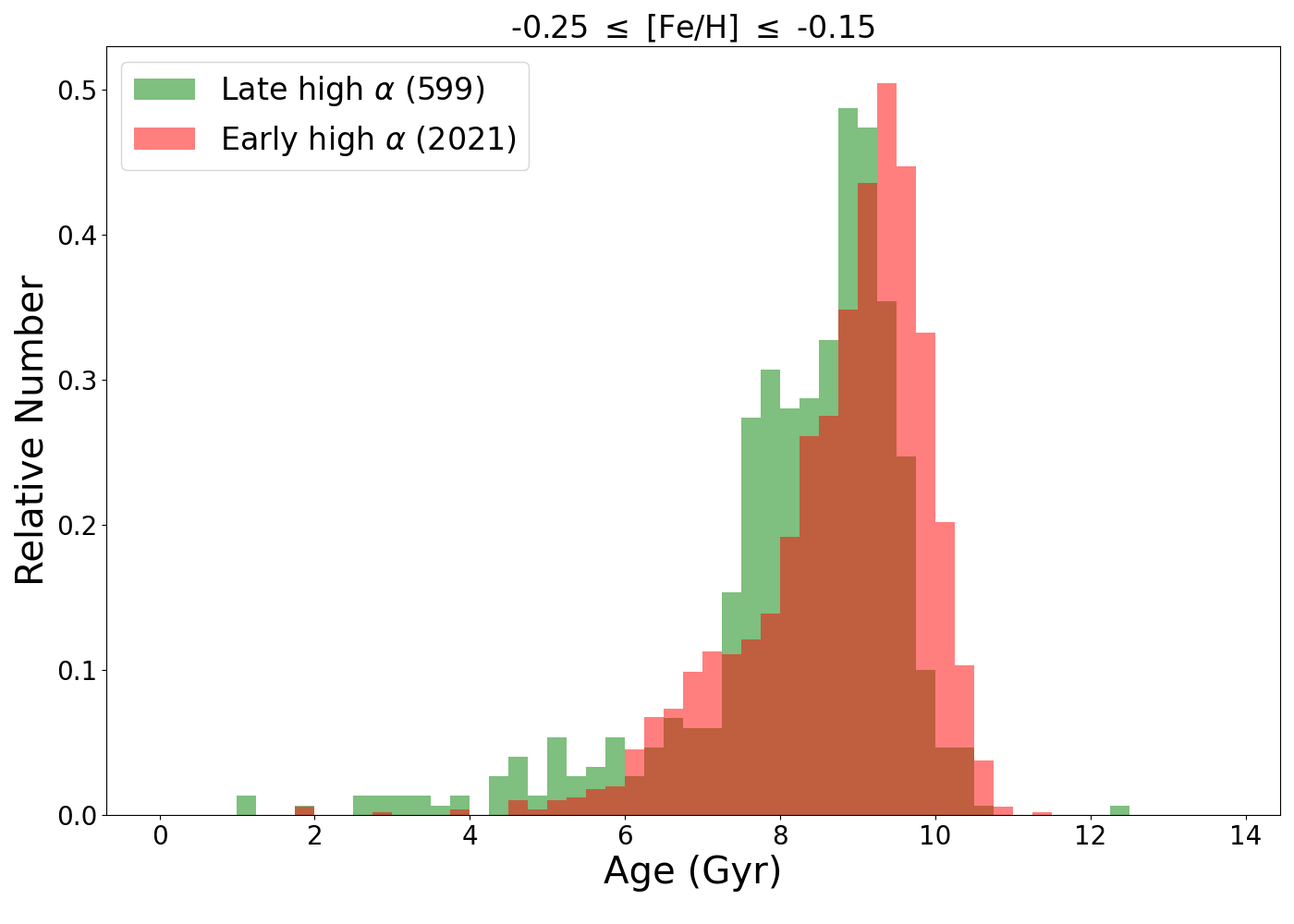}
\caption{Age distribution of stars in the late and early high $\alpha$ populations compared for the metallicity range -0.25 $\leq$ [Fe/H] $\leq$ -0.15, where the two populations overlap. The age difference between the two populations can still be seen. High eccentricity (Splash) stars are not shown as there are too few in the metallicity range for meaningful comparison.} 
\label{fig:ageoverlap} 
\end{figure}

We must apply caution to the above conclusion. The ages used again are those in the AstroNN catalogue which are corrected for AVR (Age-Velocity dispersion Relation) bias \citep[See][]{Mackereth2019}. 
Because of the likely presence of an age-metallicity relation within our sample we needed to examine the possibility that the age difference was due to comparing different metallicity segments within different populations.
Of necessity, the early and late high $\alpha$ populations are constructed such that their overlap in metallicity is small and occurs at the tail of each population distribution.
This makes comparison of whole populations at similar metallicity unworkable. To ensure that the age differences resulted from comparisons made at similar metallicities, we focused instead on a narrow metallicity band in the overlap region.  
In Fig. \ref{fig:ageoverlap}, we show similar histograms for the range of [Fe/H] where the two populations have significant overlap, -0.25 $\leq$ [Fe/H] $\leq$ -0.15. 
The mean ages are 8.76 Gyr and 8.23 Gyr for the earlier and later populations respectively, a difference of 0.53 Gyr.
A ${\chi^2}$ test \footnote{Using a bin width of 0.2 Gyr gave a ${\chi^2}$ statistic of 1469 with 58 degrees of freedom.} gives a probability of only 0.005 that these two samples are drawn from the same population.
The persistence of the population age difference in the overlap populations supports the age differences shown in Fig. \ref{fig:splash}, lending further support to the notion that the late high-$\alpha$ population is associated with a starburst caused by the gas-rich merger of GSE with the Milky Way \citep[e.g.,][]{2020MNRAS.497.1603G}.

\section{Conclusions and summary}\label{sec:conclusions}

The goal of this work is to examine the link between binarity and enhancements in the abundances of carbon and $s$-process elements.  Such a link supports a model according to which these abundance enhancements are caused by mass transfer in a binary system, whereby the atmosphere of the less massive, less evolved binary member is polluted by the material accreting from a more massive, primary AGB star. 
As discussed in the introduction, this mechanism has been used to explain the halo-resident CEMP-$s$ stars, where binarity is the norm, but  we aim to demonstrate this mechanism also explains carbon and $s$-process enhanced disk stars.

We took advantage of the availability of precision abundances for carbon and cerium for a very large sample of disk stars from APOGEE in order to determine existence of a sizeable population of carbon-enhanced stars.
We have used a robust method to determine the degree of enhancement of stars in carbon  relative to the population mean, establishing the presence of a wide range of enhancement values.
We have minimized the presence of AGB stars as far as possible from our sample and have demonstrated that any residual AGB star contamination does not bias our results. 
This approach can be extended to other elements and we have used it here to study the abundance of cerium, which is the best-represented $s$-process element amongst the APOGEE DR17 abundances.

Similar to the availability of stellar catalogues with extensive chemical abundances, there are also now catalogues with binarity information for large numbers of stars.
Thanks to the availability of two such complementary catalogues of likely binary stars, an astrometric catalogue by \citeauthor{Penoyre22} and an RV-based catalogue by \citeauthor{PriceWhelan2020}, we have been able to conduct analyses of binary incidence at the population level.

By combining these two sets of data (Figs. \ref{fig:BinarityHistos} and \ref{fig:CeBinarityHis}) we showed that there is a clear increase in binary frequency with increasing abundance of both carbon and cerium and that the increased abundances of the two are linked. 
This is precisely the behaviour expected when the less massive member of a binary system is enriched by a primary AGB star that is rich in both carbon and $s$-process elements. 
This is strong evidence that the mechanism that explains the halo CEMP-$s$ stars also explains carbon-enhanced disk stars in both the low and high $\alpha$ disks.

By looking at the distribution of carbon abundance with metallicity, we showed that there is a difference in the distribution of carbon-enhanced stars with metallicity between the low and high $\alpha$ disks.
If, as our model proposes, the cause of carbon enhancement is mass transfer then the process must still be ongoing in the low $\alpha$ disk, where the carbon-enhanced stars have a much different distribution to the carbon-normal population whereas, in the high $\alpha$ disk, the process has largely run its course. 
This allows us to set some limits on the ages of the disks.
In Sect. \ref{sec:age} we argued that a large fraction of the low $\alpha$ disk population consists of stars younger than $\sim$~5~Gyr.
Likewise, using a lower limit of 1.2 $M_{\odot}$ for the mass of the primary, we estimate a lower limit of $\sim$~6~Gyr for the ages of stars in the high-$\alpha$ disk. 
This too is in agreement with literature values \citep[e.g.]{Gallart19} 

We note the presence of a second ``zone of exclusion'' in the high $\alpha$ disk. 
There are several possible explanations for this. 
Firstly, this region represents some of the oldest stars in our sample and it may be that the secondary stars in our hypothetical binaries have also evolved beyond the RGB stage and are thus no longer included in our sample. 
Secondly, we cannot rule out selection effects related to orbital period, whereby very widely separated binaries are not detected as such by either the RV or RUWE methods that we have used.
Further analysis of this zone is warranted.

 Furthermore, we have used carbon abundances to demonstrate the existence of two separate populations within the high $\alpha$ disk, an ``early'' and a ``later'' population. 
Theoretical predictions of this clustering have been made by \cite{2020MNRAS.497.1603G} and \cite{2022arXiv220402989C}, and other works have indicated the existence of a young $\alpha$-rich (YAR) population within the thick disk (\cite{2015A&A...576L..12C}, \cite{Martig2015}, \cite{2023A&A...671A..21J}).
While it would be tempting to associate these two populations with the Splash resulting from the Gaia Sausage/Enceladus (GSE) merger, and the population born from the ensuing starburst \citep[e.g.,][]{2020MNRAS.497.1603G}, such an association based solely on the data here is premature and needs further investigation. 


\section{Acknowledgments}
The authors thank the anonymous referee for insightful and very helpful comments.
Funding  for  the  Sloan  Digital  Sky  Survey  IV  has  been  provided
by  the  Alfred  P.  Sloan  Foundation,  the  U.S.  Department  of  Energy  
Office  of  Science,  and  the  Participating  Institutions.  SDSS
acknowledges  support  and  resources  from  the  Center  for  High-
Performance  Computing  at  the  University  of  Utah.  The  SDSS
web  site  is www.sdss.org.
SDSS  is  managed  by  the  Astrophysical  Research  Consortium 
for  the  Participating  Institutions  of  the  SDSS  Collaboration 
including  the  Brazilian  Participation  Group,  the  Carnegie
Institution  for  Science,  Carnegie  Mellon  University,  the  Chilean
Participation  Group,  the  French  Participation  Group,  Harvard-
Smithsonian  Center  for  Astrophysics,  Instituto  de  Astrofísica
de  Canarias,  The  Johns  Hopkins  University,  Kavli  Institute  for
the  Physics  and  Mathematics  of  the  Universe  (IPMU)/
University  of  Tokyo,  the  Korean  Participation  Group,  Lawrence  
Berkeley  National  Laboratory,  Leibniz  Institut  für  Astrophysik 
Potsdam  (AIP),  Max-Planck-Institut  für  Astronomie 
(MPIA  Heidelberg),  Max-Planck-Institut  für  Astrophysik  (MPA  Garching),
Max-Planck-Institut  für  Extraterrestrische  Physik  (MPE),  
National  Astronomical  Observatories  of  China,  New  Mexico  State
University,  New  York  University,  University  of  Notre  Dame,  Ob-
servatório  Nacional/MCTI,  The  Ohio  State  University,  Pennsylvania  State  University,  Shanghai  Astronomical  Observatory,
United  Kingdom  Participation  Group,  Universidad  Nacional
Autónoma  de  México,  University  of  Arizona,  University  of 
Colorado  Boulder,  University  of  Oxford,  University  of  Portsmouth,
University  of  Utah,  University  of  Virginia,  University  of 
Washington,  University  of  Wisconsin,  Vanderbilt  University, 
and  Yale University.
DAGH and OZ acknowledges support provided by the Spanish 
Ministry of Economy and Competitiveness (MINECO) under grant 
AYA-2017-88254-P.
T.C.B. acknowledges partial support for this work from Grant PHY 14-30152: Physics
Frontier Center/JINA Center for the Evolution of the Elements
(JINA-CEE), awarded by the US National Science Foundation.
J.G.F-T gratefully acknowledges the grants support provided by ANID Fondecyt Iniciaci\'on No. 11220340, ANID Fondecyt Postdoc No. 3230001 (Sponsoring researcher), from two Joint Committee ESO-Government of Chile grants under the agreements 2021 ORP 023/2021 and 2023 ORP 062/2023


\section*{Data availability}
The AstroNN and radial velocity catalogues used in this paper are publicly available at the SDSS-IV DR17 website:
https: /www.sdss.org/dr17/.\\
The RUWE binaries database will be shared upon reasonable request to the author (Penoyre).\\
The ASAS-SN variable stars database is publicly available at
https://asas-sn.osu.edu/variables).\\
The catalogue of AGB stars is publicly available though Vizier at https://cdsarc.cds.unistra.fr/viz-bin/cat/J/ApJS/256/43

%
%

\bibliographystyle{aa} 
\bibliography{main} 

\begin{thebibliography}{97}
\expandafter\ifx\csname natexlab\endcsname\relax\def\natexlab#1{#1}\fi

\bibitem[{{Abate} {et~al.}(2018){Abate}, {Pols}, \& {Stancliffe}}]{Abate18}
{Abate}, C., {Pols}, O.~R., \& {Stancliffe}, R.~J. 2018, \aap, 620, A63

\bibitem[{{Abia} {et~al.}(2022){Abia}, {de Laverny}, {Romero-G{\'o}mez}, \& {Figueras}}]{Abia22}
{Abia}, C., {de Laverny}, P., {Romero-G{\'o}mez}, M., \& {Figueras}, F. 2022, \aap, 664, A45

\bibitem[{{Allen} \& {Barbuy}(2006)}]{AllenandBarbuy2006}
{Allen}, D.~M. \& {Barbuy}, B. 2006, \aap, 454, 895

\bibitem[{{Andersen} {et~al.}(2016){Andersen}, {Nordstr{\"o}m}, \& {Hansen}}]{Andersen16}
{Andersen}, J., {Nordstr{\"o}m}, B., \& {Hansen}, T.~T. 2016, IAU Focus Meeting, 29B, 158

\bibitem[{{Badenes} \& {Maoz}(2012)}]{Badenes2012}
{Badenes}, C. \& {Maoz}, D. 2012, \apjl, 749, L11

\bibitem[{{Badenes} {et~al.}(2018){Badenes}, {Mazzola}, {Thompson}, {Covey}, {Freeman}, {Walker}, {Moe}, {Troup}, {Nidever}, {Allende Prieto}, {Andrews}, {Barb{\'a}}, {Beers}, {Bovy}, {Carlberg}, {De Lee}, {Johnson}, {Lewis}, {Majewski}, {Pinsonneault}, {Sobeck}, {Stassun}, {Stringfellow}, \& {Zasowski}}]{Badenes2018}
{Badenes}, C., {Mazzola}, C., {Thompson}, T.~A., {et~al.} 2018, \apj, 854, 147

\bibitem[{{Beers} \& {Christlieb}(2005)}]{beers05}
{Beers}, T.~C. \& {Christlieb}, N. 2005, \araa, 43, 531

\bibitem[{{Belokurov} {et~al.}(2020){Belokurov}, {Sanders}, {Fattahi}, {Smith}, {Deason}, {Evans}, \& {Grand}}]{2020MNRAS.494.3880B}
{Belokurov}, V., {Sanders}, J.~L., {Fattahi}, A., {et~al.} 2020, \mnras, 494, 3880

\bibitem[{{Blanton} {et~al.}(2017){Blanton}, {Bershady}, {Abolfathi}, {Albareti}, {Allende Prieto}, {Almeida}, {Alonso-Garc{\'\i}a}, {Anders}, {Anderson}, {Andrews}, {Aquino-Ort{\'\i}z}, {Arag{\'o}n-Salamanca}, {Argudo-Fern{\'a}ndez}, {Armengaud}, {Aubourg}, {Avila-Reese}, {Badenes}, {Bailey}, {Barger}, {Barrera-Ballesteros}, {Bartosz}, {Bates}, {Baumgarten}, {Bautista}, {Beaton}, {Beers}, {Belfiore}, {Bender}, {Berlind}, {Bernardi}, {Beutler}, {Bird}, {Bizyaev}, {Blanc}, {Blomqvist}, {Bolton}, {Boquien}, {Borissova}, {van den Bosch}, {Bovy}, {Brandt}, {Brinkmann}, {Brownstein}, {Bundy}, {Burgasser}, {Burtin}, {Busca}, {Cappellari}, {Delgado Carigi}, {Carlberg}, {Carnero Rosell}, {Carrera}, {Chanover}, {Cherinka}, {Cheung}, {G{\'o}mez Maqueo Chew}, {Chiappini}, {Choi}, {Chojnowski}, {Chuang}, {Chung}, {Cirolini}, {Clerc}, {Cohen}, {Comparat}, {da Costa}, {Cousinou}, {Covey}, {Crane}, {Croft}, {Cruz-Gonzalez}, {Garrido Cuadra}, {Cunha}, {Damke}, {Darling}, {Davies}, {Dawson}, {de la Macorra}, {Dell'Agli}, {De
  Lee}, {Delubac}, {Di Mille}, {Diamond-Stanic}, {Cano-D{\'\i}az}, {Donor}, {Downes}, {Drory}, {du Mas des Bourboux}, {Duckworth}, {Dwelly}, {Dyer}, {Ebelke}, {Eigenbrot}, {Eisenstein}, {Emsellem}, {Eracleous}, {Escoffier}, {Evans}, {Fan}, {Fern{\'a}ndez-Alvar}, {Fernandez-Trincado}, {Feuillet}, {Finoguenov}, {Fleming}, {Font-Ribera}, {Fredrickson}, {Freischlad}, {Frinchaboy}, {Fuentes}, {Galbany}, {Garcia-Dias}, {Garc{\'\i}a-Hern{\'a}ndez}, {Gaulme}, {Geisler}, {Gelfand}, {Gil-Mar{\'\i}n}, {Gillespie}, {Goddard}, {Gonzalez-Perez}, {Grabowski}, {Green}, {Grier}, {Gunn}, {Guo}, {Guy}, {Hagen}, {Hahn}, {Hall}, {Harding}, {Hasselquist}, {Hawley}, {Hearty}, {Gonzalez Hern{\'a}ndez}, {Ho}, {Hogg}, {Holley-Bockelmann}, {Holtzman}, {Holzer}, {Huehnerhoff}, {Hutchinson}, {Hwang}, {Ibarra-Medel}, {da Silva Ilha}, {Ivans}, {Ivory}, {Jackson}, {Jensen}, {Johnson}, {Jones}, {J{\"o}nsson}, {Jullo}, {Kamble}, {Kinemuchi}, {Kirkby}, {Kitaura}, {Klaene}, {Knapp}, {Kneib}, {Kollmeier}, {Lacerna}, {Lane}, {Lang}, {Law},
  {Lazarz}, {Lee}, {Le Goff}, {Liang}, {Li}, {Li}, {Lian}, {Lima}, {Lin}, {Lin}, {Bertran de Lis}, {Liu}, {de Icaza Lizaola}, {Long}, {Lucatello}, {Lundgren}, {MacDonald}, {Deconto Machado}, {MacLeod}, {Mahadevan}, {Geimba Maia}, {Maiolino}, {Majewski}, {Malanushenko}, {Malanushenko}, {Manchado}, {Mao}, {Maraston}, {Marques-Chaves}, {Masseron}, {Masters}, {McBride}, {McDermid}, {McGrath}, {McGreer}, {Medina Pe{\~n}a}, {Melendez}, {Merloni}, {Merrifield}, {Meszaros}, {Meza}, {Minchev}, {Minniti}, {Miyaji}, {More}, {Mulchaey}, {M{\"u}ller-S{\'a}nchez}, {Muna}, {Munoz}, {Myers}, {Nair}, {Nandra}, {Correa do Nascimento}, {Negrete}, {Ness}, {Newman}, {Nichol}, {Nidever}, {Nitschelm}, {Ntelis}, {O'Connell}, {Oelkers}, {Oravetz}, {Oravetz}, {Pace}, {Padilla}, {Palanque-Delabrouille}, {Alonso Palicio}, {Pan}, {Parejko}, {Parikh}, {P{\^a}ris}, {Park}, {Patten}, {Peirani}, {Pellejero-Ibanez}, {Penny}, {Percival}, {Perez-Fournon}, {Petitjean}, {Pieri}, {Pinsonneault}, {Pisani}, {Poleski}, {Prada}, {Prakash}, {Queiroz},
  {Raddick}, {Raichoor}, {Barboza Rembold}, {Richstein}, {Riffel}, {Riffel}, {Rix}, {Robin}, {Rockosi}, {Rodr{\'\i}guez-Torres}, {Roman-Lopes}, {Rom{\'a}n-Z{\'u}{\~n}iga}, {Rosado}, {Ross}, {Rossi}, {Ruan}, {Ruggeri}, {Rykoff}, {Salazar-Albornoz}, {Salvato}, {S{\'a}nchez}, {Aguado}, {S{\'a}nchez-Gallego}, {Santana}, {Santiago}, {Sayres}, {Schiavon}, {da Silva Schimoia}, {Schlafly}, {Schlegel}, {Schneider}, {Schultheis}, {Schuster}, {Schwope}, {Seo}, {Shao}, {Shen}, {Shetrone}, {Shull}, {Simon}, {Skinner}, {Skrutskie}, {Slosar}, {Smith}, {Sobeck}, {Sobreira}, {Somers}, {Souto}, {Stark}, {Stassun}, {Stauffer}, {Steinmetz}, {Storchi-Bergmann}, {Streblyanska}, {Stringfellow}, {Su{\'a}rez}, {Sun}, {Suzuki}, {Szigeti}, {Taghizadeh-Popp}, {Tang}, {Tao}, {Tayar}, {Tembe}, {Teske}, {Thakar}, {Thomas}, {Thompson}, {Tinker}, {Tissera}, {Tojeiro}, {Hernandez Toledo}, {de la Torre}, {Tremonti}, {Troup}, {Valenzuela}, {Martinez Valpuesta}, {Vargas-Gonz{\'a}lez}, {Vargas-Maga{\~n}a}, {Vazquez}, {Villanova}, {Vivek}, {Vogt},
  {Wake}, {Walterbos}, {Wang}, {Weaver}, {Weijmans}, {Weinberg}, {Westfall}, {Whelan}, {Wild}, {Wilson}, {Wood-Vasey}, {Wylezalek}, {Xiao}, {Yan}, {Yang}, {Ybarra}, {Y{\`e}che}, {Zakamska}, {Zamora}, {Zarrouk}, {Zasowski}, {Zhang}, {Zhao}, {Zheng}, {Zheng}, {Zhou}, {Zhou}, {Zhu}, {Zoccali}, \& {Zou}}]{Blanton2017}
{Blanton}, M.~R., {Bershady}, M.~A., {Abolfathi}, B., {et~al.} 2017, \aj, 154, 28

\bibitem[{{Bohm-Vitense} \& {Johnson}(1985)}]{bohmvitense85}
{Bohm-Vitense}, E. \& {Johnson}, H.~R. 1985, \apj, 293, 288

\bibitem[{{Bondi} \& {Hoyle}(1944)}]{Bondi}
{Bondi}, H. \& {Hoyle}, F. 1944, \mnras, 104, 273

\bibitem[{{Bovy}(2015)}]{Bovy2015}
{Bovy}, J. 2015, \apjs, 216, 29

\bibitem[{{Bovy} {et~al.}(2014){Bovy}, {Nidever}, {Rix}, {Girardi}, {Zasowski}, {Chojnowski}, {Holtzman}, {Epstein}, {Frinchaboy}, {Hayden}, {Rodrigues}, {Majewski}, {Johnson}, {Pinsonneault}, {Stello}, {Allende Prieto}, {Andrews}, {Basu}, {Beers}, {Bizyaev}, {Burton}, {Chaplin}, {Cunha}, {Elsworth}, {Garc{\'\i}a}, {Garc{\'\i}a-Her{\'n}andez}, {Garc{\'\i}a P{\'e}rez}, {Hearty}, {Hekker}, {Kallinger}, {Kinemuchi}, {Koesterke}, {M{\'e}sz{\'a}ros}, {Mosser}, {O'Connell}, {Oravetz}, {Pan}, {Robin}, {Schiavon}, {Schneider}, {Schultheis}, {Serenelli}, {Shetrone}, {Silva Aguirre}, {Simmons}, {Skrutskie}, {Smith}, {Stassun}, {Weinberg}, {Wilson}, \& {Zamora}}]{Bovy2014}
{Bovy}, J., {Nidever}, D.~L., {Rix}, H.-W., {et~al.} 2014, \apj, 790, 127

\bibitem[{{Busso} {et~al.}(1995){Busso}, {Lambert}, {Beglio}, {Gallino}, {Raiteri}, \& {Smith}}]{busso1995}
{Busso}, M., {Lambert}, D.~L., {Beglio}, L., {et~al.} 1995, \apj, 446, 775

\bibitem[{{Chamandy} {et~al.}(2020){Chamandy}, {Blackman}, {Frank}, {Carroll-Nellenback}, \& {Tu}}]{Chamandy}
{Chamandy}, L., {Blackman}, E.~G., {Frank}, A., {Carroll-Nellenback}, J., \& {Tu}, Y. 2020, \mnras, 495, 4028

\bibitem[{{Charbonnel} {et~al.}(1998){Charbonnel}, {Brown}, \& {Wallerstein}}]{charbonnel98}
{Charbonnel}, C., {Brown}, J.~A., \& {Wallerstein}, G. 1998, \aap, 332, 204

\bibitem[{{Chen} {et~al.}(2020){Chen}, {Ivanova}, \& {Carroll-Nellenback}}]{Chen20}
{Chen}, Z., {Ivanova}, N., \& {Carroll-Nellenback}, J. 2020, \apj, 892, 110

\bibitem[{{Chiappini} {et~al.}(2015){Chiappini}, {Anders}, {Rodrigues}, {Miglio}, {Montalb{\'a}n}, {Mosser}, {Girardi}, {Valentini}, {Noels}, {Morel}, {Minchev}, {Steinmetz}, {Santiago}, {Schultheis}, {Martig}, {da Costa}, {Maia}, {Allende Prieto}, {de Assis Peralta}, {Hekker}, {Theme{\ss}l}, {Kallinger}, {Garc{\'\i}a}, {Mathur}, {Baudin}, {Beers}, {Cunha}, {Harding}, {Holtzman}, {Majewski}, {M{\'e}sz{\'a}ros}, {Nidever}, {Pan}, {Schiavon}, {Shetrone}, {Schneider}, \& {Stassun}}]{2015A&A...576L..12C}
{Chiappini}, C., {Anders}, F., {Rodrigues}, T.~S., {et~al.} 2015, \aap, 576, L12

\bibitem[{{Chiavassa} {et~al.}(2020){Chiavassa}, {Kravchenko}, {Millour}, {Schaefer}, {Schultheis}, {Freytag}, {Creevey}, {Hocd{\'e}}, {Morand}, {Ligi}, {Kraus}, {Monnier}, {Mourard}, {Nardetto}, {Anugu}, {Le Bouquin}, {Davies}, {Ennis}, {Gardner}, {Labdon}, {Lanthermann}, {Setterholm}, \& {ten Brummelaar}}]{Chiavassa2020}
{Chiavassa}, A., {Kravchenko}, K., {Millour}, F., {et~al.} 2020, \aap, 640, A23

\bibitem[{{Cody} {et~al.}(2014){Cody}, {Stauffer}, {Baglin}, {Micela}, {Rebull}, {Flaccomio}, {Morales-Calder{\'o}n}, {Aigrain}, {Bouvier}, {Hillenbrand}, {Gutermuth}, {Song}, {Turner}, {Alencar}, {Zwintz}, {Plavchan}, {Carpenter}, {Findeisen}, {Carey}, {Terebey}, {Hartmann}, {Calvet}, {Teixeira}, {Vrba}, {Wolk}, {Covey}, {Poppenhaeger}, {G{\"u}nther}, {Forbrich}, {Whitney}, {Affer}, {Herbst}, {Hora}, {Barrado}, {Holtzman}, {Marchis}, {Wood}, {Medeiros Guimar{\~a}es}, {Lillo Box}, {Gillen}, {McQuillan}, {Espaillat}, {Allen}, {D'Alessio}, \& {Favata}}]{Cody}
{Cody}, A.~M., {Stauffer}, J., {Baglin}, A., {et~al.} 2014, \aj, 147, 82

\bibitem[{{Conroy} {et~al.}(2022){Conroy}, {Weinberg}, {Naidu}, {Buck}, {Johnson}, {Cargile}, {Bonaca}, {Caldwell}, {Chandra}, {Han}, {Johnson}, {Speagle}, {Ting}, {Woody}, \& {Zaritsky}}]{2022arXiv220402989C}
{Conroy}, C., {Weinberg}, D.~H., {Naidu}, R.~P., {et~al.} 2022, arXiv e-prints, arXiv:2204.02989

\bibitem[{{Cristallo} {et~al.}(2016){Cristallo}, {Piersanti}, \& {Straniero}}]{FRUITY}
{Cristallo}, S., {Piersanti}, L., \& {Straniero}, O. 2016, in Journal of Physics Conference Series, Vol. 665, Journal of Physics Conference Series, 012019

\bibitem[{{Crossfield} {et~al.}(2019){Crossfield}, {Lothringer}, {Flores}, {Mills}, {Freedman}, {Valverde}, {Miles}, {Guo}, \& {Skemer}}]{Crossfield19}
{Crossfield}, I.~J.~M., {Lothringer}, J.~D., {Flores}, B., {et~al.} 2019, \apjl, 871, L3

\bibitem[{{Cseh} {et~al.}(2018){Cseh}, {Lugaro}, {D'Orazi}, {de Castro}, {Pereira}, {Karakas}, {Moln{\'a}r}, {Plachy}, {Szab{\'o}}, {Pignatari}, \& {Cristallo}}]{Cseh2018}
{Cseh}, B., {Lugaro}, M., {D'Orazi}, V., {et~al.} 2018, \aap, 620, A146

\bibitem[{{Cseh} {et~al.}(2022){Cseh}, {Vil{\'a}gos}, {Roriz}, {Pereira}, {D'Orazi}, {Karakas}, {So{\'o}s}, {Drake}, {Junqueira}, \& {Lugaro}}]{Cseh2022}
{Cseh}, B., {Vil{\'a}gos}, B., {Roriz}, M.~P., {et~al.} 2022, \aap, 660, A128

\bibitem[{{de Castro} {et~al.}(2016){de Castro}, {Pereira}, {Roig}, {Jilinski}, {Drake}, {Chavero}, \& {Sales Silva}}]{deCastro16}
{de Castro}, D.~B., {Pereira}, C.~B., {Roig}, F., {et~al.} 2016, \mnras, 459, 4299

\bibitem[{{Eisenstein} {et~al.}(2011){Eisenstein}, {Weinberg}, {Agol}, {Aihara}, {Allende Prieto}, {Anderson}, {Arns}, {Aubourg}, {Bailey}, {Balbinot}, {Barkhouser}, {Beers}, {Berlind}, {Bickerton}, {Bizyaev}, {Blanton}, {Bochanski}, {Bolton}, {Bosman}, {Bovy}, {Brandt}, {Breslauer}, {Brewington}, {Brinkmann}, {Brown}, {Brownstein}, {Burger}, {Busca}, {Campbell}, {Cargile}, {Carithers}, {Carlberg}, {Carr}, {Chang}, {Chen}, {Chiappini}, {Comparat}, {Connolly}, {Cortes}, {Croft}, {Cunha}, {da Costa}, {Davenport}, {Dawson}, {De Lee}, {Porto de Mello}, {de Simoni}, {Dean}, {Dhital}, {Ealet}, {Ebelke}, {Edmondson}, {Eiting}, {Escoffier}, {Esposito}, {Evans}, {Fan}, {Femen{\'\i}a Castell{\'a}}, {Dutra Ferreira}, {Fitzgerald}, {Fleming}, {Font-Ribera}, {Ford}, {Frinchaboy}, {Garc{\'\i}a P{\'e}rez}, {Gaudi}, {Ge}, {Ghezzi}, {Gillespie}, {Gilmore}, {Girardi}, {Gott}, {Gould}, {Grebel}, {Gunn}, {Hamilton}, {Harding}, {Harris}, {Hawley}, {Hearty}, {Hennawi}, {Gonz{\'a}lez Hern{\'a}ndez}, {Ho}, {Hogg}, {Holtzman},
  {Honscheid}, {Inada}, {Ivans}, {Jiang}, {Jiang}, {Johnson}, {Jordan}, {Jordan}, {Kauffmann}, {Kazin}, {Kirkby}, {Klaene}, {Knapp}, {Kneib}, {Kochanek}, {Koesterke}, {Kollmeier}, {Kron}, {Lampeitl}, {Lang}, {Lawler}, {Le Goff}, {Lee}, {Lee}, {Leisenring}, {Lin}, {Liu}, {Long}, {Loomis}, {Lucatello}, {Lundgren}, {Lupton}, {Ma}, {Ma}, {MacDonald}, {Mack}, {Mahadevan}, {Maia}, {Majewski}, {Makler}, {Malanushenko}, {Malanushenko}, {Mandelbaum}, {Maraston}, {Margala}, {Maseman}, {Masters}, {McBride}, {McDonald}, {McGreer}, {McMahon}, {Mena Requejo}, {M{\'e}nard}, {Miralda-Escud{\'e}}, {Morrison}, {Mullally}, {Muna}, {Murayama}, {Myers}, {Naugle}, {Neto}, {Nguyen}, {Nichol}, {Nidever}, {O'Connell}, {Ogando}, {Olmstead}, {Oravetz}, {Padmanabhan}, {Paegert}, {Palanque-Delabrouille}, {Pan}, {Pandey}, {Parejko}, {P{\^a}ris}, {Pellegrini}, {Pepper}, {Percival}, {Petitjean}, {Pfaffenberger}, {Pforr}, {Phleps}, {Pichon}, {Pieri}, {Prada}, {Price-Whelan}, {Raddick}, {Ramos}, {Reid}, {Reyle}, {Rich}, {Richards}, {Rieke},
  {Rieke}, {Rix}, {Robin}, {Rocha-Pinto}, {Rockosi}, {Roe}, {Rollinde}, {Ross}, {Ross}, {Rossetto}, {S{\'a}nchez}, {Santiago}, {Sayres}, {Schiavon}, {Schlegel}, {Schlesinger}, {Schmidt}, {Schneider}, {Sellgren}, {Shelden}, {Sheldon}, {Shetrone}, {Shu}, {Silverman}, {Simmerer}, {Simmons}, {Sivarani}, {Skrutskie}, {Slosar}, {Smee}, {Smith}, {Snedden}, {Stassun}, {Steele}, {Steinmetz}, {Stockett}, {Stollberg}, {Strauss}, {Szalay}, {Tanaka}, {Thakar}, {Thomas}, {Tinker}, {Tofflemire}, {Tojeiro}, {Tremonti}, {Vargas Maga{\~n}a}, {Verde}, {Vogt}, {Wake}, {Wan}, {Wang}, {Weaver}, {White}, {White}, {Wilson}, {Wisniewski}, {Wood-Vasey}, {Yanny}, {Yasuda}, {Y{\`e}che}, {York}, {Young}, {Zasowski}, {Zehavi}, \& {Zhao}}]{Eisenstein2011}
{Eisenstein}, D.~J., {Weinberg}, D.~H., {Agol}, E., {et~al.} 2011, \aj, 142, 72

\bibitem[{{Fischer} \& {Marcy}(1992)}]{Fischer1992}
{Fischer}, D.~A. \& {Marcy}, G.~W. 1992, \apj, 396, 178

\bibitem[{{Gallart} {et~al.}(2019){Gallart}, {Bernard}, {Brook}, {Ruiz-Lara}, {Cassisi}, {Hill}, \& {Monelli}}]{Gallart19}
{Gallart}, C., {Bernard}, E.~J., {Brook}, C.~B., {et~al.} 2019, Nature Astronomy, 3, 932

\bibitem[{{Gao} {et~al.}(2017){Gao}, {Zhao}, {Yang}, \& {Gao}}]{2017MNRAS.469L..68G}
{Gao}, S., {Zhao}, H., {Yang}, H., \& {Gao}, R. 2017, \mnras, 469, L68

\bibitem[{{Garc{\'\i}a-Hern{\'a}ndez} {et~al.}(2007){Garc{\'\i}a-Hern{\'a}ndez}, {Garc{\'\i}a-Lario}, {Plez}, {Manchado}, {D'Antona}, {Lub}, \& {Habing}}]{GH07}
{Garc{\'\i}a-Hern{\'a}ndez}, D.~A., {Garc{\'\i}a-Lario}, P., {Plez}, B., {et~al.} 2007, \aap, 462, 711

\bibitem[{{Garc{\'\i}a-Hern{\'a}ndez} {et~al.}(2013){Garc{\'\i}a-Hern{\'a}ndez}, {Zamora}, {Yag{\"u}e}, {Uttenthaler}, {Karakas}, {Lugaro}, {Ventura}, \& {Lambert}}]{2013A&A...555L...3G}
{Garc{\'\i}a-Hern{\'a}ndez}, D.~A., {Zamora}, O., {Yag{\"u}e}, A., {et~al.} 2013, \aap, 555, L3

\bibitem[{{Garc{\'\i}a P{\'e}rez} {et~al.}(2016){Garc{\'\i}a P{\'e}rez}, {Allende Prieto}, {Holtzman}, {Shetrone}, {M{\'e}sz{\'a}ros}, {Bizyaev}, {Carrera}, {Cunha}, {Garc{\'\i}a-Hern{\'a}ndez}, {Johnson}, {Majewski}, {Nidever}, {Schiavon}, {Shane}, {Smith}, {Sobeck}, {Troup}, {Zamora}, {Weinberg}, {Bovy}, {Eisenstein}, {Feuillet}, {Frinchaboy}, {Hayden}, {Hearty}, {Nguyen}, {O'Connell}, {Pinsonneault}, {Wilson}, \& {Zasowski}}]{garcia16}
{Garc{\'\i}a P{\'e}rez}, A.~E., {Allende Prieto}, C., {Holtzman}, J.~A., {et~al.} 2016, \aj, 151, 144

\bibitem[{{Girardi}(2016)}]{Girardi}
{Girardi}, L. 2016, \araa, 54, 95

\bibitem[{{Grand} {et~al.}(2020){Grand}, {Kawata}, {Belokurov}, {Deason}, {Fattahi}, {Fragkoudi}, {G{\'o}mez}, {Marinacci}, \& {Pakmor}}]{2020MNRAS.497.1603G}
{Grand}, R. J.~J., {Kawata}, D., {Belokurov}, V., {et~al.} 2020, \mnras, 497, 1603

\bibitem[{{Guinan} \& {Engle}(2009)}]{Guinan2009}
{Guinan}, E.~F. \& {Engle}, S.~G. 2009, in The Ages of Stars, ed. E.~E. {Mamajek}, D.~R. {Soderblom}, \& R.~F.~G. {Wyse}, Vol. 258, 395--408

\bibitem[{{Han} {et~al.}(1995){Han}, {Eggleton}, {Podsiadlowski}, \& {Tout}}]{han95}
{Han}, Z., {Eggleton}, P.~P., {Podsiadlowski}, P., \& {Tout}, C.~A. 1995, \mnras, 277, 1443

\bibitem[{{Hansen} {et~al.}(2016{\natexlab{a}}){Hansen}, {Nordstr{\"o}m}, {Hansen}, {Kennedy}, {Placco}, {Beers}, {Andersen}, {Cescutti}, \& {Chiappini}}]{Hansen16}
{Hansen}, C.~J., {Nordstr{\"o}m}, B., {Hansen}, T.~T., {et~al.} 2016{\natexlab{a}}, \aap, 588, A37

\bibitem[{{Hansen} {et~al.}(2016{\natexlab{b}}){Hansen}, {Andersen}, {Nordstr{\"o}m}, {Beers}, {Placco}, {Yoon}, \& {Buchhave}}]{Hansen16a}
{Hansen}, T.~T., {Andersen}, J., {Nordstr{\"o}m}, B., {et~al.} 2016{\natexlab{b}}, \aap, 586, A160

\bibitem[{{Hayden} {et~al.}(2015){Hayden}, {Bovy}, {Holtzman}, {Nidever}, {Bird}, {Weinberg}, {Andrews}, {Majewski}, {Allende Prieto}, {Anders}, {Beers}, {Bizyaev}, {Chiappini}, {Cunha}, {Frinchaboy}, {Garc{\'\i}a-Her{\'n}andez}, {Garc{\'\i}a P{\'e}rez}, {Girardi}, {Harding}, {Hearty}, {Johnson}, {M{\'e}sz{\'a}ros}, {Minchev}, {O'Connell}, {Pan}, {Robin}, {Schiavon}, {Schneider}, {Schultheis}, {Shetrone}, {Skrutskie}, {Steinmetz}, {Smith}, {Wilson}, {Zamora}, \& {Zasowski}}]{Hayden2015}
{Hayden}, M.~R., {Bovy}, J., {Holtzman}, J.~A., {et~al.} 2015, \apj, 808, 132

\bibitem[{{Heber} {et~al.}(1993){Heber}, {Bade}, {Jordan}, \& {Voges}}]{Heber}
{Heber}, U., {Bade}, N., {Jordan}, S., \& {Voges}, W. 1993, \aap, 267, L31

\bibitem[{{Holl} {et~al.}(2022){Holl}, {Fabricius}, {Portell}, {Lindegren}, {Panuzzo}, {Bernet}, {Casta{\~n}eda}, {Jevardat de Fombelle}, {Audard}, {Ducourant}, {Harrison}, {Evans}, {Busso}, {Sozzetti}, {Gosset}, {Arenou}, {De Angeli}, {Riello}, {Eyer}, {Rimoldini}, {Gavras}, {Mowlavi}, {Nienartowicz}, {Lecoeur-Ta{\"\i}bi}, {Garc{\'\i}a-Lario}, \& {Pourbaix}}]{Holl2022}
{Holl}, B., {Fabricius}, C., {Portell}, J., {et~al.} 2022, arXiv e-prints, arXiv:2212.11971

\bibitem[{{Holtzman} {et~al.}(2018){Holtzman}, {Hasselquist}, {Shetrone}, {Cunha}, {Allende Prieto}, {Anguiano}, {Bizyaev}, {Bovy}, {Casey}, {Edvardsson}, {Johnson}, {J{\"o}nsson}, {Meszaros}, {Smith}, {Sobeck}, {Zamora}, {Chojnowski}, {Fernandez-Trincado}, {Garcia-Hernandez}, {Majewski}, {Pinsonneault}, {Souto}, {Stringfellow}, {Tayar}, {Troup}, \& {Zasowski}}]{Holtzman2018}
{Holtzman}, J.~A., {Hasselquist}, S., {Shetrone}, M., {et~al.} 2018, \aj, 156, 125

\bibitem[{{Indebetouw} {et~al.}(2005){Indebetouw}, {Mathis}, {Babler}, {Meade}, {Watson}, {Whitney}, {Wolff}, {Wolfire}, {Cohen}, {Bania}, {Benjamin}, {Clemens}, {Dickey}, {Jackson}, {Kobulnicky}, {Marston}, {Mercer}, {Stauffer}, {Stolovy}, \& {Churchwell}}]{Indebetouw}
{Indebetouw}, R., {Mathis}, J.~S., {Babler}, B.~L., {et~al.} 2005, \apj, 619, 931

\bibitem[{{Jayasinghe} {et~al.}(2021){Jayasinghe}, {Kochanek}, {Stanek}, {Shappee}, {Holoien}, {Thompson}, {Prieto}, {Dong}, {Pawlak}, {Pejcha}, {Pojmanski}, {Otero}, {Hurst}, \& {Will}}]{ASASSN}
{Jayasinghe}, T., {Kochanek}, C.~S., {Stanek}, K.~Z., {et~al.} 2021, \mnras, 503, 200

\bibitem[{{Jofr{\'e}} {et~al.}(2023){Jofr{\'e}}, {Jorissen}, {Aguilera-G{\'o}mez}, {Van Eck}, {Tayar}, {Pinsonneault}, {Zinn}, {Goriely}, \& {Van Winckel}}]{2023A&A...671A..21J}
{Jofr{\'e}}, P., {Jorissen}, A., {Aguilera-G{\'o}mez}, C., {et~al.} 2023, \aap, 671, A21

\bibitem[{{J{\"o}nsson} {et~al.}(2020){J{\"o}nsson}, {Holtzman}, {Allende Prieto}, {Cunha}, {Garc{\'\i}a-Hern{\'a}ndez}, {Hasselquist}, {Masseron}, {Osorio}, {Shetrone}, {Smith}, {Stringfellow}, {Bizyaev}, {Edvardsson}, {Majewski}, {M{\'e}sz{\'a}ros}, {Souto}, {Zamora}, {Beaton}, {Bovy}, {Donor}, {Pinsonneault}, {Poovelil}, \& {Sobeck}}]{Jonsson2020}
{J{\"o}nsson}, H., {Holtzman}, J.~A., {Allende Prieto}, C., {et~al.} 2020, \aj, 160, 120

\bibitem[{{Jorissen} {et~al.}(2016){Jorissen}, {Van Eck}, {Van Winckel}, {Merle}, {Boffin}, {Andersen}, {Nordstr{\"o}m}, {Udry}, {Masseron}, {Lenaerts}, \& {Waelkens}}]{jorissen16}
{Jorissen}, A., {Van Eck}, S., {Van Winckel}, H., {et~al.} 2016, \aap, 586, A158

\bibitem[{{Karakas} {et~al.}(2022){Karakas}, {Cinquegrana}, \& {Joyce}}]{Karakas22}
{Karakas}, A.~I., {Cinquegrana}, G., \& {Joyce}, M. 2022, \mnras, 509, 4430

\bibitem[{{Karakas} \& {Lattanzio}(2014)}]{Karakas2014}
{Karakas}, A.~I. \& {Lattanzio}, J.~C. 2014, \pasa, 31, e030

\bibitem[{{Karakas} {et~al.}(2002){Karakas}, {Lattanzio}, \& {Pols}}]{Karakas02}
{Karakas}, A.~I., {Lattanzio}, J.~C., \& {Pols}, O.~R. 2002, \pasa, 19, 515

\bibitem[{{Keenan}(1942)}]{keenan42}
{Keenan}, P.~C. 1942, \apj, 96, 101

\bibitem[{{Keenan} \& {Morgan}(1941)}]{KM1941}
{Keenan}, P.~C. \& {Morgan}, W.~W. 1941, \apj, 94, 501

\bibitem[{{Lee} {et~al.}(2013){Lee}, {Beers}, {Masseron}, {Plez}, {Rockosi}, {Sobeck}, {Yanny}, {Lucatello}, {Sivarani}, {Placco}, \& {Carollo}}]{lee13}
{Lee}, Y.~S., {Beers}, T.~C., {Masseron}, T., {et~al.} 2013, \aj, 146, 132

\bibitem[{{Leung} \& {Bovy}(2019)}]{Leung2019}
{Leung}, H.~W. \& {Bovy}, J. 2019, \mnras, 489, 2079

\bibitem[{{Liebert} {et~al.}(1994){Liebert}, {Schmidt}, {Lesser}, {Stepanian}, {Lipovetsky}, {Chaffe}, {Foltz}, \& {Bergeron}}]{Liebert}
{Liebert}, J., {Schmidt}, G.~D., {Lesser}, M., {et~al.} 1994, \apj, 421, 733

\bibitem[{{Liu} {et~al.}(2017){Liu}, {Ruchti}, {Feltzing}, \& {Primas}}]{Liu2017}
{Liu}, C., {Ruchti}, G., {Feltzing}, S., \& {Primas}, F. 2017, \aap, 601, A31

\bibitem[{{Lodders}(2019)}]{Lodders2019}
{Lodders}, K. 2019, arXiv e-prints, arXiv:1912.00844

\bibitem[{{Lucatello} {et~al.}(2005){Lucatello}, {Tsangarides}, {Beers}, {Carretta}, {Gratton}, \& {Ryan}}]{lucatello05}
{Lucatello}, S., {Tsangarides}, S., {Beers}, T.~C., {et~al.} 2005, \apj, 625, 825

\bibitem[{{Maben} {et~al.}(2023){Maben}, {Kumar}, {Reddy}, {Campbell}, \& {Zhao}}]{Maben2023}
{Maben}, S., {Kumar}, Y.~B., {Reddy}, B.~E., {Campbell}, S.~W., \& {Zhao}, G. 2023, \mnras, 525, 4554

\bibitem[{{Mackereth} {et~al.}(2019){Mackereth}, {Bovy}, {Leung}, {Schiavon}, {Trick}, {Chaplin}, {Cunha}, {Feuillet}, {Majewski}, {Martig}, {Miglio}, {Nidever}, {Pinsonneault}, {Aguirre}, {Sobeck}, {Tayar}, \& {Zasowski}}]{Mackereth2019}
{Mackereth}, J.~T., {Bovy}, J., {Leung}, H.~W., {et~al.} 2019, \mnras, 489, 176

\bibitem[{{Mackereth} {et~al.}(2017){Mackereth}, {Bovy}, {Schiavon}, {Zasowski}, {Cunha}, {Frinchaboy}, {Garc{\'\i}a Perez}, {Hayden}, {Holtzman}, {Majewski}, {M{\'e}sz{\'a}ros}, {Nidever}, {Pinsonneault}, \& {Shetrone}}]{Mackereth2017}
{Mackereth}, J.~T., {Bovy}, J., {Schiavon}, R.~P., {et~al.} 2017, \mnras, 471, 3057

\bibitem[{{Majewski} {et~al.}(2017){Majewski}, {Schiavon}, {Frinchaboy}, {Allende Prieto}, {Barkhouser}, {Bizyaev}, {Blank}, {Brunner}, {Burton}, {Carrera}, {Chojnowski}, {Cunha}, {Epstein}, {Fitzgerald}, {Garc{\'\i}a P{\'e}rez}, {Hearty}, {Henderson}, {Holtzman}, {Johnson}, {Lam}, {Lawler}, {Maseman}, {M{\'e}sz{\'a}ros}, {Nelson}, {Nguyen}, {Nidever}, {Pinsonneault}, {Shetrone}, {Smee}, {Smith}, {Stolberg}, {Skrutskie}, {Walker}, {Wilson}, {Zasowski}, {Anders}, {Basu}, {Beland}, {Blanton}, {Bovy}, {Brownstein}, {Carlberg}, {Chaplin}, {Chiappini}, {Eisenstein}, {Elsworth}, {Feuillet}, {Fleming}, {Galbraith-Frew}, {Garc{\'\i}a}, {Garc{\'\i}a-Hern{\'a}ndez}, {Gillespie}, {Girardi}, {Gunn}, {Hasselquist}, {Hayden}, {Hekker}, {Ivans}, {Kinemuchi}, {Klaene}, {Mahadevan}, {Mathur}, {Mosser}, {Muna}, {Munn}, {Nichol}, {O'Connell}, {Parejko}, {Robin}, {Rocha-Pinto}, {Schultheis}, {Serenelli}, {Shane}, {Silva Aguirre}, {Sobeck}, {Thompson}, {Troup}, {Weinberg}, \& {Zamora}}]{majewski17}
{Majewski}, S.~R., {Schiavon}, R.~P., {Frinchaboy}, P.~M., {et~al.} 2017, \aj, 154, 94

\bibitem[{{Marigo} \& {Girardi}(2007)}]{Marigo2007}
{Marigo}, P. \& {Girardi}, L. 2007, in Astronomical Society of the Pacific Conference Series, Vol. 374, From Stars to Galaxies: Building the Pieces to Build Up the Universe, ed. A.~{Vallenari}, R.~{Tantalo}, L.~{Portinari}, \& A.~{Moretti}, 33

\bibitem[{{Martig} {et~al.}(2016){Martig}, {Fouesneau}, {Rix}, {Ness}, {M{\'e}sz{\'a}ros}, {Garc{\'\i}a-Hern{\'a}ndez}, {Pinsonneault}, {Serenelli}, {Silva Aguirre}, \& {Zamora}}]{Martig2016}
{Martig}, M., {Fouesneau}, M., {Rix}, H.-W., {et~al.} 2016, \mnras, 456, 3655

\bibitem[{{Martig} {et~al.}(2015){Martig}, {Rix}, {Silva Aguirre}, {Hekker}, {Mosser}, {Elsworth}, {Bovy}, {Stello}, {Anders}, {Garc{\'\i}a}, {Tayar}, {Rodrigues}, {Basu}, {Carrera}, {Ceillier}, {Chaplin}, {Chiappini}, {Frinchaboy}, {Garc{\'\i}a-Hern{\'a}ndez}, {Hearty}, {Holtzman}, {Johnson}, {Majewski}, {Mathur}, {M{\'e}sz{\'a}ros}, {Miglio}, {Nidever}, {Pan}, {Pinsonneault}, {Schiavon}, {Schneider}, {Serenelli}, {Shetrone}, \& {Zamora}}]{Martig2015}
{Martig}, M., {Rix}, H.-W., {Silva Aguirre}, V., {et~al.} 2015, \mnras, 451, 2230

\bibitem[{{Masseron} \& {Gilmore}(2015)}]{Masseron2015}
{Masseron}, T. \& {Gilmore}, G. 2015, \mnras, 453, 1855

\bibitem[{{Mazzola} {et~al.}(2020){Mazzola}, {Badenes}, {Moe}, {Koposov}, {Kounkel}, {Kratter}, {Covey}, {Walker}, {Thompson}, {Andrews}, {Freeman}, {Anguiano}, {Carlberg}, {De Lee}, {Frinchaboy}, {Lewis}, {Majewski}, {Nidever}, {Nitschelm}, {Price-Whelan}, {Roman-Lopes}, {Stassun}, \& {Troup}}]{Mazzola2020}
{Mazzola}, C.~N., {Badenes}, C., {Moe}, M., {et~al.} 2020, \mnras, 499, 1607

\bibitem[{{McClure}(1984)}]{mcclure84}
{McClure}, R.~D. 1984, \apjl, 280, L31

\bibitem[{{McClure} {et~al.}(1980){McClure}, {Fletcher}, \& {Nemec}}]{mcclure80}
{McClure}, R.~D., {Fletcher}, J.~M., \& {Nemec}, J.~M. 1980, \apjl, 238, L35

\bibitem[{{McClure} \& {Woodsworth}(1990)}]{mcclure90}
{McClure}, R.~D. \& {Woodsworth}, A.~W. 1990, \apj, 352, 709

\bibitem[{{McMillan}(2017)}]{McMillan2017}
{McMillan}, P.~J. 2017, \mnras, 465, 76

\bibitem[{{Moe} {et~al.}(2019){Moe}, {Kratter}, \& {Badenes}}]{Moe19}
{Moe}, M., {Kratter}, K.~M., \& {Badenes}, C. 2019, \apj, 875, 61

\bibitem[{{Moedas} {et~al.}(2022){Moedas}, {Deal}, {Bossini}, \& {Campilho}}]{Moedas2022}
{Moedas}, N., {Deal}, M., {Bossini}, D., \& {Campilho}, B. 2022, \aap, 666, A43

\bibitem[{{Nidever} {et~al.}(2014){Nidever}, {Bovy}, {Bird}, {Andrews}, {Hayden}, {Holtzman}, {Majewski}, {Smith}, {Robin}, {Garc{\'\i}a P{\'e}rez}, {Cunha}, {Allende Prieto}, {Zasowski}, {Schiavon}, {Johnson}, {Weinberg}, {Feuillet}, {Schneider}, {Shetrone}, {Sobeck}, {Garc{\'\i}a-Hern{\'a}ndez}, {Zamora}, {Rix}, {Beers}, {Wilson}, {O'Connell}, {Minchev}, {Chiappini}, {Anders}, {Bizyaev}, {Brewington}, {Ebelke}, {Frinchaboy}, {Ge}, {Kinemuchi}, {Malanushenko}, {Malanushenko}, {Marchante}, {M{\'e}sz{\'a}ros}, {Oravetz}, {Pan}, {Simmons}, \& {Skrutskie}}]{Nidever2014}
{Nidever}, D.~L., {Bovy}, J., {Bird}, J.~C., {et~al.} 2014, \apj, 796, 38

\bibitem[{{Nidever} {et~al.}(2015){Nidever}, {Holtzman}, {Allende Prieto}, {Beland}, {Bender}, {Bizyaev}, {Burton}, {Desphande}, {Fleming}, {Garc{\'\i}a P{\'e}rez}, {Hearty}, {Majewski}, {M{\'e}sz{\'a}ros}, {Muna}, {Nguyen}, {Schiavon}, {Shetrone}, {Skrutskie}, {Sobeck}, \& {Wilson}}]{Nidever2015}
{Nidever}, D.~L., {Holtzman}, J.~A., {Allende Prieto}, C., {et~al.} 2015, \aj, 150, 173

\bibitem[{{Norris} {et~al.}(1997){Norris}, {Ryan}, \& {Beers}}]{norris97}
{Norris}, J.~E., {Ryan}, S.~G., \& {Beers}, T.~C. 1997, \apj, 488, 350

\bibitem[{{Penoyre} {et~al.}(2022){Penoyre}, {Belokurov}, \& {Evans}}]{Penoyre22}
{Penoyre}, Z., {Belokurov}, V., \& {Evans}, N.~W. 2022, \mnras, 513, 5270

\bibitem[{{Price-Whelan} {et~al.}(2020){Price-Whelan}, {Hogg}, {Rix}, {Beaton}, {Lewis}, {Nidever}, {Almeida}, {Badenes}, {Barba}, {Beers}, {Carlberg}, {De Lee}, {Fern{\'a}ndez-Trincado}, {Frinchaboy}, {Garc{\'\i}a-Hern{\'a}ndez}, {Green}, {Hasselquist}, {Longa-Pe{\~n}a}, {Majewski}, {Nitschelm}, {Sobeck}, {Stassun}, {Stringfellow}, \& {Troup}}]{PriceWhelan2020}
{Price-Whelan}, A.~M., {Hogg}, D.~W., {Rix}, H.-W., {et~al.} 2020, \apj, 895, 2

\bibitem[{{Sackmann} \& {Boothroyd}(1992)}]{Sackmann1992}
{Sackmann}, I.~J. \& {Boothroyd}, A.~I. 1992, \apjl, 392, L71

\bibitem[{{Salaris} {et~al.}(2015){Salaris}, {Pietrinferni}, {Piersimoni}, \& {Cassisi}}]{Salaris15}
{Salaris}, M., {Pietrinferni}, A., {Piersimoni}, A.~M., \& {Cassisi}, S. 2015, \aap, 583, A87

\bibitem[{{Schiavon} {et~al.}(2017){Schiavon}, {Zamora}, {Carrera}, {Lucatello}, {Robin}, {Ness}, {Martell}, {Smith}, {Garc{\'\i}a-Hern{\'a}ndez}, {Manchado}, {Sch{\"o}nrich}, {Bastian}, {Chiappini}, {Shetrone}, {Mackereth}, {Williams}, {M{\'e}sz{\'a}ros}, {Allende Prieto}, {Anders}, {Bizyaev}, {Beers}, {Chojnowski}, {Cunha}, {Epstein}, {Frinchaboy}, {Garc{\'\i}a P{\'e}rez}, {Hearty}, {Holtzman}, {Johnson}, {Kinemuchi}, {Majewski}, {Muna}, {Nidever}, {Nguyen}, {O'Connell}, {Oravetz}, {Pan}, {Pinsonneault}, {Schneider}, {Schultheis}, {Simmons}, {Skrutskie}, {Sobeck}, {Wilson}, \& {Zasowski}}]{Schiavon2017}
{Schiavon}, R.~P., {Zamora}, O., {Carrera}, R., {et~al.} 2017, \mnras, 465, 501

\bibitem[{{Si} {et~al.}(2015){Si}, {Li}, {Luo}, {Tu}, {Shi}, {Zhang}, {Wei}, {Zhao}, {Wu}, {Wu}, \& {Zhao}}]{Si15}
{Si}, J.-M., {Li}, Y.-B., {Luo}, A.~L., {et~al.} 2015, Research in Astronomy and Astrophysics, 15, 1671

\bibitem[{{Sperauskas} {et~al.}(2016){Sperauskas}, {Za{\v{c}}s}, {Schuster}, \& {Deveikis}}]{Sperauskas}
{Sperauskas}, J., {Za{\v{c}}s}, L., {Schuster}, W.~J., \& {Deveikis}, V. 2016, \apj, 826, 85

\bibitem[{{Spoo} {et~al.}(2022){Spoo}, {Tayar}, {Frinchaboy}, {Cunha}, {Myers}, {Donor}, {Majewski}, {Bizyaev}, {Garc{\'\i}a-Hern{\'a}ndez}, {J{\"o}nsson}, {Lane}, {Pan}, {Longa-Pe{\~n}a}, \& {Roman-Lopes}}]{Spoo22}
{Spoo}, T., {Tayar}, J., {Frinchaboy}, P.~M., {et~al.} 2022, \aj, 163, 229

\bibitem[{{Staff} {et~al.}(2016){Staff}, {De Marco}, {Macdonald}, {Galaviz}, {Passy}, {Iaconi}, \& {Low}}]{staff16}
{Staff}, J.~E., {De Marco}, O., {Macdonald}, D., {et~al.} 2016, \mnras, 455, 3511

\bibitem[{{Stancliffe}(2021)}]{Stancliffe21}
{Stancliffe}, R.~J. 2021, \mnras, 505, 5554

\bibitem[{{Starkenburg} {et~al.}(2014){Starkenburg}, {Shetrone}, {McConnachie}, \& {Venn}}]{starkenburg14}
{Starkenburg}, E., {Shetrone}, M.~D., {McConnachie}, A.~W., \& {Venn}, K.~A. 2014, \mnras, 441, 1217

\bibitem[{{Straniero} {et~al.}(2006){Straniero}, {Gallino}, \& {Cristallo}}]{Straniero}
{Straniero}, O., {Gallino}, R., \& {Cristallo}, S. 2006, \nphysa, 777, 311

\bibitem[{{Suh}(2021)}]{2021ApJS..256...43S}
{Suh}, K.-W. 2021, \apjs, 256, 43

\bibitem[{{Udry} {et~al.}(1998){Udry}, {Jorissen}, {Mayor}, \& {Van Eck}}]{udry98}
{Udry}, S., {Jorissen}, A., {Mayor}, M., \& {Van Eck}, S. 1998, \aaps, 131, 25

\bibitem[{{Wachlin} {et~al.}(2011){Wachlin}, {Miller Bertolami}, \& {Althaus}}]{Wachlin2011}
{Wachlin}, F.~C., {Miller Bertolami}, M.~M., \& {Althaus}, L.~G. 2011, \aap, 533, A139

\bibitem[{{Whitehouse} {et~al.}(2018){Whitehouse}, {Farihi}, {Green}, {Wilson}, \& {Subasavage}}]{Whitehouse}
{Whitehouse}, L.~J., {Farihi}, J., {Green}, P.~J., {Wilson}, T.~G., \& {Subasavage}, J.~P. 2018, \mnras, 479, 3873

\bibitem[{{Yamashita}(1975)}]{Yamashita}
{Yamashita}, Y. 1975, \pasj, 27, 325

\bibitem[{{Yoon} {et~al.}(2018){Yoon}, {Beers}, {Dietz}, {Lee}, {Placco}, {Da Costa}, {Keller}, {Owen}, \& {Sharma}}]{Yoon18}
{Yoon}, J., {Beers}, T.~C., {Dietz}, S., {et~al.} 2018, \apj, 861, 146

\bibitem[{{Yuan} {et~al.}(2015){Yuan}, {Liu}, {Xiang}, {Huang}, \& {Chen}}]{Yuan2015}
{Yuan}, H., {Liu}, X., {Xiang}, M., {Huang}, Y., \& {Chen}, B. 2015, in IAU General Assembly, Vol.~29, 2254968

\bibitem[{{Zamora} {et~al.}(2009){Zamora}, {Abia}, {Plez}, {Dom{\'\i}nguez}, \& {Cristallo}}]{zamora2009}
{Zamora}, O., {Abia}, C., {Plez}, B., {Dom{\'\i}nguez}, I., \& {Cristallo}, S. 2009, \aap, 508, 909

\end{thebibliography}

\begin{appendix}
\section{Discounting of IRAC magnitudes for AGB elimination}\label{App:IRAC}
\begin{figure*}
\centering
\includegraphics[width= 0.95\textwidth]{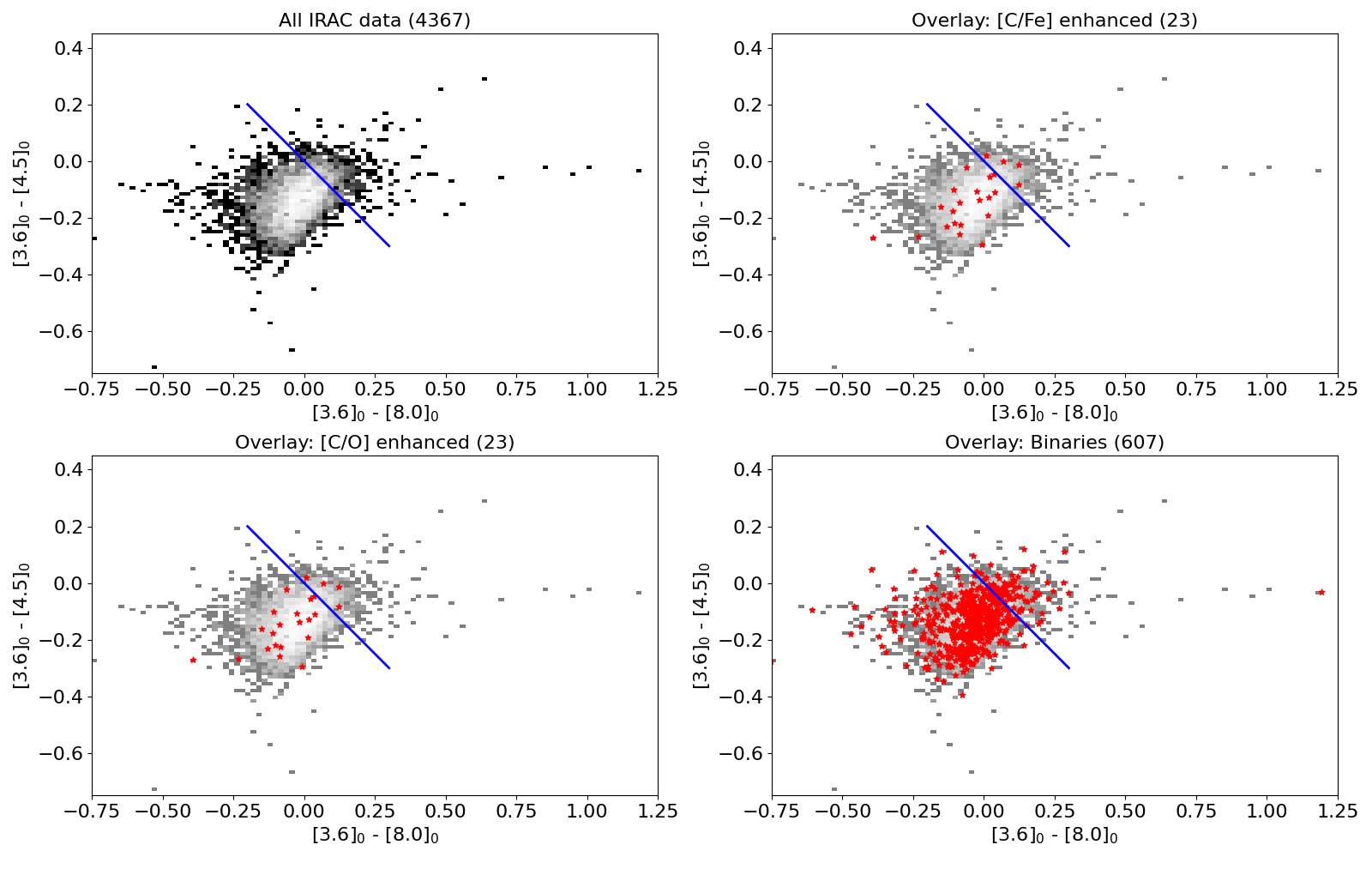}
\caption{Plots of pre-filtered APOGEE DR17 stars with IRAC magnitudes available on dereddened colour excess diagrams. Upper left plot shows the base population with blue line indicating potential AGB stars to the right of the line. Lower left plot is overlaid with [C/O] enhanced stars as discussed in Sect. \ref{sec:crich}.  Upper right plot is overlaid with stars having high [C/Fe] on a similar basis. Lower right plot is overlaid with binaries. In the three latter cases, the number of stars in the overlaid plots follows the split of the underlying population to within 1.1$\sigma$}
\label{fig:IRACAssess}
\end{figure*}

There is a potential fifth method that could be used that is also based on mid-IR colour excess using Spitzer IRAC magnitudes at 3.6, 4.5 and 8.0 microns.  
However, only about 8\% of stars in APOGEE DR17 have IRAC magnitudes. 
To limit the stellar sample to just these stars would produce an unusably small data set.
Instead, we consider the question of whether ignoring the IRAC data altogether introduces any bias in the sample. 
To this end, all stars with IRAC magnitudes were plotted on colour excess diagrams with stars that are high in [C/Fe], high in [C/O] and are binaries.
These are segregated into potential AGB and non-AGB using the two-axis IR colour excess criteria from \cite{Schiavon2017}. 
In each case, the highlighted number of stars in the AGB sample and non-AGB sample is consistent with the numbers of the underlying IRAC population in each category (see Fig. \ref{fig:IRACAssess} and Table \ref{tab:IRACstats}). 
Thus, at least for this sample, there seems to be no significant bias in carbon abundance or binarity introduced by position on the IRAC colour-excess diagram.
For this reason, the authors have elected not to use the IRAC method to eliminate potential AGB stars. 
Some contamination of the stellar sample with AGB stars probably remains, but that does not appear to bias the distributions of abundances or binarity properties of the sample.

\begin{table}[h!]
\caption{Distribution of stars with IRAC magnitudes}
    \centering
    \resizebox{\columnwidth}{!}{
    \begin{tabular}{l r r r} 
         {} & Count & Percentage & $\sigma$ \% \\
         \hline
         {Base population AGB stars}& 539 & 12.3\% &\\
         {Base population non-AGB stars}& 3,828 & 87.7\% & \\
         & & &\\
         {High [C/Fe] stars in AGB region}& 4 & 17.4\% & 7.9\%\\
         {High [C/Fe] stars in non-AGB region}& 19 & 82.6\% & \\
         & & &\\
         {High [C/O] stars in AGB region}&  4  & 17.4\% & 7.9\%\\
         {High [C/O] stars in non-AGB region}& 19 & 82.6\% \\
         & & &\\
         {Binary stars in AGB region}&  84 & 13.8\% & 1.4\%\\
         {Binary stars in non-AGB region}&  523 & 86.2\% &\\
         & & &\\
    \end{tabular}
    } 
    \tablefoot{Base population consists of stars in initial sample that all have IRAC magnitudes and are divided into AGB and non-AGB based on IR excess. Only about 8\% of stars in the parent DR17 catalogue have IRAC magnitudes. Preliminary filtering on \lg, $T_{\rm eff}$ and other criteria as explained reduce this percentage to 4,367 stars or 7.4\% in this initial sample. The last column gives the standard deviation based on an assumed uniform binomial distribution and expressed as a percentage of the total count. The number of high [C/Fe], high [C/O]  and binaries, split between AGB and non-AGB follows the distribution of the base population to within 1.1$\sigma$. This demonstrates that no significant bias is being introduced by ignoring the Mid IR excess method.}
    \label{tab:IRACstats}
\end{table}
\FloatBarrier

\section{Carbon enhancement vs. oxygen depletion}\label{app:OFe}
\begin{figure*}
    \centering
    \includegraphics[width= 0.95\textwidth]{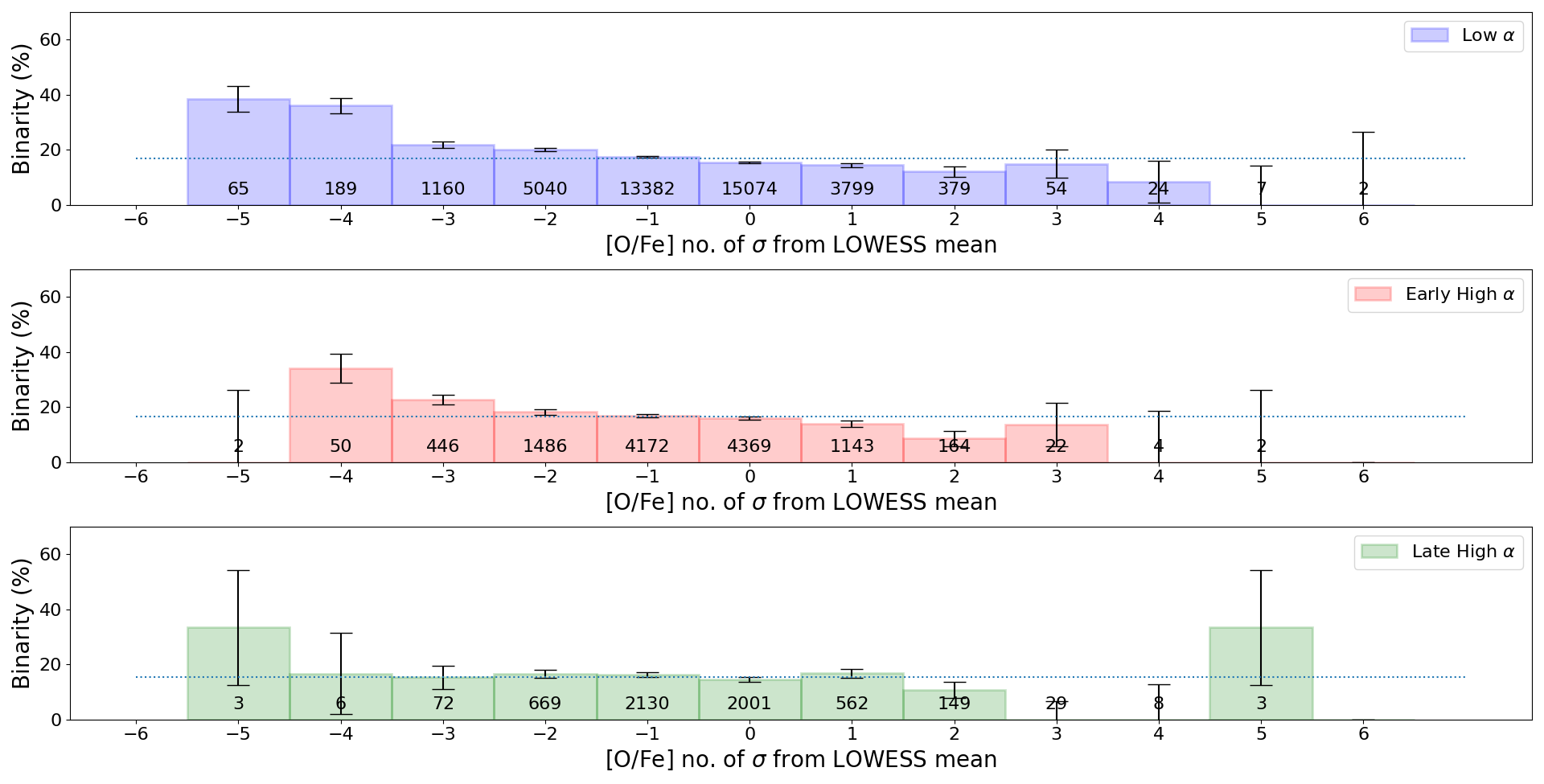}
    \caption{Analogous to Fig. \ref{fig:BinarityHistos}, we show binary fraction against number of standard deviations  of [O/Fe] abundance from the mean for stars of the same population and similar [Fe/H]. There is a narrower range of deviation in each population and only very weak trends of increasing binary fraction with [O/Fe] depletion.}
    \label{fig:OFeBinarityHistos}
\end{figure*}

In Secs. \ref{sec:conclusions} and \ref{sec:bintrends}, we discuss the possibility that high [c/O] deviation from the population mean for a given metallicity could, in certain circumstances, be caused in [C/Fe]-poor stars by extreme depletion in [O/Fe]. for completeness, here we show in Fig. \ref{fig:OFeBinarityHistos} binary fraction against degree of [O/Fe] deviation from mean for the population and metallicity bin. We note that there is both less variation in [O/Fe] and only a very weak trend of increasing binary fraction with [O/Fe] depletion.  

\end{appendix}

\end{document}